\documentclass[fleqn,usenatbib]{mnras}
\usepackage[T1]{fontenc}
\DeclareRobustCommand{\VAN}[3]{#2}
\let\VANthebibliography\thebibliography
\def\thebibliography{\DeclareRobustCommand{\VAN}[3]{##3}\VANthebibliography}
\usepackage{graphicx}	% Including figure files
\usepackage{amsmath}	% Advanced maths commands
\usepackage{amssymb}	% Extra maths symbols
\usepackage{xcolor}
\usepackage{float}
\usepackage{newtxtext,newtxmath}
\title[Long-term profile evolution]{The Thousand-Pulsar-Array programme on MeerKAT - XII.\newline Discovery of long-term pulse profile evolution in 7 young pulsars}

\author[Basu et al.]{A. Basu,$^{1}$\thanks{E-mail: avishek.basu@manchester.ac.uk}
P. Weltevrede,$^{1}$
M.~J.~Keith,$^{1}$
S.~Johnston,$^2$
A.~Karastergiou,$^{3}$
L.~S.~Oswald,$^{3,4}$ 
B. Posselt,$^{3,5}$ \newauthor
X.~Song$^{6}$ 
and A.~D.~Cameron$^{7,8}$
\\ \\
$^{1}$Jodrell Bank Centre for Astrophysics, School of Physics and Astronomy, University of Manchester, Manchester, UK, M13 9PL\\
$^2$Australia Telescope National Facility, CSIRO Space and Astronomy, PO~Box~76, Epping NSW~1710, Australia\\
$^{3}$Department of Astrophysics, University of Oxford, Denys Wilkinson Building, Keble Road, Oxford OX1 3RH, UK\\
$^{4}$Magdalen College, University of Oxford, Oxford OX1 4AU, UK\\
$^{5}$Department of Astronomy \& Astrophysics, Pennsylvania State University, 525 Davey Lab, 16802 University Park, PA, USA\\ 
$^{6}$ASTRON, The Netherlands Institute for Radio Astronomy, Oude Hoogeveensedijk 4, 7991 PD, Dwingeloo, The Netherlands\\
$^{7}$ Centre for Astrophysics and Supercomputing, Swinburne University of Technology, PO Box 218, VIC 3122, Australia.\\
$^{8}$ ARC Centre of Excellence for Gravitational Wave Discovery (OzGrav), Swinburne University of Technology, PO Box 218, VIC 3122, Australia.\\
}

\date{Accepted 2024 February 05. Received 2024 February 05; in original form 2023 September 13}

\pubyear{2023}

\begin{document}
\label{firstpage}
\pagerange{\pageref{firstpage}--\pageref{lastpage}}
\maketitle

% Abstract of the paper
\begin{abstract}
A number of pulsars are known to have profile evolution on timescales of months, often correlated with spin-down rate changes. Here, we present the first result from 3 years of monitoring observations from MeerKAT as part of the Thousand Pulsar Array programme. This programme obtains high-fidelity pulse profiles for $\sim$ 500 pulsars, which enabled the detection of subtle changes in seven sources not previously known to exhibit long-term profile evolution. A 2D Gaussian convolution is used to highlight correlated emission variability in both the pulse phase and observing epoch direction. Simulations show that for one additional source the observed profile variability is likely to originate from stochastic single-pulse shape variability (jitter). We find that it is common for long-term profile variability to be associated with changes in polarization fractions, but not with polarisation position angle (PA) changes. PA changes are expected if emission height changes or precession is responsible for the profile variability. PSR J1741$-$3927 is the only pulsar in our sample that shows correlated PA variability, and this is associated 
with orthogonal polarization mode activity. For the six
%the rest, without correlated PA variability, 
other pulsars limits on possible emission height changes and impact angle changes are derived. These limits are consistent with the small changes in the total intensity profile shape. None of the sources show detectable spin-down variability correlated with the emission changes, which are thought to be driven by magnetospheric current fluctuations. Therefore the absence of correlated spin-down rate variability allows upper limits to be placed on changes in the magnetospheric charge density. 
\end{abstract}

% Select between one and six entries from the list of approved keywords.
% Don't make up new ones.
\begin{keywords}
methods: data analysis -- methods: statistical
-- techniques: polarimetric -- stars: neutron -- pulsars: individual: PSR J0729$-$1448, J1121$-$5444, J1141$-$3322, J1705$-$3950, J1741$-$3927, J1844+1454, J1919+0951 and J1919+0021
%keyword1 -- keyword2 -- keyword3
\end{keywords}

%%%%%%%%%%%%%%%%%%%%%%%%%%%%%%%%%%%%%%%%%%%%%%%%%%

%%%%%%%%%%%%%%%%% BODY OF PAPER %%%%%%%%%%%%%%%%%%

\section{Introduction} \label{intro}
Pulsars are often considered to be objects with extremely stable rotation and profile shapes. Although often true, this cannot be generalized for all pulsars. A small fraction of pulsars, especially the young to middle-aged part of the population, and also some old recycled/millisecond pulsars,  is known to exhibit long-term correlated structures in their timing residuals known as timing noise \citep{HLK+2010, Pathasarathy+2019}. This indicates some unmodelled rotational variations. Various physical origins for timing noise have been proposed, which includes the recovery from unseen glitches\footnote{Glitches are sudden rapid spin up of pulsars.}\citep{Alpar+1986}, multiple micro-glitches \citep{cordes+down+1985,Janssen+Stappers+2006} or turbulence in the interior superfluid of neutron stars \citep{Melatos+2014}.

From the early days of pulsar astronomy pulsars are known to be variable. Changes in rotation rate of pulsars were known to give rise to timing noise. But also the radio pulse profile shape of some pulsars were known to be variable. For example, {\it mode-changing} is a well-known phenomenon, where the pulse shapes change between two or more different shapes \citep{Backer+modechanging1970, Lyne+1971} with a timescale of $\sim$ few tens of pulse periods to a few hours. {\it Nulling} \citep{Backer+1970}, where the emission suddenly ceases for a few or many pulse periods before resuming back to its normal emission mode, is a phenomenon which could 
%possibly 
be considered an extreme form of mode changing. Apart from these variations on a timescale of at least a few pulse periods, every single pulse is different in its shape and intensity (e.g. \citealt{Helfand+Manchester+Taylor+1975,Rathnasree+Rankin+1995}) which is known as {\it jitter}. These phenomena highlight the dynamic nature of the pulsar magnetosphere and the radio emission mechanism. 

In the last couple of decades, much longer time scale profile shape changes have been discovered in pulsars. For some sources, the profile changes are associated with correlated changes in the spin-down rate.
For example, the intermittent pulsar B1931+24 \citep{Kramer+Lyne+2006} was discovered to exhibit two distinct emission states where it behaved like a regular pulsar for $\sim$ 5-10 days (known as the ``on'' state of the pulsar) with a 50\% higher spin down rate than the ``off'' state for which the radio emission is switched off. The switch happens within 10 seconds and the pulsar maintains radio silence for the next 25-35 days. The change in the magnetospheric current, which impacts both the emission and torque was identified as a cause of such correlated change in the spin-down and emission. Later, \citet{Lyne+2010} discovered six more pulsars which exhibit correlated spin-down and profile shape changes. They concluded that also for these pulsars the magnetosphere switches abruptly between two states. In some pulsars the variabilities are associated with the central component of the pulse profile. The central part of the profile is often associated to core emission, as opposed to the more peripheral conal emission  \citet{Rankin1990core, Rankin1993cone, Rankin+2022}.
However, more complex behaviour than a simple two-state system was reported in PSR B0919+06 \citep{PSW+2015}. 
In contrast, a slow magnetospheric change was seen in PSR J0738$-$4042, where the evolution of the pulse profile was hypothesised to be triggered by an asteroid's interaction with the pulsar magnetosphere \citep{Brook+2014}.
Furthermore, long timescale emission variability was established in millisecond pulsars \citep{brook+nanograv}. This relied on Gaussian process regression, a technique proven to be useful for the analysis of Murriyang, Parkes 64m radio-telescope data \citep{Brook+2016}, as well as for the analysis of Jodrell Bank Observatory data \citep{Shaw+2022}.

%A complementary example of a 

The different emission states suggest a pulsar magnetosphere can have different meta-stable states that transition stochastically between local minima of some effective potential \citep{Cordes+2013}. According to this formalism, the presence of two states is considered common, but for a few sources, a larger number of states are possible. The existence of several such meta-stable magnetospheric configurations has been speculated in \citet{Timokhin+2010} to be strongly dependent on the physical conditions prevailing near the polar cap particle cascade zone and the ``Y'' point\footnote{Y-point is the region where the current sheet from the outer magnetosphere reaches the last closed field line.}, of which the latter regulates the size of the magnetosphere and currents flowing within it. The combined effect of the different magnetosphere sizes and current density distribution further determines the total energy of the system whose local minima may correspond to various observable meta-stable states.

PSR B1828$-$11 is the first source discovered with quasi-periodic changes in the profile shape that are correlated with spin-down rate variability \citep{Stairs+lyne+2000}, which was initially thought to be due to the free precession in the pulsar. Later, the discovery of different emission modes in this pulsar was used as an argument against the free precession model, hence the magnetospheric state change interpretation was favoured \citep{Stairs+Lyne+2019}. Though the classical free precession model was not enough to explain the observations, \citet{Jones+2012} propose a hybrid model involving both precession and magnetospheric state changes. The magnetosphere is delicately balanced between two states and the state changes are regulated by the precession effects. \citet{Kerr+2016} arrived at a similar conclusion after a detailed analysis of periodic modulation in 7 sources identified in a sample of 151 young energetic pulsars, where also they propose the magnetic stress supported by the superconducting core could provide the ellipticity required for the precession. This model removed the mutual exclusiveness between magnetospheric processes and precession as explanations for correlated spin-down and profile variability. By modelling the spin-down ($\dot \nu$) time-series of PSR B1828$-$11 \citet{AJP+2017} concluded that the precession model along with the external electromagnetic torque is somewhat favoured over the pure magnetospheric switching model for this pulsar. 

The presence of timing noise implies the total torque acting on a neutron star must vary over time, which can have contributions both from the interior and exterior of a pulsar (its magnetosphere)  \citep{ABH+2022}. The intrinsic fluctuation in the vortex-mediated process inside neutron stars leads to internal torque fluctuations \citep{Alpar+1986} and the external torque can be regulated by fluctuation in the current density as discussed above. Therefore the sources showing correlated emission and rotation variability may have a much larger contribution from the magnetospheric current fluctuations towards their timing noise, which also impacts their emission properties. 
%The previous work by \citealt{Lyne+2010, Lyne+2013} indicates that many pulsars might have similar behaviour but difficult to detect such correlated variations due to observational limitations (like sensitivity and observational cadence). 
However, the magnitude of the current fluctuations is not known a priori. A tiny fluctuation in current may lead to a strong change in the emission process due to non-linear plasma processes,
%associated with the pulsar radio emission) 
but not in the spin-down torque or maybe both. 
%Hence, it becomes reasonable to search for any subtle changes in the emission from pulsars using telescopes with much higher sensitivity.

%{\bf write a sentence to bridge between the two.}
In order to detect potentially subtle changes in the emission state, we make use of the highly sensitive MeerKAT telescope. 
%The observations have a span of $\sim$ 3 years on $\sim$ 500 pulsars with somewhat variable cadence.
Its sensitivity and capability to perform sensitive polarisation measurements are used to search for correlated emission and spin-down rate changes. In this paper, we present data of 7 sources with long-term profile evolution, which were not known to exhibit such variations. For one additional source, apparent profile variability is associated with single-pulse shape variability. A follow-up publication is planned which will discuss the implications of similar analysis on a much bigger sample of pulsars.
In Sec. \ref{obs} we describe the monitoring programme, in Sec.\ref {analysis} we explain the analysis methodology and in Secs. \ref {result} and \ref{discussion} present and discuss the results of the analysis. This is followed with the conclusions in Sec. \ref{conclude}.

\section{Observations}\label{obs}
\begin{table}
\begin{center}
\begin{tabular}{llllll}
\hline \hline
PSR J & PSR B& $P$ (s) &$\mathrm{N_{obs}}$& $\Tilde{T}_\mathrm{int}$ (min) & $\widetilde{\Delta T}$ (days)\\
\hline \hline
J0729$-$1448 & $-$        &0.25 &42  & 1.5  & 30 \\
J1121$-$5444 & B1119$-$54 &0.53  &43  & 2.0  & 26 \\
J1141$-$3322 & $-$        &0.29  &45  & 1.5  & 25 \\
J1705$-$3950 & $-$        &0.31  &50  & 2.0  & 27 \\
J1741$-$3927 & B1737$-$39 &0.51  &12  & 1.8  & 27 \\
J1844+1454 & B1842$+$14 &0.37  &44  & 1.5  & 27 \\
J1916+0951 & B1914+09   &0.27  &45  & 1.5  & 27 \\
J1919+0021 & B1917+00   &1.27  &45  & 2.0  & 26 \\
\hline
\end{tabular}
\end{center}
\caption{In this table we summarise details of the observations reported in this paper. The first and second column shows the J and B name of the pulsars respectively. The third column shows the pulse period ($P$) of the pulsars in sec.
%, these values are taken from the \href{https://www.atnf.csiro.au/research/pulsar/psrcat/}{ATNF pulsar catalogue} \citep{ATNF}. 
The fourth column indicates the number of epochs over which the observations were done with a median integration length $\Tilde{T}_\mathrm{int}$, shown in the fifth column, and median cadence $\widetilde{\Delta T}$, shown in the sixth column.\label{obstab}}
\end{table}

The Thousand Pulsar Array (TPA) programme \citep{Johnston+2020} is a part of the large pulsar monitoring project using MeerKAT, called MeerTime\citep{Bailes+2019}. The TPA has observed over a thousand
%young 
%pulsars and $\sim$500 of them are monitored regularly at a cadence of at least once every month.
pulsars and $\sim$500 of them are monitored
with somewhat variable cadence, typically at least once every month.
%The observations have a span of $\sim$ 3 years on $\sim$ 500 pulsars 
The TPA sample excludes millisecond pulsars. Their astrometric positions are confined below a declination of $+20^\circ$.
%within the southern hemisphere. 
A broader discussion of the TPA project and its scientific goals 
%and its first results 
can be found in the paper by \citet{Johnston+2020}. The results reported in this paper constitute of 8 sources 
%out of $\sim$500 pulsars which are monitored using MeerKAT.
%in a sub-array mode. 
%Pulsars with slow profile evolution were identified,
%that can be resolved by the ∼ monthly observing cadence. The eight
for which an initial visual inspection suggested slow profile evolution not reported before in the literature. It will be shown that for one of them, the evolution is likely caused by single pulse variability.

The MeerKAT telescope is an array of 64 dishes located in the Karoo region of South Africa. To improve the efficiency of the TPA's pulsar monitoring programme, 
%In the sub-array mode of observations, 
the available antennas are divided into two groups, each containing up to 32 antennas. The signals of the telescopes in each sub array are coherently combined to produce a tied-array beam. Therefore, the two sub arrays allows two different patches of the sky to be observed at the same time. 
%Despite the 
%cost of a 
%factor of 2
%$\sqrt{2}$ 
%degradation in sensitivity, it allows us 
The degradation in sensitivity by a factor of 2 is counterbalanced by the opportunity to observe almost twice the number of pulsars compared to what is achievable with the 
%could have been observed using the 
full array.
%in a given amount of time. 
A detailed discussion on the effectiveness of the observing strategy for the TPA programme 
%has been presented in the paper by
can be found in \citet{Song+2021}, which proposes a methodology to design an observing programme aiming to be sensitive for pulse profile variability.
%obtain an estimate of integration time in order to detect changes in pulse profile at a 10\% level. 

The observations presented here are obtained with the L-band receiver. Signals covering a bandwidth of 775~MHz centred at 1283~MHz were recorded with full polarisation information. The large bandwidth in combination with a low receiver temperature of 18 K
%in the phased array mode 
makes MeerKAT a very sensitive telescope to observe high-fidelity pulse profiles. Our analysis makes use of the integrated pulse profiles produced from the channelized data after de-dispersion.
%time series was collapsed to a single channel keeping the polarisation information intact. 
The number of bins across the pulse profile was chosen optimally to enhance the signal-to-noise (S/N) ratio per bin, without compromising details of the pulse profile morphology \citep{Song+2023}.
In Table \ref{obstab} we have summarised the details of the observations for the 8 sources reported in this paper.

\section{Analysis}\label{analysis}

The analysis was performed using a combination of 
%the {\sc Python} interface of the package 
{\sc psrchive}\footnote{\href{http://psrchive.sourceforge.net/}{http://psrchive.sourceforge.net}} \citep{hvsm04}
%. The difference maps were generated using the software package 
and {\sc psrsalsa}\footnote{\href{https://github.com/weltevrede/psrsalsa}{https://github.com/weltevrede/psrsalsa}} \citep{psrsalsa+2016}, using the following  procedure.

\subsection{Emission variability and pulse jitter simulation}\label{ppv}
The variability in the shape of the pulse profile has been studied by investigating the structures in the difference map (as shown in left-hand side panel of Fig.~\ref{fig:0729diffmap}, which will be explained in more detail in Sec.~\ref{0729}) produced from the time and frequency averaged 
%{\bf [There is a comment here that should be resolved.]}
pulse profiles, as well as various polarisation products.
In order to obtain the total intensity difference map, the observed pulse profiles are first aligned using cross-correlation with a noise-free template of the pulse profile obtained from the Gaussian process models as described in \citet{Posselt+2023}. Furthermore, the profiles are scaled with respect to the noise-free template. The scaled flux density of profile bin $i$ is
\begin{equation}\label{normalisation_scheme}
I^\mathrm{norm}_i = \left ( \frac{\sum_j T_j}{\max \{T\}} \right) \left (\frac{I_i}{\sum_j I_j} \right) = f_s I_i,
\end{equation}
where $T$ is the noise-free profile template, and the index $j$ runs over all the on-pulse phase bins as determined from the template. 
This scaling makes the average flux density of each observed profile identical, with an average set such that the peak intensity is $\sim1$. The actual resulting peak flux density depends on the profile shape, but if the profile shape is identical to that of the template, the peak flux density would be exactly $1$.

After scaling, the median observed profile is subtracted from each observation, and the resulting profile residuals are stacked resulting in what we will refer to as the ``difference map''. This is used to reveal potential systematic profile evolution.
Care should be taken to distinguish these systematic long-term variations from more stochastic variability arising from pulse-to-pulse jitter.
%(see also Section \ref{intro}). 
To quantify the expected nature of jitter-induced variability, 
%as a control set to the difference map constructed from the real profiles obtained from observations over multiple epochs, 
we also simulate a difference map constructed from profiles affected by the observed single-pulse jitter.
%that can account for the jitter variations. 
The input for this simulation is the sequence of single pulses observed with MeerKAT, as used by \citet{Song+2023}. For all sources studied here, this corresponds to the longest available observation.

In order to consider time-dependent single pulse phenomena such as mode-changing, nulling, (slowly) drifting subpulses, we have adopted two different methods to obtain simulated difference maps. One of them we refer to as the \textit{individual pulse method}, and the other as \textit{block method}. These will be discussed in more detail in Sec.~\ref{IPmethod} and \ref{Bmethod} respectively.
Comparison of the observed and simulated difference maps allows systematic slow profile evolution to be distinguished from profile variability arising from single-pulse variability within the relatively short observations.

\subsubsection{Jitter Simulation: Individual pulse method}\label{IPmethod}
%Even without phenomena like mode-changes, 
The individual pulses from a pulsar 
%will have 
shows stochastic variations in their shape and intensity \citep{Helfand+Manchester+Taylor+1975, Rathnasree+Rankin+1995}. Averaging over a certain number of pulses (usually a few hundred to thousands) is required  to achieve a stable pulse profile. Therefore, a finite observation duration may, or may, not record enough number of pulses for a given pulsar to achieve a stable pulse shape. Therefore it becomes crucial to identify if any profile variability seen is associated with jitter or genuine long-term variation of the profile. Hence, a simulation to capture variations that arise from jitter 
%A second type of jitter simulation 
was performed under the assumption this variability has no memory in the sense that the emission variability is uncorrelated in consecutive pulses. This allows us to randomly select pulses from the full single-pulse observation, without the requirement that they should be in sequence. In order to simulate the expected effect of jitter, difference maps were constructed from the relatively long MeerKAT observations analysed in \citet{Song+2023}. 
To capture the effect of short timescale pulse shape variability, we replace each observation with a pulse profile obtained from averaging a random selection of pulses from this longer dataset. Each observation is replaced with a simulation of equal duration.
%equal to the actual observations at every epoch. 
The obtained profiles are processed in the same way as the actual observations, including alignment and scaling as discussed in Sec.~\ref{ppv}. This results in difference maps which can be compared 
%{\bf qualitatively 
visually, in terms of the magnitude of differences and the appearance of any emerging structures, to those constructed from the actual observations. Therefore, this method helps in identifying the variations that might arise from the uncorrelated jitter process between pulses.

%{\bf doing both is important to disentangle the various time-scale associated with the profile change... also in block method all block are not all not independent.}

\subsubsection{Jitter Simulation: Block method}\label{Bmethod}
Emission variabilities like mode-changing, nulling and drifting subpulses are associated with timescales of typically tens to hundreds of pulse periods, which can be shorter than the length of our observations. Hence, when a pulsar which exhibits such phenomena is observed, its profile shape is affected in a way %This can be stochastic in nature, or itcould 
which depends on the proportion of the emission states captured. Since the variability can have memory, in the sense that if one pulse is in a different emission state, the next is more likely to also be in a different emission state,
%To quantify the latter, 
the block method as explained here, can be more applicable.
In this method, a block of consecutive pulses is chosen with an equal length compared to the original observation. The consecutive pulses are kept in the original order, but with a random start pulse number.
%the pulses are randomly chosen as sequences of consecutive pulses, 

Otherwise, the performed analysis is identical to that in Sec.~\ref{IPmethod}. Therefore, this simulation helps in identifying structures that arise from short-term correlated changes in the emission process. 
%It is to be noted that this method also has a limitation due to the single pulse stack being not infinitely long, the blocks are not completely independent of each other.
A drawback of this method is that for a finite data set, the selected blocks are not completely independent despite a random start pulse number. Therefore it is beneficial to use both jitter simulations side-by-side when judging the possible effect of pulse jitter on the difference maps.

\subsection{Polarisation}\label{polanalysis}
Full Stokes data were recorded for our observations which enables us to study the variability in the polarised signal from the pulsar. Therefore, difference maps were not only produced in Stokes $I$ (total intensity), but also in $L/I$, (where $L$ is the linear polarized intensity), $V/I$ (where Stokes $V$ is the circular polarization) and $\psi$ (position angle of the linear polarization).
Following \citet{wk74,EW+2001},  a bias correction on $L$ was applied. 
To construct the $\psi$ difference map, care must be taken to consider the effect of Faraday rotation. This is quantified with the rotation measure, RM, which is dominated by an ISM contribution.
De-Faraday rotation was performed on every data set with the RM measured from the MeerKAT data \citep{Posselt+2023}. The position angle has a strong dependence on rotational phase. To compute the difference map in $\psi$, and highlight any variations
%in the polarisation position angle
over time, the typical rotational phase dependence obtained from combining the profiles from the individual observations is subtracted. After aligning and scaling of the profiles (see Sec. \ref{ppv}), for each rotational phase bin $j$ the median Stokes parameters $Q_j$ and $U_j$ are determined. This corresponds to the typical position angle swing $\psi_j =0.5 \tan^{-1} (U_j/Q_j)$. To subtract this from the observations, the Stokes parameters $Q$ and $U$ of each bin in each observation are rotated by $2\psi_j$.

Since the ISM contribution to the RM, as well as the ionospheric component, is in general time-dependent, the performed de-Faraday rotation with a fixed RM will not be perfect. The effect is that there can be offsets in $\psi$ from epoch to epoch. 
Since the offsets are constant in rotational phase, any left-over effect of Faraday rotation can be removed by ensuring the phase-averaged $\psi$ is zero in each observation. To do this, the on-pulse\footnote{The on-pulse region was determined using an identical procedure as explained in \cite{Song+2023}.} averaged $Q$ and $U$, after subtraction of the typical position angle swing, were determined for each profile. The corresponding phase-averaged $\psi$ was subtracted analogous to how the typical rotational phase dependent $\psi$ was subtracted.

\subsection{Sensitivity for profile polarization variations}
\label{SectPolarizationVariationSensitivity}
Especially when systematic profile variation is detected in Stokes $I$, but not in $L/I$ and $V/I$, it is useful to set limits on what this implies for possible changes in polarization. The following procedure is adopted.

Each profile is modelled analytically with a number of von Mises components. This is done for all Stokes parameters separately. This allows the pulsar signal to be subtracted from each observation, leaving noise, any remaining low-level RFI and some profile variability likely caused by jitter. As will be explained in the following, a simulated polarized signal will be added to these residuals, allowing us to judge under what assumptions about the profile variability a detectable signature in the $L/I$ and $V/I$ difference maps are to be expected.

To model the profile variability, each profile in Stokes $I$ is modelled with von Mises components: most are to describe the non-varying pulse profile (analytic profile A), and some to describe the varying component of the emission (analytic profile B). 
Each individual profiles (in Stokes $I$) is fitted as a linear combination of profiles A and B. 

The sensitivity to detect variability in polarization is quantified with the parameters $f_{L/I}$ and $f_{V/I}$. 
These represent the fractional difference in the polarization fraction between the non-varying and varying component of the emission. So when they are unity, $L/I$ and $V/I$ will remain constant despite the changes in Stokes $I$. The parameters $f_{L/I}$ and $f_{V/I}$ are used to simulate changing profile shapes in full polarization. The simulated profiles are added to the profile residual (noise) which was obtained after subtracting the modelled pulse shape. The resulting simulated polarized profiles with noise are used to calculate difference maps with the same methodology as used for the actual data. From visual inspection of the obtained difference maps limits on the required $f_{L/I}$ and $f_{V/I}$ to detect changes in polarization are found.

\subsection{Timing and $\dot \nu$ variations}\label{timing}
To quantify the spin-down evolution, the observed profiles were first cross-correlated in the frequency domain with their noise-free template 
%$T$ 
\citep{taylor1992} to measure the topocentric time of arrivals (TOAs).
%using the code {\sc pat} of {\sc prschive}. 
A phase-coherent timing solution was obtained by fitting the measured TOAs using the pulsar timing package {\sc tempo2}\footnote{\href{https://bitbucket.org/psrsoft/tempo2}{https://bitbucket.org/psrsoft/tempo2}} \citep{tempo2I}. Initial timing solutions were obtained from \citet{JSD+2021}, and from the work related to the initial timing results of the full TPA monitoring data set (Keith et al. in prep). In addition, for PSR J0729$-$1448 we have derived the timing solution from a Jodrell Bank Observatory data set.
%However, 
After fitting for a spin period and spin-down rate,
%even after fitting for the rotational parameters given in Equation \ref{timingmodel}, the residuals clearly exhibit 
most pulsar have systematic red-noise patterns in their residuals. This implies un-modelled rotational behaviour, which is also known as timing-noise.
Noise modelling is carried out simultaneously with the fitting for the timing model parameters, and done using \textsc{run\_enterprise} \citep{run_enterprise}, which is based on the \textsc{enterprise} framework \citep{evt+19}, and uses \textsc{tempo2} to fit the pulsar timing parameters (the spin frequency and spin-down rate). 
%\textbf{MJK: Which parameters did you fit for?}
Following the model of \citet{Lentati+2014} the red noise is assumed to be a power law in the frequency domain, with power spectral density characterised by spectral index, $\gamma$, and amplitude, $A_{\rm{red}}$, and given by
\begin{equation}\label{timingnoisepower}
    P(f) = \frac{A_{\rm{red}}^2}{12\pi^2} \left(\frac{f}{\rm{yr}^{-1}} \right )^{-\gamma}.
\end{equation}
White noise is modelled using the  EFAC and EQUAD parameters,  which are a multiplicative factor and quadrature additive to the TOA uncertainties.
Both the red noise and the white noise parameters were sampled with the Markov-chain Monte Carlo (MCMC) technique using {\sc emcee} \citep{Foreman_Mackey_2013}.

The time-variable spin-down rate, $\dot \nu (t)$ can be derived from the second derivative of the red noise model, $r(t)$, by
\[
\dot{\nu}(t) = \mathrm{F1}+\mathrm{F2} t-\mathrm{F0}\,\ddot{r}(t),
\]
where $\mathrm{F0}$, $\mathrm{F1}$ and $\mathrm{F2}$ are the maximum likelihood values of the Taylor coefficients in spin frequency from the pulsar rotational ephemeris.
We compute both $\dot \nu (t)$ and its error analytically using the method described in \citet{kn23} as implemented in the \texttt{make\_pulsar\_plots} utility from \textsc{run\_enterprise}.

\subsection{Correlation Analysis}\label{correlation-analysis}
%\subsection{Profile shape parameters} 
\label{shapeparam}
In order to quantify the correlation between emission changes, as identified in the difference maps, and the spin-down rate of the pulsar, we first 
%try to 
capture the emission changes with a shape parameter.
This could be the ratio of amplitudes of various profile components, the width of the profile at a certain level compared to the peak flux density, %from the peak 
or 
%even 
the (normalised) peak flux density itself. The shape parameter is a single number, quantifying an aspect of the shape of the profile at each observing epoch.
Following the same methodology as used in \cite{Song+2021}, the shape parameters are derived from an analytic description of the pulse profile in the form of a sum of a set of von Mises functions. By perturbing the observed profiles with the off-pulse noise level, errors in the shape parameters are deduced.

To quantify the correlation between emission state changes and spin-down variability a
Pearson's correlation analysis is performed between the $\dot \nu$ (Sec.~\ref{timing}) and shape parameter 
%(Section \ref{shapeparam}) 
time series. It is also essential to capture the uncertainty arising from other uncorrelated processes such as pulse jitter, which is not correlated with time. Therefore, we have also established a confidence in the measured correlation coefficient by randomising the order of both  time series 10$^{5}$ times. The width of the measured distribution from this uncorrelated process is quoted as the uncertainty of the correlation coefficients quoted in Sec.~\ref{result}.

\subsection{Difference map smoothing}\label{diffsmooth}
Gaussian process regression has been used extensively in the literature to smooth difference maps \citep{Brook+2016, Brook+2018_ApJ, Shaw+2022}. 
This enables the prediction of intensity variations between observation epochs, resulting in a map of profile variability as a function of time.
Here we follow the technique as presented in \citet{Shaw+2022}. The Matern covariance function is used with two hyperparameters and a white noise parameter to model the noise. The resulting maps can be found in Sec. \ref{apdxA} of the online supplementary material.
%Appendix \ref{apdxA}.

A drawback of this treatment is that every individual rotational phase bin is considered independently. So the emission is assumed to be correlated from epoch to epoch, but not necessarily in the rotational phase direction. Because the emission changes occur simultaneously in different phase bins, this method is not optimal in suppressing the Gaussian noise. In addition, because of the relatively short overall time span studied, there is not necessarily enough information in a given bin to constrain the hyperparameters reliably. To overcome this, we found that a 2-dimensional Gaussian convolution of the difference maps does well in highlighting the profile variability by utilising the correlated changes in both the rotational phase direction and from epoch to epoch.

Before the convolution is applied, the difference map is uniformly sampled in time.
This is done by creating an array of profiles at daily epochs. At each epoch, the closest observed pulse profile is used.
%by using pulse profiles of an epoch closest to the desired epochs. 
The resulting map is convolved with a 2D Gaussian. The time scales in rotational phase and date are chosen such as to highlight the correlated structures seen in the original difference map. Unlike the Gaussian process regression, this choice was made through visual inspection. The smoothed maps are presented in Sec.~\ref{result}. Any conclusions we draw are compatible with the raw difference maps, as well as those obtained with Gaussian process regression (both of which can be found in Sec.~\ref{apdxA} of online supplementary material.
%Appendix \ref{apdxA}).

\section{Results}\label{result}

Here we present the results for the analysis of 8 individual sources using the techniques described in Sec.~\ref{analysis}. For each source, we show the $\dot \nu$ and shape parameter variation 
%(different variables for every source) 
in the same figure together with the total intensity difference maps. We also segregate the observation epochs based on the emission variability for every individual pulsar to construct average pulse profiles with the aim to show the differences in profile shape and polarisation properties. The figures showing the polarisation difference maps both from data and the jitter simulations can be found in Sec.~\ref{apdxI} of online supplementary material.
%Appendix \ref{apdxI}.

\subsection{PSR J0729$-$1448}\label{0729}

\begin{figure*}
    \centering
    \includegraphics[height=0.56\hsize,angle=90]{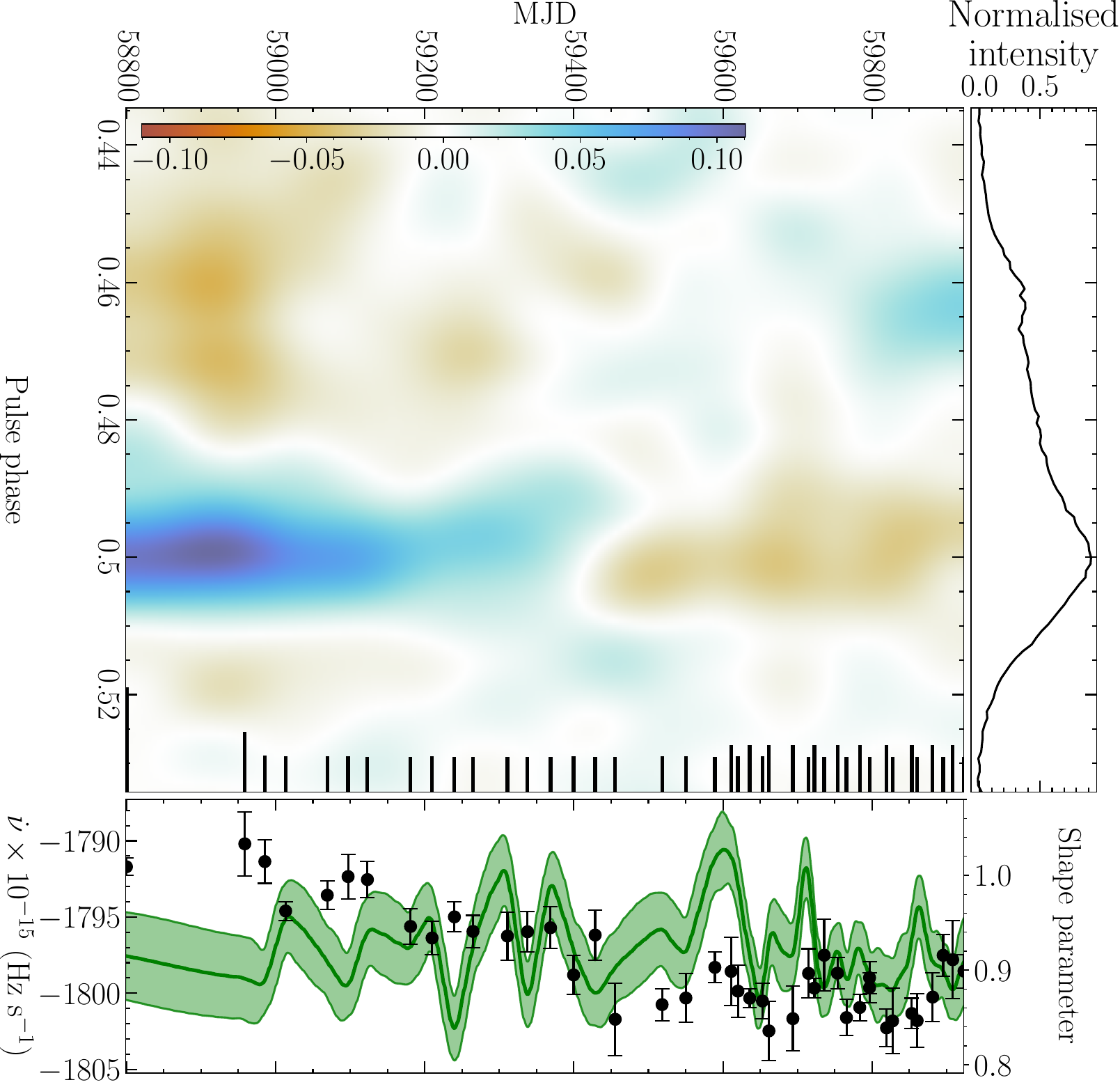}
    \hspace{0.02\hsize}
    \includegraphics[width=0.41\hsize,angle=0]{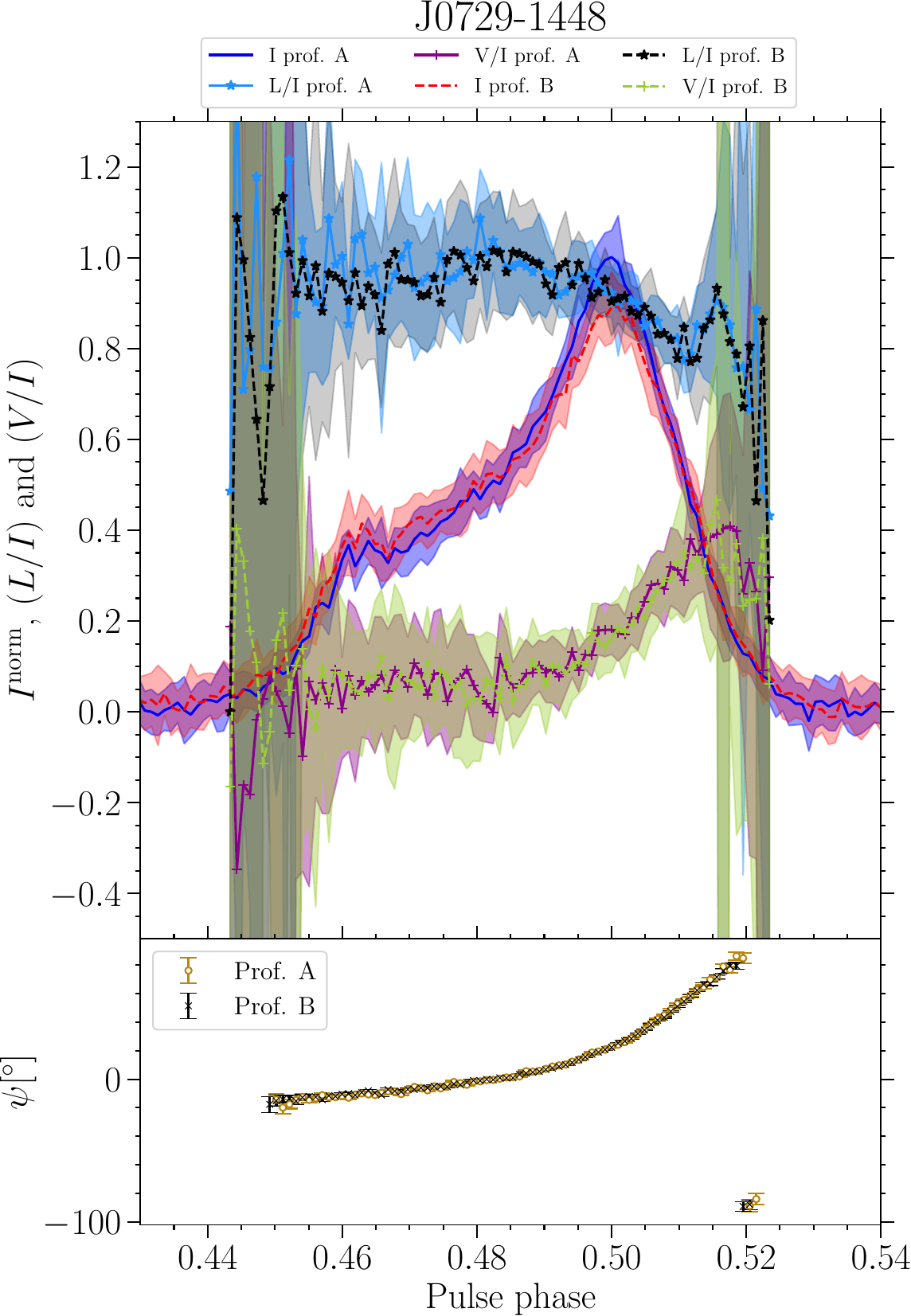}
    \caption{    \label{fig:0729diffmap}
\textit{Left-hand figure:} 
%shows 
the smoothed difference map of the total intensity of pulsar J0729$-1448$, showing the on-pulse region. The right-hand ticks in the main panel of the figure indicate the observation epochs, and the size is a proxy for the integration time. The upper panel 
%of the same figure 
shows the average pulse profile. The right-hand side panel 
%of the figure 
shows the time evolution of $\dot \nu$ (the solid line) and the temporal evolution of the shape parameter shown by the black dots with error bars. For this source the shape parameter corresponds to the peak amplitude of the normalised pulse profile. \textit{Right-hand figure:} 
%shows 
the total intensity of profile A (solid blue line), and profile B (dashed red line). For this pulsar profile A and B have been obtained by averaging over all observations before and after MJD 59400 respectively. The linear polarization fraction of profile A is shown as the sky blue solid line with $\star$ marks and that of profile B as the black dashed line with $\star$ marks. The circular polarisation fraction of profile A is shown as the purple solid line with $+$ marks and that of profile B as the green dashed line with $+$ marks. The lower panel of the same figure shows the polarisation position angle of profile A and B by the hollow circles and cross marks respectively.}
%See the main text for a detailed discussion, including the definition of the difference between profiles A and B. }
\end{figure*}

The upper left panel of  Fig. \ref{fig:0729diffmap}  shows the smoothed total intensity difference map for the pulsar. %J0729$-$1448. 
The intensity of the main peak (centred at phase 0.50) decays over time relative to the leading and trailing edge of the profile. Such a systematic feature is absent in the jitter simulated total intensity difference map shown in Fig.~\ref{J0729alldiffs} of the online supplementary material, reinforcing the conclusion that the profile of this pulsar changes gradually over time. 
%A visual inspection of the difference map clearly shows a 
%transition occurring at $\sim 59400$ MJD. 
The amplitude of the main peak is initially $\sim 10\%$ 
larger than the median value before transitioning (at MJD $\sim 59400$) to a state where it is lower than the median total intensity. A visual inspection of the single-pulse data does not reveal any obvious evolution from observation to observation. 

An average profile is constructed by combining all the observations before and after MJD 59400, labelled as profile A and B respectively in Fig. \ref{fig:0729diffmap}. 
%The total intensity pulse profile for both emission modes clearly reflects the conclusion drawn from the difference map in Fig. \ref{fig:0729diffmap}, where 
As expected, the peak amplitude of profile A (corresponding to the blue region of the difference map) is greater compared to the peak amplitude of profile B (corresponding to the brown region in the difference map). Looking at the $L/I$, $V/I$ and $\psi$ profiles, 
%for both A and B in Fig.~\ref{fig:0729diffmap} 
there is no indication of associated changes in polarization. 
%A similar
The same conclusion is drawn by looking at the $L/I$, $V/I$ and $\psi$ difference maps (shown in Fig. \ref{fig:0729poldiff} of the online supplementary material), which do no not show any evidence for polarization evolution on a timescale comparable to that seen for the changes in Stokes $I$. 
Given that the radio emission is very highly linearly polarized, and following the methodology in Sec.~\ref{SectPolarizationVariationSensitivity}, we find that $L/I$ should decrease by $\sim30\%$ ($f_{L/I}=0.7$) for one of the emission states to leave a detectable signature in the difference map. The low circular polarization implies that one of the states should be $\sim3$ times stronger polarized to leave a detectable $V/I$ signature.

The profile variability can be captured by a shape parameter, taken to be the peak amplitude of the area normalised profiles. Its time evolution is shown in the right panel of the left-hand figure in Fig. \ref{fig:0729diffmap} (points with error bars). This confirms that if the profile variation were to be periodic, we would only have observed at most half a cycle in $\sim3$ years of data.

The spin-down rate evolution (solid line in the same panel as the shape parameter in Fig. \ref{fig:0729diffmap}) does not show significant variability on a timescale similar to the profile evolution. Indeed, the correlation coefficient as computed using the method described in Sec.~\ref{correlation-analysis} is consistent with zero, implying no significant detection of a correlation between the spin-down rate and the shape parameter.

Assuming the variability is associated with core emission, the pulse profile of this source could be a core-cone triple, with the trailing conal outrider weak and conflated with the core component. 
%The pulse profile of this source may be classified under the core-cone triple class with the trailing outrider weak and conflated. 
%A detailed summary of cone/conal emission pattern classification can be found in \citet{Rankin+2022}. 
\cite{MR+2011} suggest that this asymmetry in the pulse profile component configuration is due to aberration/retardation (A/R) effects \citep{BCW+91} in this pulsar with a relatively large spin-down energy loss rate.
%However, the asymmetry\footnote{This asymmetry in the average pulse profile may be due to the aberration/retardation (A/R) effects \citep{BCW+91} (previously studied for many different sources \citep{MR+2011})} makes it hard to disentangle the core and cone components uniquely. And possibly the long-term variability is associated with the core component of the emission beam.}

\subsection{PSR J1121$-$5444 (B1119$-$54)}\label{1121}

\begin{figure*}
    \centering
    \includegraphics[height=0.56\hsize,angle=90]{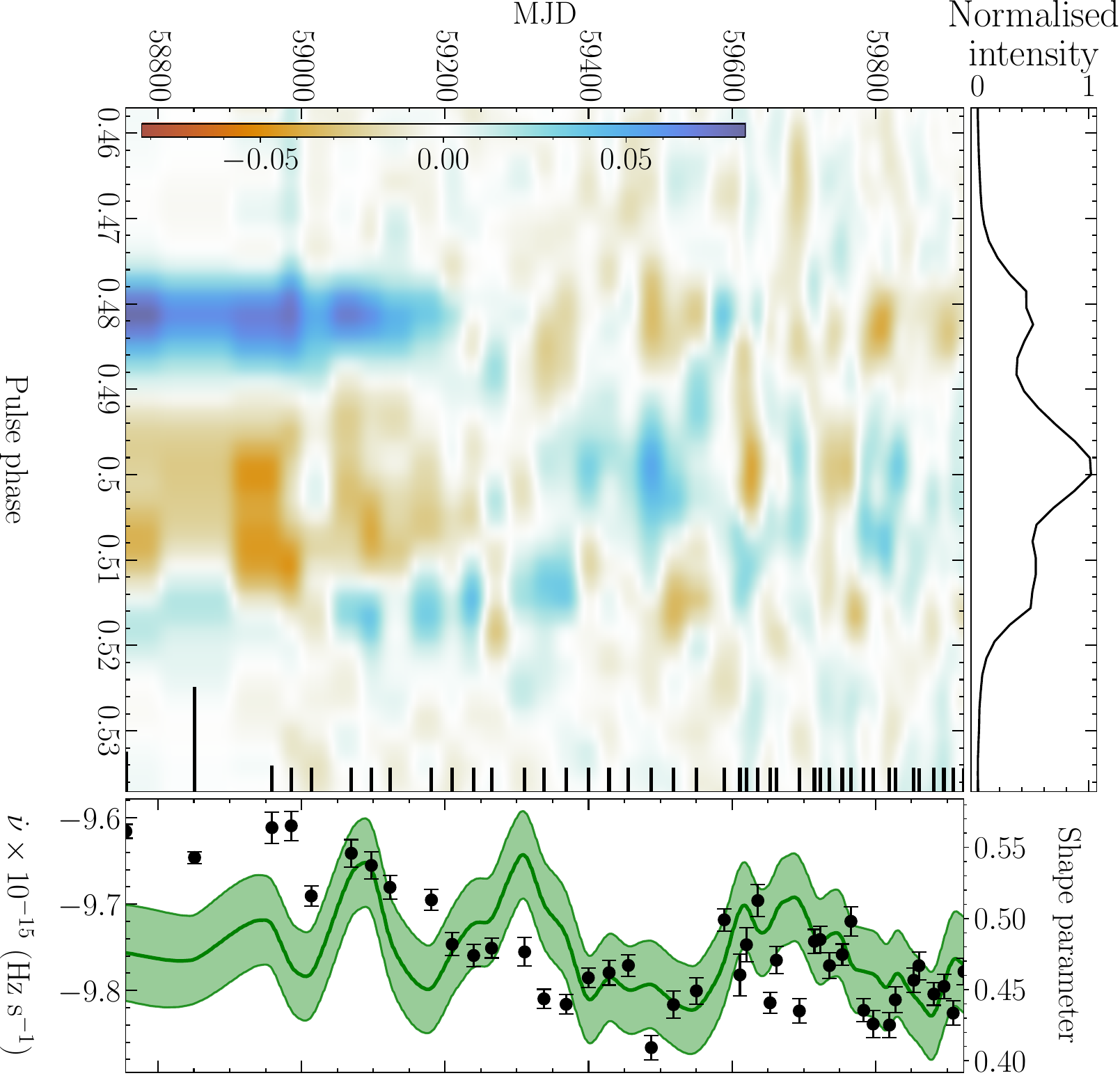}
    \hspace{0.02\hsize}
    \includegraphics[width=0.41\hsize,angle=0]{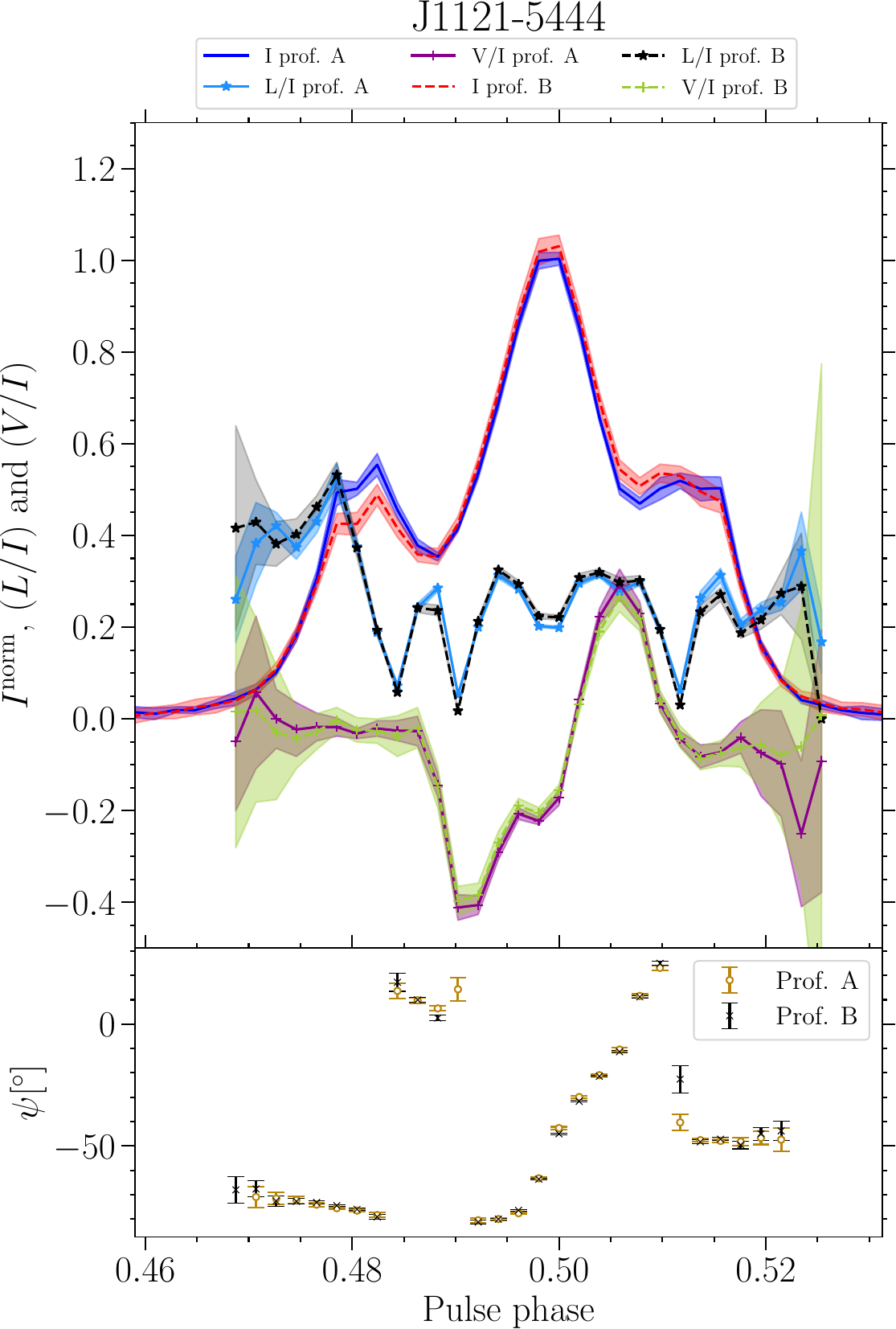}
    \caption{    \label{fig:1121diffmap}
Provides the same information as in Fig. \ref{fig:0729diffmap}, but for the pulsar J1121$-5444$. However, for this pulsar, the shape parameter is given by the ratio of amplitudes at pulse phase 0.50 and 0.48. Profile A has been obtained by averaging all the observations made before MJD 59400 and all the later epochs were averaged to obtain profile B.}
\end{figure*}

PSR J1121$-$5444 has three profile components. 
The difference map of Stokes $I$, shown in Fig.\ref{fig:1121diffmap}, shows a strong 
temporal evolution such that the leading component was relatively strong compared to the main peak until MJD $\sim 59200$.
In the later epochs the peak amplitude ratio appears to be much more variable on short timescales. This cannot be attributed to changes in observing cadence or the length of the observations. At these later epochs the profile variations become more
%, such characteristic emission features disappear and the variations become more 
stochastic in nature, thereby resembling the expectations for single-pulse jitter
%-dominated variation as shown in 
(see Fig. \ref{J1121alldiffs} of online supplementary material). 
%However, the absence of a long-term profile evolution feature in the jitter-simulated difference map makes this pulsar a candidate exhibiting a long-term profile evolution. 
The relative change of the peak amplitudes is $\sim10\%$.
%in the peak amplitudes of the leading and main profile component. 
%In order to identify if the emission feature explained above originates from the short-term emission variability, we investigate the single pulse stack recorded at every epoch. 
The single-pulse data reveals quasi-periodic relatively slow intensity modulation (as also reported in \citealt{Song+2023}).
%Though the leading component exhibits 
%periodic burst-like emission features, 
%We do not find a higher rate of occurrence of such bursts in the epochs before the transition from emission modes B to A. This implies the long-term feature seen in Fig. \ref{fig:1121diffmap} is independent of short-term variations.
No systematic evolution of these properties is identified, including any potential change in the repetition period of the modulation.

To highlight differences in the average emission properties, the observations are divided into two groups resulting in profiles A and B. All observations before MJD 59200 are referred to as profile A, and all later epochs as profile B.
%The average profile of $I$, $L/I$, $V/I$ and $\psi$ swing for both the emission modes are shown in Fig. \ref{fig:1121modesep}. 
The right-hand side of Fig. \ref{fig:1121diffmap} shows that profile A has a relatively strong leading component relative to the saddle region between the central and trailing components. At phase $\sim 0.485$ and $0.515$ the degree of linear polarisation appears to change slightly. The difference map of $L/I$ (second panel of Fig. \ref{fig:1121poldiff} in the online supplementary material) indeed shows small, $\sim3\%$,  changes in $L/I$ synchronised with the variability seen in Stokes $I$. Similarly, very weak patterns at comparable epochs can be seen in the $V/I$ difference map at phases $\sim0.49$ and $0.51$. There is no noticeable change in the $\psi$ difference map. 
 
We construct a shape parameter by taking the ratio of the amplitude of the components at phases 0.50 and 0.48. This captures the profile variability well, with a single significant transition over the $\sim3$ years of data analysed (shown in the right panel of the left-hand side plot of Fig. \ref{fig:1121diffmap} as black points with errorbars). The Pearson's correlation coefficient
between the evolution of the shape parameter 
%(A$_1$/A$_2$) 
and $\dot \nu$ is
%gives a coefficient of 
0.3$\pm$0.1.

We note, however, that a significant contribution to the correlation is the shorter timescale structure, rather than the single transition seen as the main profile variability.
Observations of more transitions will be required to make the correlation convincing. 

Although the variability in the leading component dominates the difference map, there is a hint of correlated variability in the trailing component (as shown in Fig. \ref{fig:1121diffmap} between pulse phase 0.54 and 0.52). However, the variability in the middle and trainling components is not significant when compared to the jitter
simulated total intensity difference map (Fig. \ref{J1121alldiffs} of online supplementary material). This makes it inconclusive if the variability is related to a change in relative intensity of core and conal emission. As argued in \citet{Rankin+2022}, 
%[MNRAS 514, 3202 (2022; see arXiv:2206.07739 for a single PDF)] \textcolor{red}{(TRUE?)}, 
the leading component is associated with conal emission.
% The correlated change in the conal component may be real or an artefact due to normalisation. In that case, the variation may purely be attributed to the core component.

\subsection{PSR J1141$-$3322} \label{1141}
\begin{figure*}
    \centering
    \includegraphics[height=0.56\hsize,angle=90]{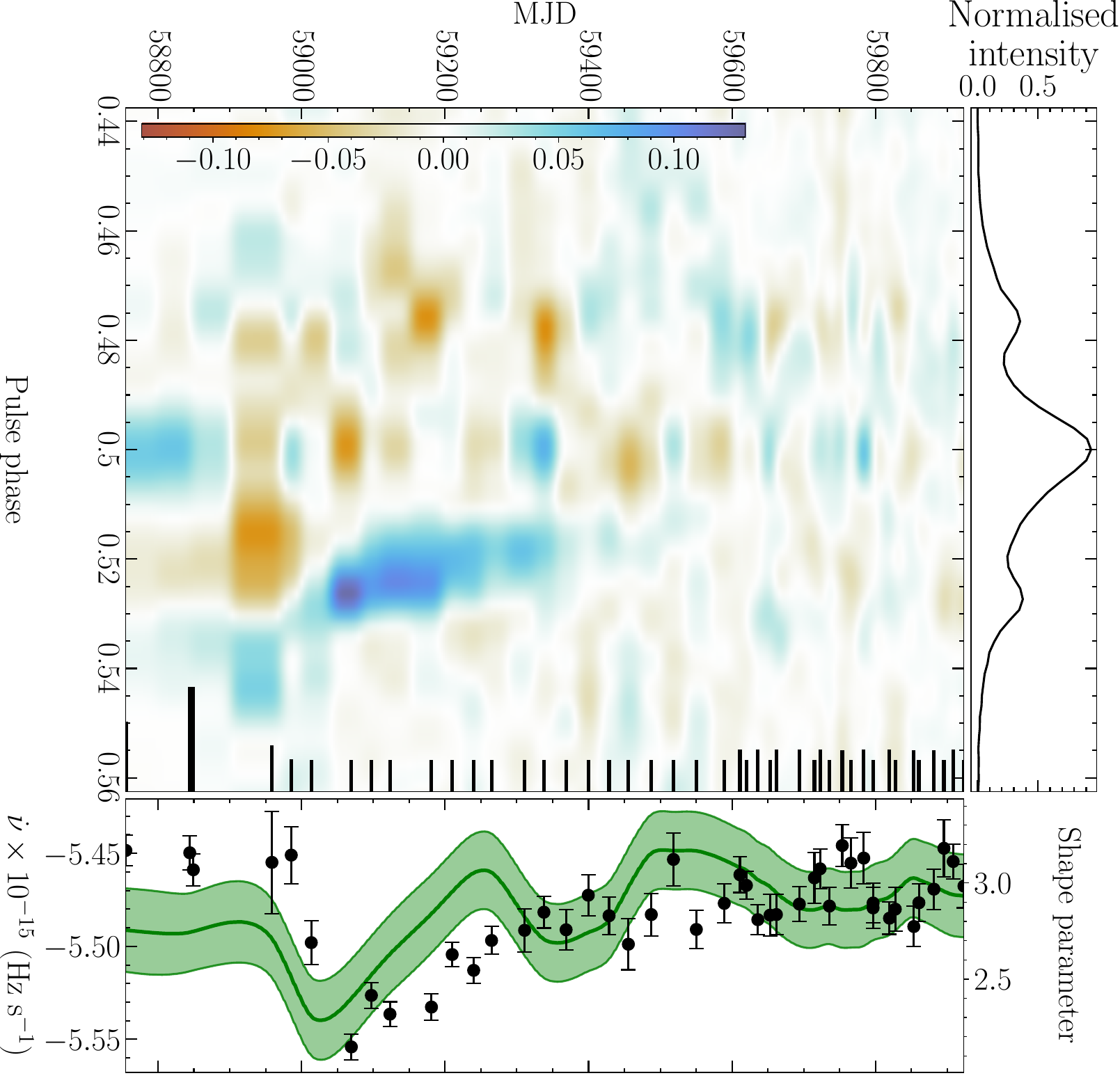}
    \hspace{0.02\hsize}
    \includegraphics[width=0.41\hsize,angle=0]{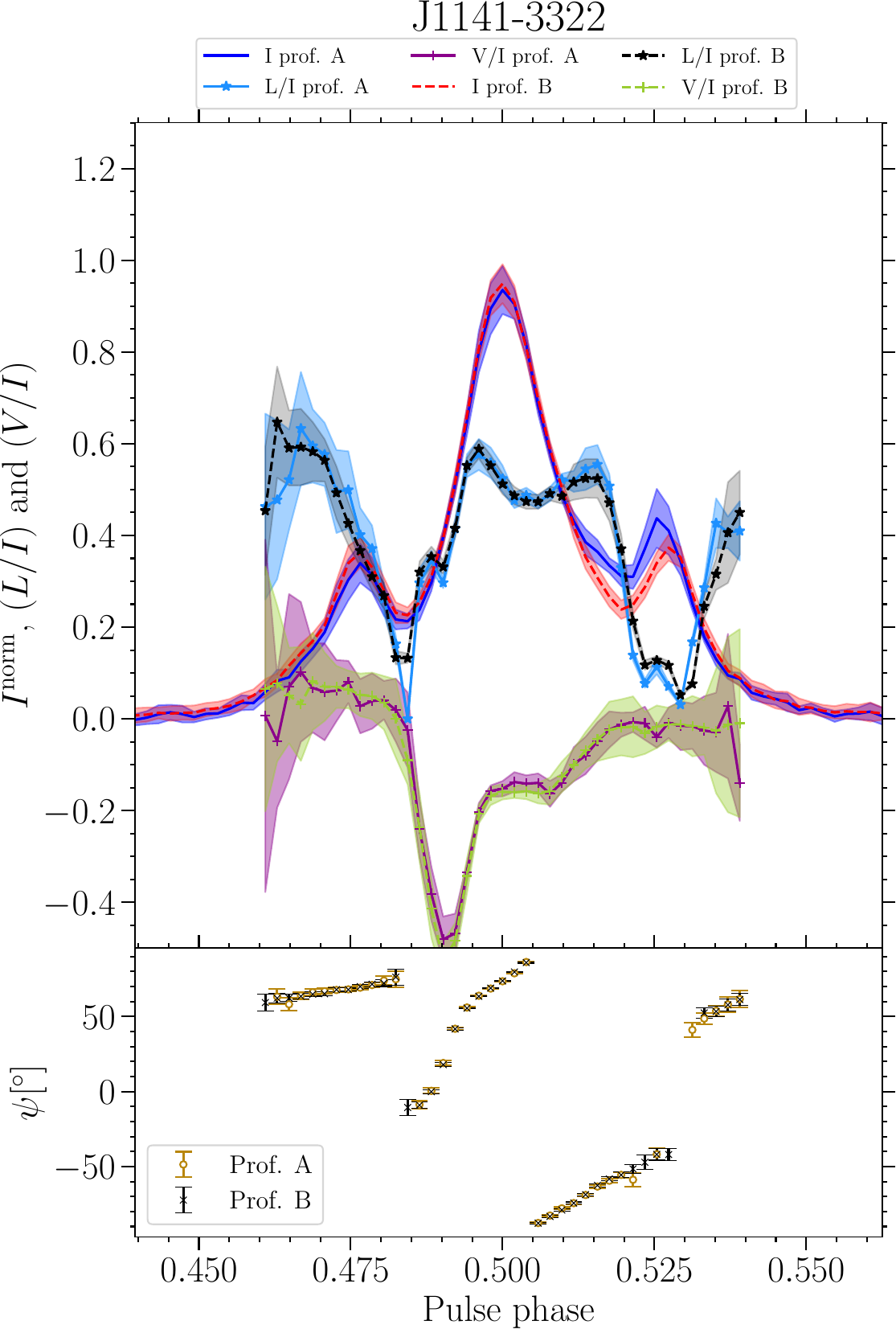}
    \caption{    \label{fig:1141diffmap}
Provides the same information as in Fig. \ref{fig:0729diffmap}, but for the pulsar J1141$-$3322. The ratio of amplitudes of the main peak and the trailing component is used as the shape parameter for this pulsar. Profile A has been constructed by averaging all the profiles observed between MJD 59000 and 59400, all the rest of the observations were averaged to produce profile B.}
\end{figure*}

Pulsar J1141$-$3322 also has three profile components. The difference map shown in left panel of the left-hand side Fig.~\ref{fig:1141diffmap} clearly indicates a different emission state of the pulsar between MJD $\sim59000$ and $\sim59400$, where the emission in between the main peak and the trailing component is relatively strong  (at phase 0.52, corresponding to a $\sim10\%$ increase relative to the amplitude of the main component). Such variations are absent in the jitter-simulations (shown in Fig. \ref{J1141alldiffs} of online supplementary material). This extra emission gradually drifts in pulse phase towards the main peak of the profile before disappearing. 

All the observations between MJD 59000 to 59400 are combined to construct an average profile A and the rest of the epochs are used to produce average profile B, shown in the right hand side of Fig.~\ref{fig:1141diffmap}. The extra emission at phase 0.52 captured in the difference map 
%(Fig. \ref{fig:1141diffmap}), 
is evident in the shape of profile A in Fig. \ref{fig:1141diffmap}.
%, where the intensity in the saddle region between the central and trailing peak for profile A is greater than profile B. 
The peak position of the trailing component in profile A is shifted closer to the main peak compared to that in profile B.

The $L/I$ of profile A near the saddle region between phase $\sim 0.52$ and 0.53 is smaller than in profile B, indicating an anti-correlated change in the linear polarisation fraction with the total intensity. 
%After which for a very narrow phase range 
At a phase $\gtrsim 0.53$ the pattern reverses, leading to a correlated change between $L/I$ and total intensity. The change in $L/I$ in the same phase range is also visible in the $L/I$ difference map shown in the second panel of Fig. \ref{fig:1141poldiff} of the online supplementary material. 
%Though there is no significant 
The long-term variation in total intensity in the saddle region between the first and central components is much less significant (over a narrow phase range $\sim 0.48$), but at this phase a marginal dip in $L/I$  can be seen in the right-hand panel of Fig. \ref{fig:1141diffmap}. This is also present in $L/I$ difference map shown in Fig. \ref{fig:1141poldiff}. The change in the $L/I$  happens in tandem with the total intensity variation (can be seen by comparing the first and the second panel of Fig. \ref{fig:1141poldiff}) leading to the conclusion of the simultaneous correlated change in both the total intensity and linear polarisation fraction. The $V/I$ difference map (shown in the third panel of Fig. \ref{fig:1141poldiff}) shows a correlated change such that when the 
extra emission is seen in total intensity, a blue structure is seen at the same time.
However, we do not find any evidence for evolution of the polarisation position angle.

The single-pulse data reveals that this source is highly variable with many short nulls occurring in each observation. The nulling properties and overall appearance of the single-pulse variability do not appear to be different during the period of significant pulse profile evolution. 

We find the ratio of amplitudes of the main peak and trailing component is an appropriate shape parameter, which is shown by the black points with error bars in the right panel of the left-hand side Fig.~\ref{fig:1141diffmap}.
In order to account for the drift of the extra emission, the components of the analytic model used to determine the shape parameter (see Sec.~ \ref{shapeparam}) are allowed to shift in phase. The evolution of the shape parameter, hence the pattern seen in the difference map, appears to be correlated with the change in the spin-down rate. 
%Both the amplitude of the third component and the $\dot \nu$ increase and follow almost a similar trend in recovery. 
The correlation coefficient between the spin-down rate and the shape parameter is 0.3$\pm$0.1 following the method described in Sec.~\ref{correlation-analysis}. 
This indicates continued monitoring is required to establish if this correlation is significant
%implying a mild correlation between the emission and spin-down rate. Therefore, this pulsar appears to belong to the category of pulsars where both the spin-down rate and the emission undergo a simultaneous correlated change. Continued monitoring will be required 
and to see if the profile variability represents (quasi-periodic) state switching.

\begin{figure*}
    \centering
    \includegraphics[height=0.56\hsize,angle=90]{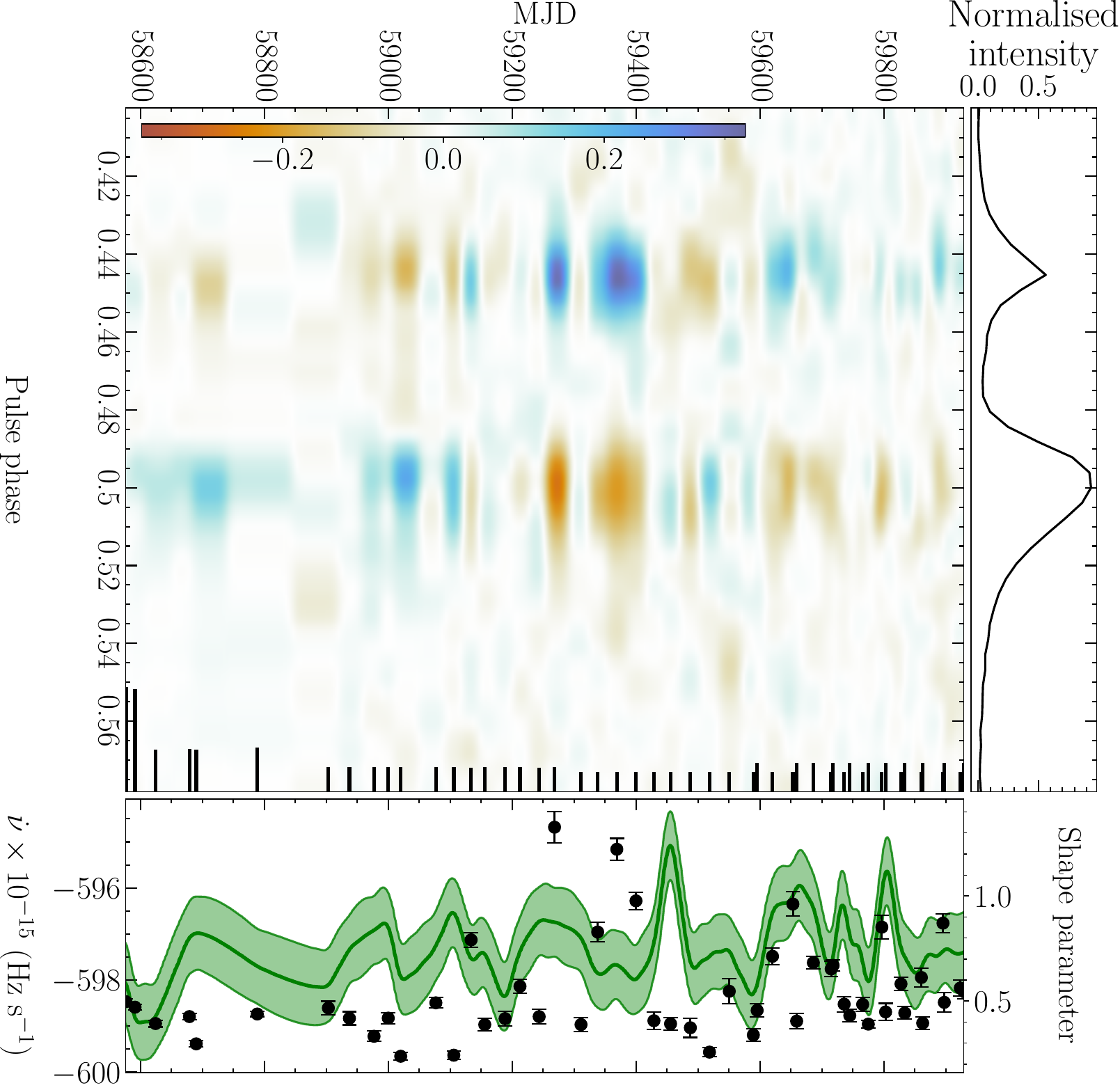}
    \hspace{0.02\hsize}
    \includegraphics[width=0.41\hsize,angle=0]{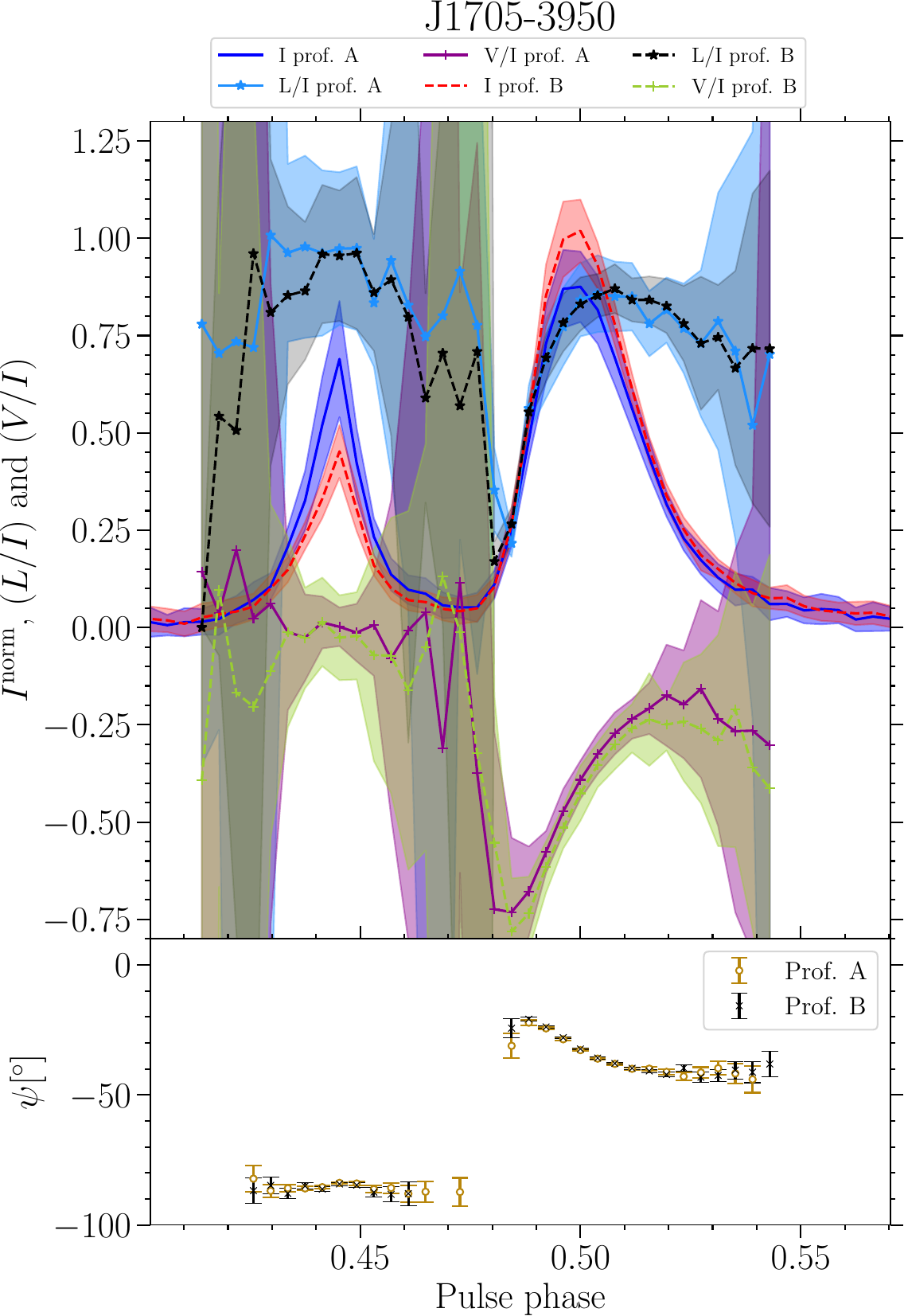}
    \caption{    \label{fig:1705diffmap}
Provides the same information as in Fig. \ref{fig:0729diffmap}, but for the pulsar J1705$-$3950. In this case, the amplitude ratio of the leading and trailing component represent the shape parameter, which has also been used to categorise the emission state in either profile A or profile B. Epochs where the shape parameter is smaller than the median have been combined to produce profile A. The others are combined to construct profile B.}
\end{figure*}

The main variability in Fig. ~\ref{fig:1141diffmap} is associated with the trailing component of the profile, and the saddle region in between the main and trailing components. If the variability is associated with an intensity ratio change of conal versus core emission, correlated variability in the leading component can be expected, which is not observed. The spectral index of the outer components is flat relative to the central component (compare with the 436 MHz profile of \citep{Lorimer+1994}\footnote{The pulse profile can be found in the \href{https://psrweb.jb.man.ac.uk/epndb/}{EPN database}.}. This is consistent with an interpretation of the profile being a core-cone triple (see also \citealt{Rankin1993cone, Rankin+2022}).

\subsection{PSR J1705$-$3950}\label{1705}
The difference map for PSR J1705$-$3950, shown in left panel of the left-hand side Fig.~\ref{fig:1705diffmap}, exhibits structures
corresponding to switches between states where the relative amplitude of the two profile components changes. These switches are rapid and are only occasionally marginally resolved by the cadence of our observations. The jitter simulations (Fig. \ref{J1705alldiffs} of the online supplementary material) show that variability with similar amplitude (the peak amplitude changes with $\sim 40\%$) can be expected from pulse shape variability on a timescale shorter than the length of a single observation, but only when blocks of consecutive pulses are used. This suggests there might be a link between short and longer timescale pulse shape variability in this source. However, the longer event around MJD 59350 (three consecutive observations) seems unlikely to be a chance occurrence, suggesting the existence of a profile variability timescale that extends to $\sim 2$ months. 

The amplitude ratio of the leading and trailing components captures the features in the difference map.
%and hence the appropriate shape parameter for this pulsar. 
The correlation analysis reveals no correlation between the spin-down rate and the shape parameter (see Fig.~\ref{fig:1705diffmap}).
All the observations with a shape parameter smaller and greater than the median shape parameter were combined to obtain profiles A and B respectively, as shown in the right-hand panels of Fig. \ref{fig:1705diffmap}.
Apart from the expected difference in Stokes $I$, we do not find evidence for any systematic temporal evolution in the polarised emission for this source. The structures seen in the polarization difference maps (shown in Fig.~\ref{fig:1705poldiff} of the online supplementary material) seem uncorrelated with the Stokes $I$ evolution, and are consistent with the expectations from pulse jitter (see Figs.~\ref{fig:1705poldiff_ipm} and \ref{fig:1705poldiff_bm} of online supplementary material).
Like PSR J0729$-$1148, for which also no correlated polarization evolution is observed, this pulsar is also highly linearly polarized. The methodology in Sec.~\ref{SectPolarizationVariationSensitivity} shows that $L/I$ should be much lower (by $\sim50\%$; $f_{L/I}=0.5$) for one of the emission states to expect detectable correlated variability. The circular polarization is required to change with a comparable percentage.

\begin{figure*}
    \centering
    \includegraphics[height=0.45\hsize,angle=0]{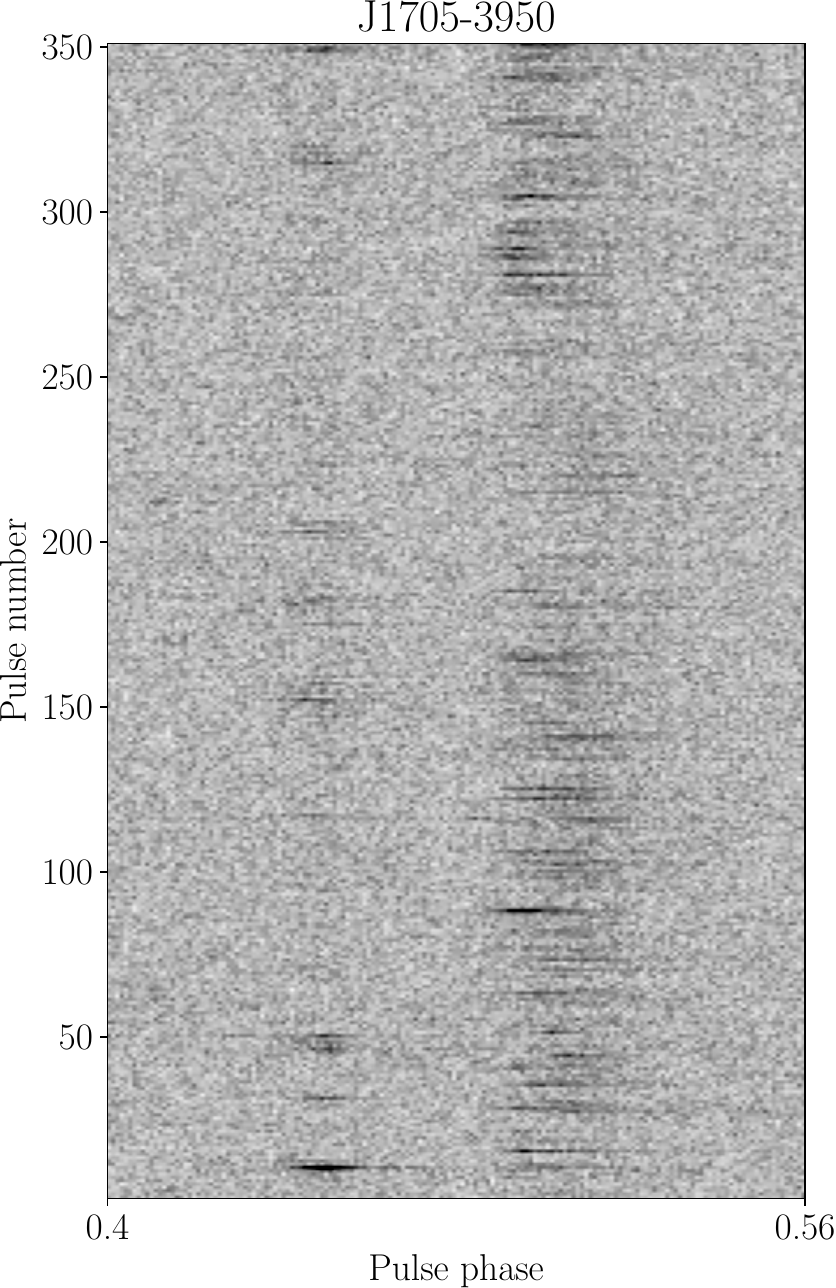}
    \hspace{0.0\hsize}
    \includegraphics[height=0.45\hsize,angle=0]{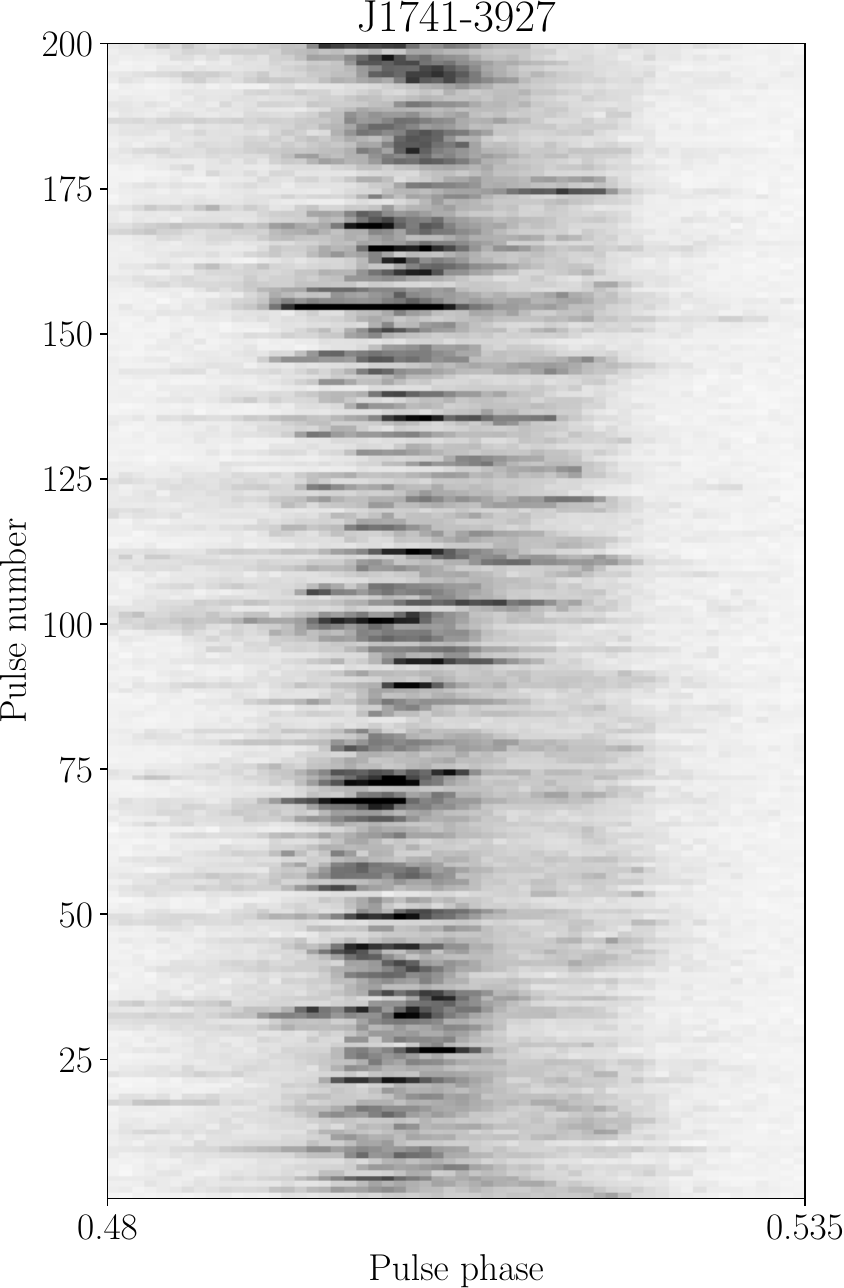}
     \hspace{0.0\hsize}
    \includegraphics[height=0.45\hsize,angle=0]{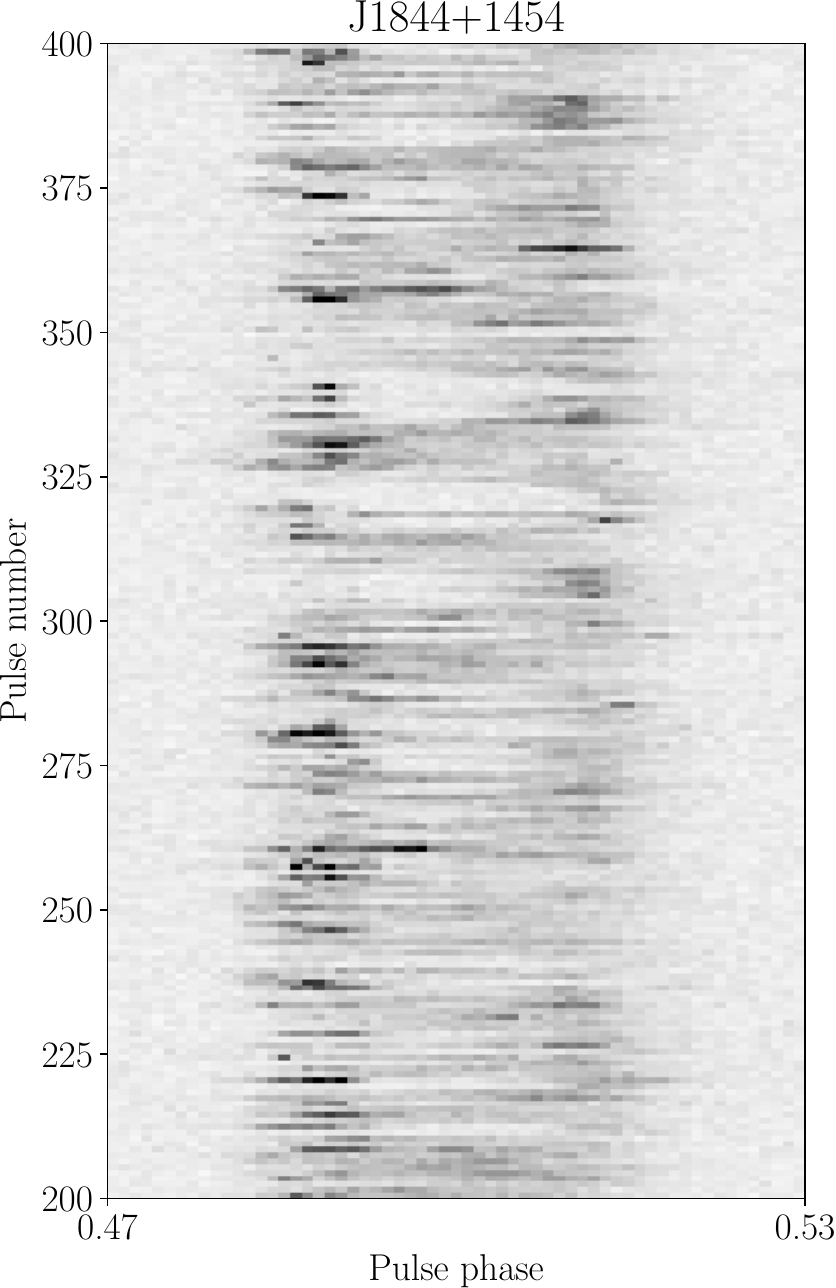}
     \hspace{0.0\hsize}
    \includegraphics[height=0.45\hsize,angle=0]{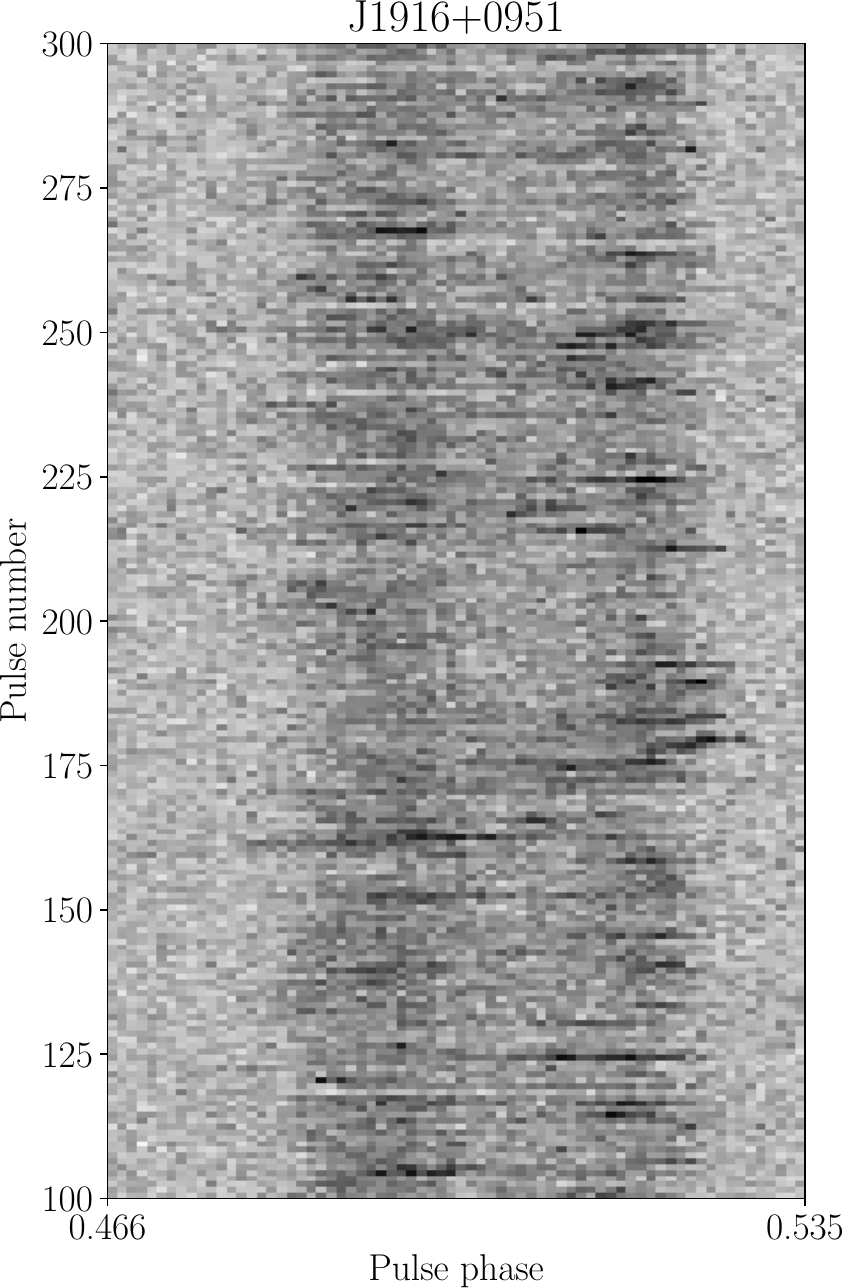}
     \hspace{0.0\hsize}
    \includegraphics[height=0.45\hsize,angle=0]{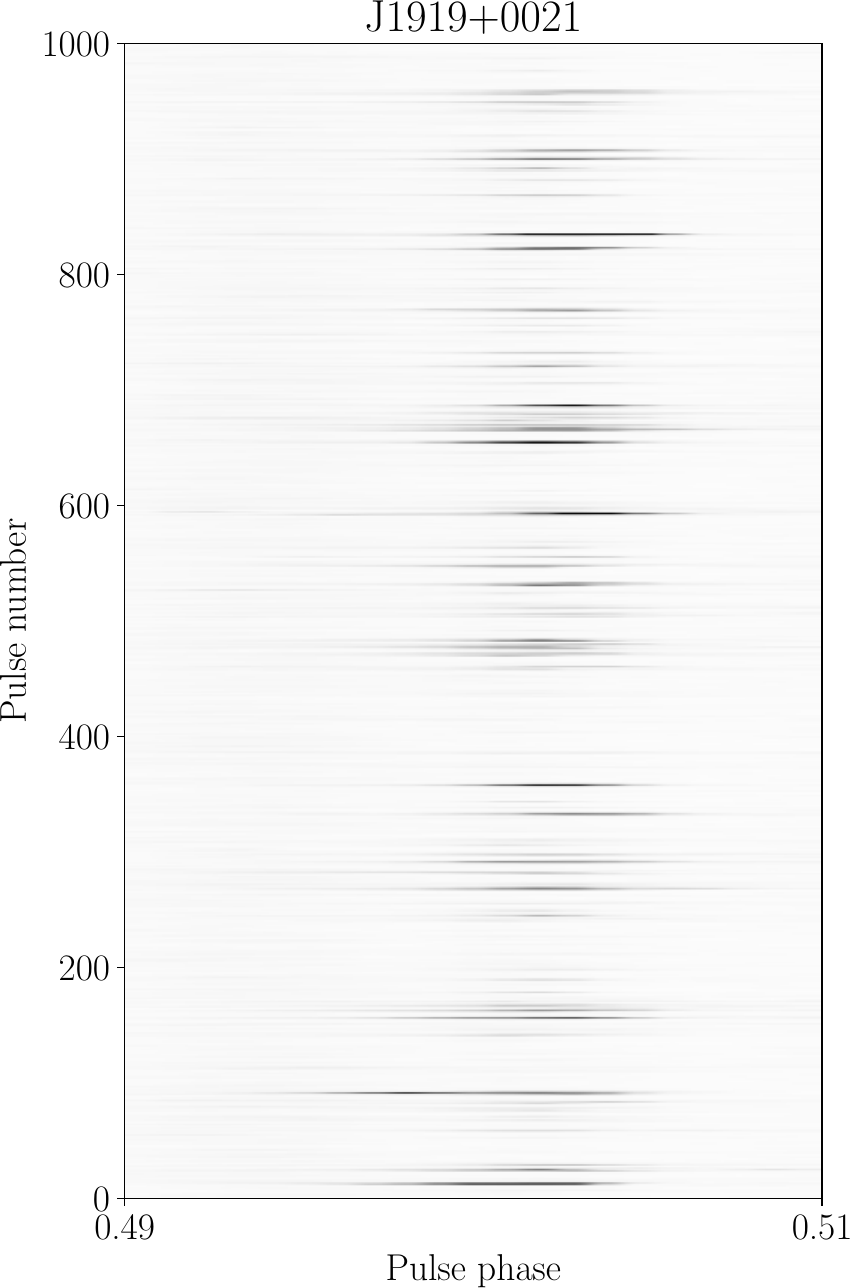}
       
    \caption{Pulse stacks of five pulsars showing their single pulses. The dynamic range of the colour scale is adjusted so that the maximum colour corresponds to 1/3th (PSRs J1705$-$3950, J1741$-$3927, J1844+1454, J1916+0951) and 1/5th (PSR J1919+0021) of the maximum flux density of the data shown. This helps in making weaker features prominent. \label{fig:singlepulse}}
\end{figure*}

The single-pulses of PSR J1705$-$3950 are highly variable with bursts of activity for both components of the profile. There are periods of up to $\sim100$ pulses where either of the two profile components can be weak, especially the leading component (see left-hand panel of Fig.~\ref{fig:singlepulse}). This means long observations are required to get a stable profile, as confirmed with the jitter simulation when taking blocks of consecutive pulses (see Fig. \ref{J1705alldiffs} of online supplementary material). However, no evidence was found for the burst activity to be correlated with the observed profile evolution.

The intensity variation in the two profile components is anti-correlated in Fig. ~\ref{fig:1705diffmap}. If intensity ratio changes of core and conal emission is responsible, the two components cannot be two sides of a single cone. Instead, the fact that the PA inflection point is associated with the trailing component may suggest it is the core component, which means that the trailing part of the cone is absent or too weak to detect.
%The bright component in the average pulse profile seems to fall with the PA inflection point and maybe the core component of the emission. Also, a very faint variability can be seen near the trailing edge of the profile, especially a few bright variations in the vicinity of $\sim$ MJD 59600 may imply the trailing emission is associated with conal emission from the pulsar. Therefore this pulsar may be categorised as a core-cone triple profile, with variations in the core component or may be correlated variations in the conal component. Since all the profiles are normalised it is not possible to disentangle the variability uniquely to a particular component of the emission beam.}

%A number of such pulsars in \citet{Mitra+Rankin2002} showed similar variations too weak to form a discernible profile feature.}

\begin{figure*}
    \centering
    \includegraphics[height=0.56\hsize,angle=90]{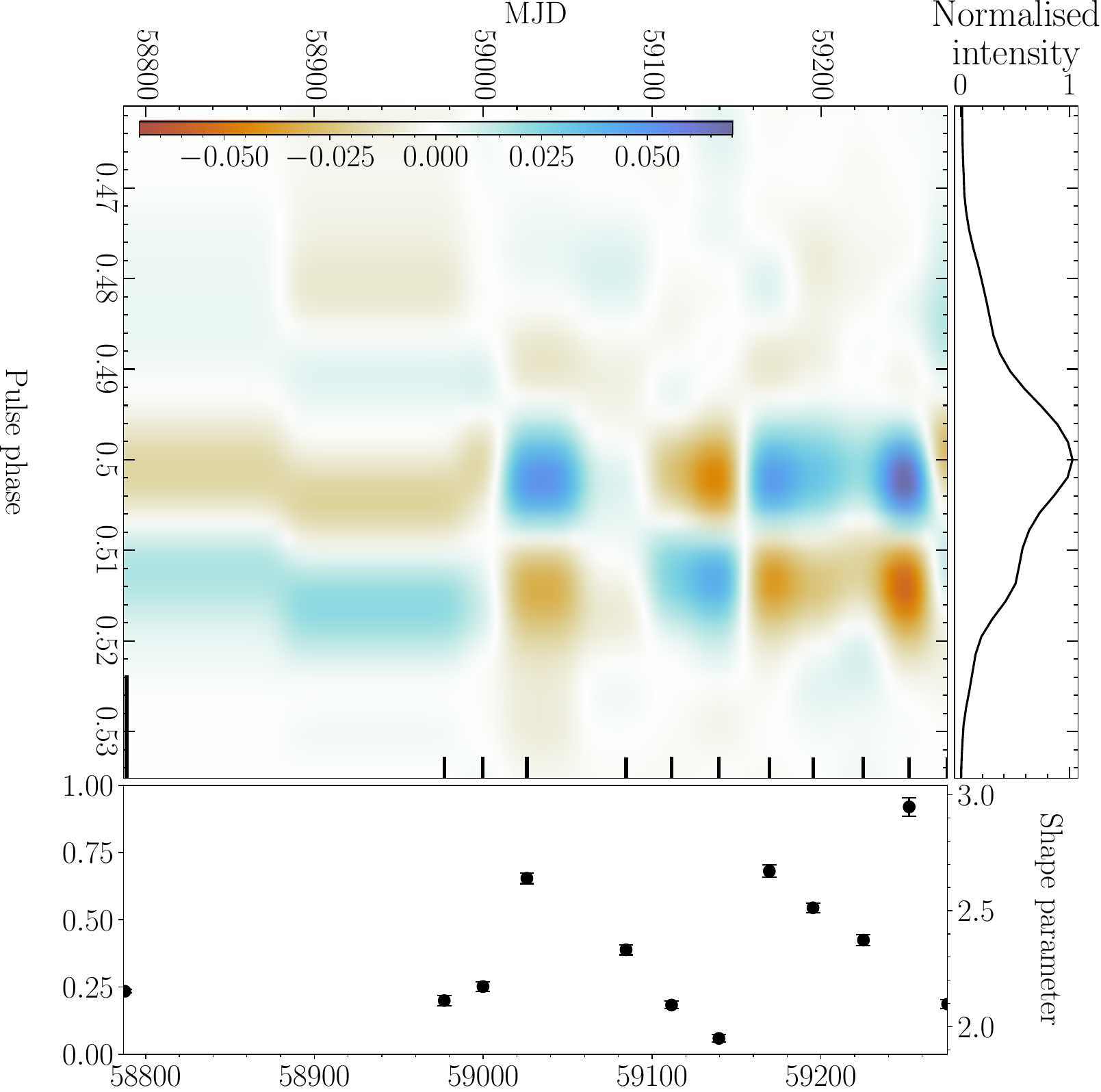}
    \hspace{0.02\hsize}
    \includegraphics[width=0.41\hsize,angle=0]{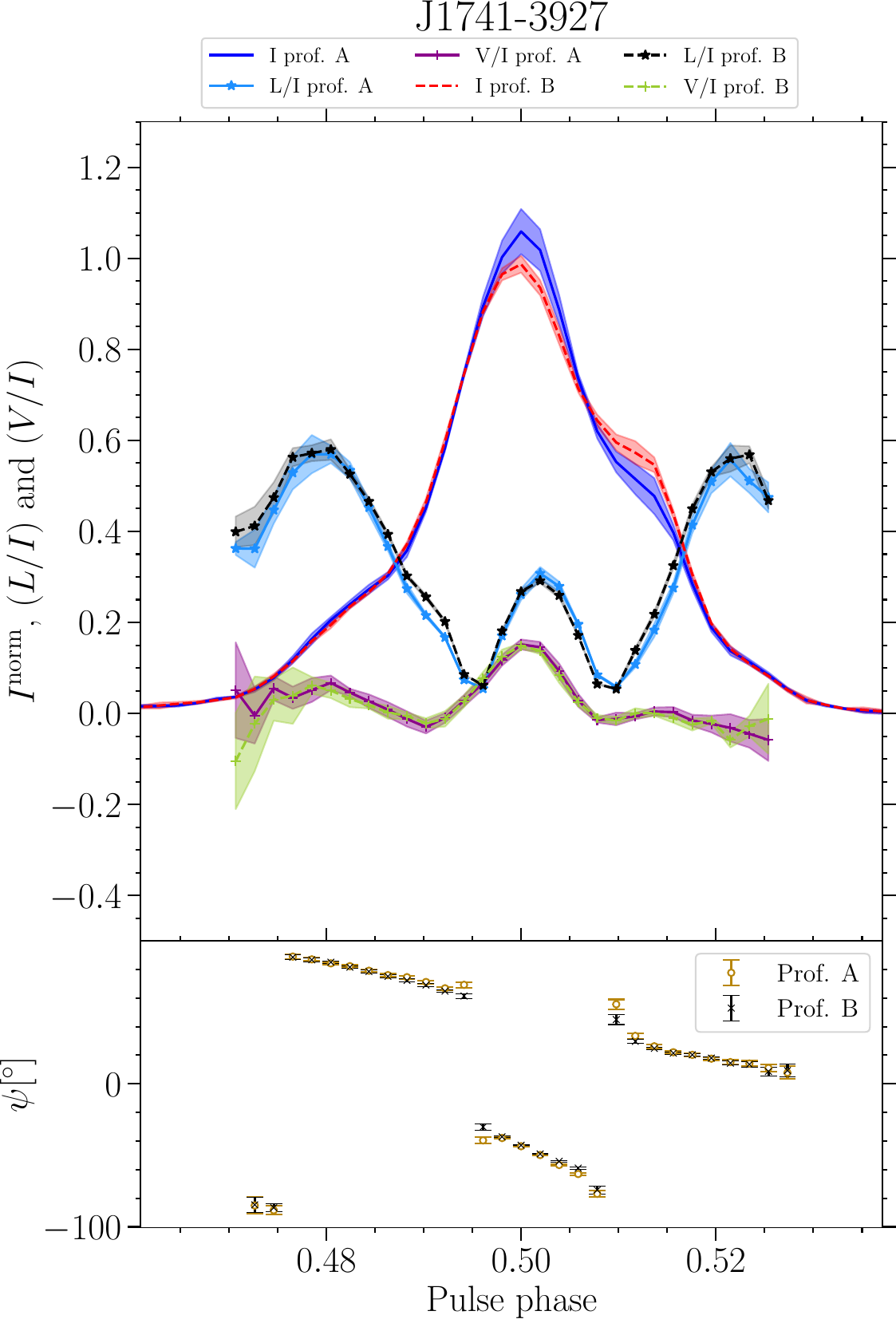}
    \caption{    \label{fig:1741diffmap}
Provides the same information as in Fig. \ref{fig:0729diffmap}, but for the pulsar J1741$-$3927. The ratio of the flux densities at pulse phase 0.50 and 0.515 is used as the shape parameter. All the epochs for which the central component is brighter and weaker than the median were combined in profiles A and B respectively. We do not find any variation in the spin-down rate in this source, hence, the $\dot \nu$ evolution is not shown in the figure.}
\end{figure*}

\subsection{PSR J1741$-$3927 (B1737$-$39)}\label{1741}
The total intensity variability for this source 
%is shown by the difference map in 
(shown in the left-hand side of the left panel of Fig. \ref{fig:1741diffmap}) shows some long-term features absent in the 
%difference map obtained from 
jitter simulations 
%by sampling pulses randomly (shown in the fourth panel from the top in 
(Fig. \ref{J1741alldiffs} of online supplementary material). 
Especially the stable structure spanning over $\sim 4$ months from MJD $\sim 59150$ to MJD $\sim$ 59300 in the difference map of Fig. \ref{fig:1741diffmap} suggests systematic profile evolution such that the intensity ratio of the main peak and the shoulder component 
%corresponding to a shoulder in the profile 
centred at a pulse phase of 0.515 oscillates. 
The ratio of flux densities at pulse phase 0.50 and 0.515 is taken to be the shape parameter, 
%which seems to capture the variability seen in the difference map. However, the variation in the intensities is found to be uncorrelated with the spin-down rate.
but no correlation with the spin-down rate could be identified.

This pulsar shows single-pulse variability on a timescale $\gtrsim10$ pulses (see Fig.~\ref{fig:singlepulse}), although it is inconclusive 
%that
whether it is periodic \citep{Song+2023}. This makes the profile relatively unstable in short observations. No clear changes in single-pulse behaviour could be associated with the observed profile variability.

All the epochs for which the central component is brighter/weaker than the median were combined as
profiles A and B respectively in the right panels of Fig. \ref{fig:1741diffmap}.
In this figure, the central component of profile B has indeed a lower amplitude (by $\sim10\%$), but a stronger shoulder at phase 0.515.
The emission in profile B is somewhat more linearly polarized at the shoulder (pulse phase 0.515), as well as at phase 0.49, corresponding to the leading edge of the central component. Indeed, the second panel of  Fig. \ref{fig:1741poldiff} of the online supplementary material shows some marginal evidence for faint structures at these phases which switch simultaneously with the Stokes $I$ profile variability.
No evidence of any correlated change in $V/I$ is found. 

\begin{figure}
    \centering
    \includegraphics[scale=.58]{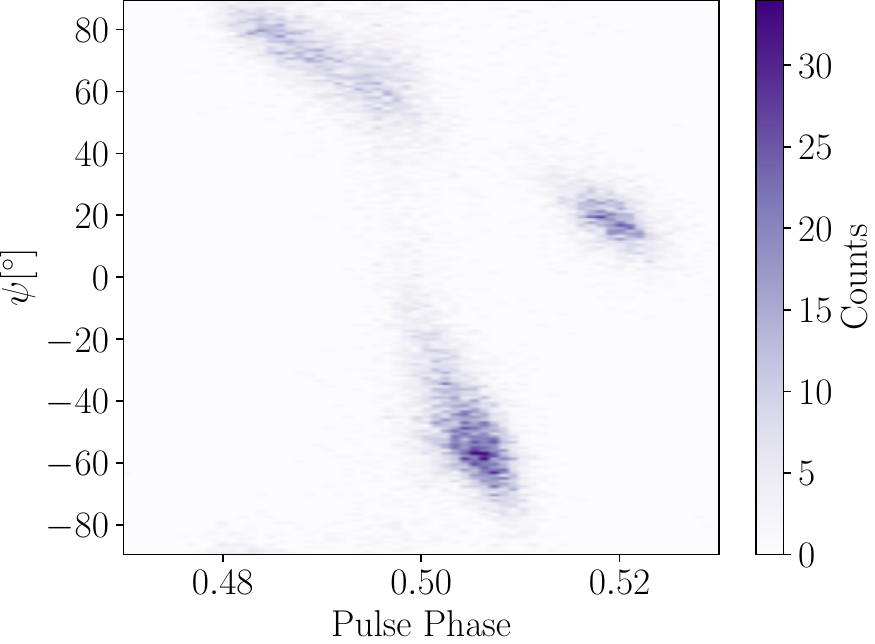}
    \caption{Histogram of the position angle as a function of pulse phase for PSR~J1741$-$3927
    %. This was obtained from a
    for the 1055 pulses long TPA observation on 2019-10-31.
        \label{fig:1741padist}
}
\end{figure}

The bottom panel of Fig. \ref{fig:1741poldiff} reveals correlated changes by $\sim 10^\circ$ in the polarisation position angle along with the change in the total intensity (shown in Fig. \ref{fig:1741poldiff} in the vicinity of pulse phase  $\sim 0.50$ and $0.51$). This coincides with the phases where the pulsar exhibits orthogonal polarisation mode (OPM) jumps. The OPM jumps are evident in the bottom-right panel of Fig. \ref{fig:1741diffmap} as discrete $\sim 90^\circ$ discontinuities in the PA swing, which as expected is also where $L/I$ drops to zero. 
Given that the PA variability over long timescales is much less than $90^\circ$, switches between the dominating OPM cannot be the full explanation. 
However, apart from OPM activity, the PA histogram shown in Fig. \ref{fig:1741padist} shows evidence for PA variability at the single-pulse level. The PA distribution is relatively wide between phases $\sim$0.50 and 0.51, despite this being the most intense part of the profile where the measurement uncertainties should be relatively low. So evidently in that phase range, the intrinsic PA is variable on both single-pulse timescales and the longer timescales probed by the profile monitoring from epoch to epoch.

\begin{figure*}
    \centering
    \includegraphics[height=0.56\hsize,angle=90]{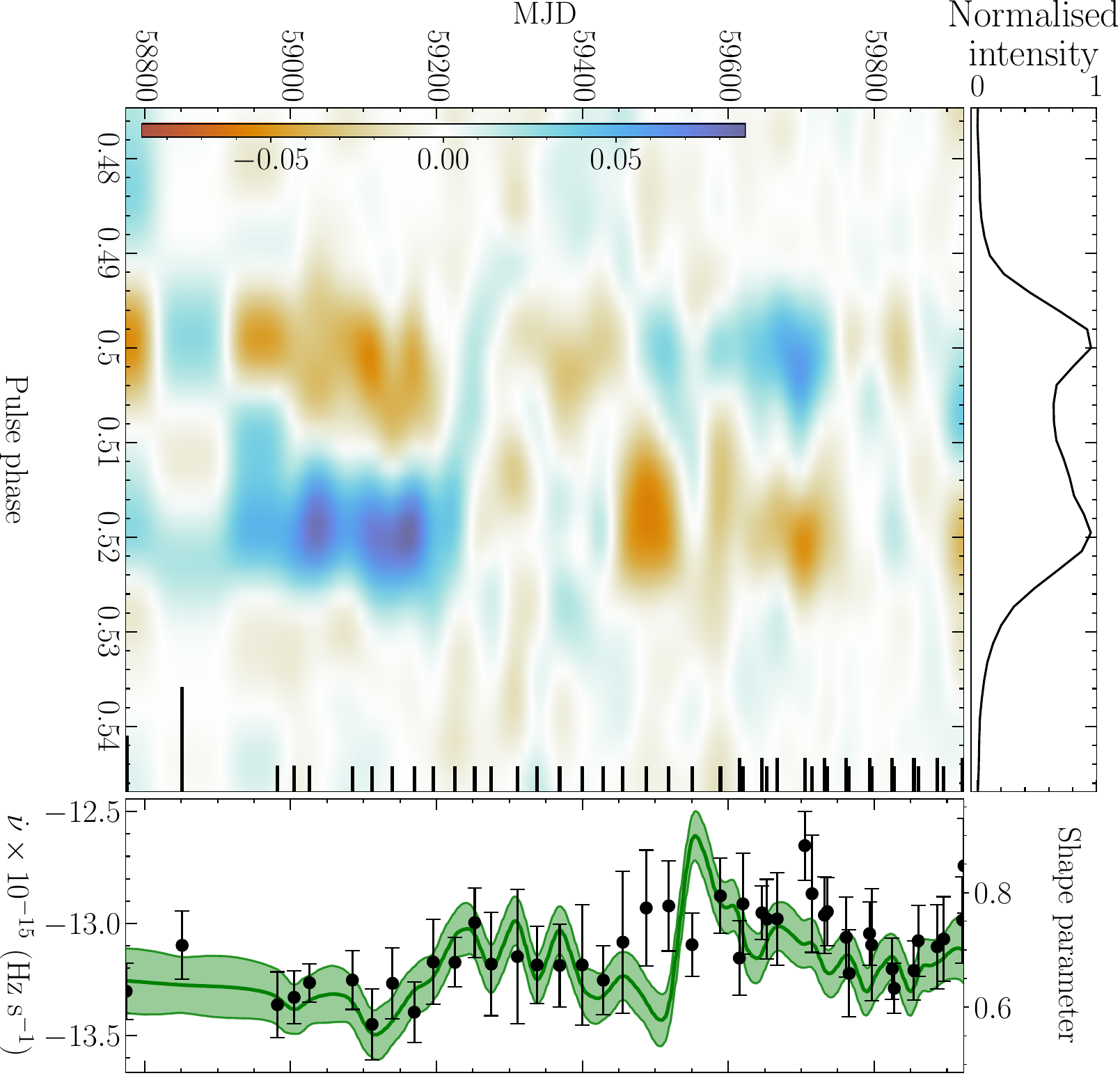}
    \hspace{0.02\hsize}
    \includegraphics[width=0.41\hsize,angle=0]{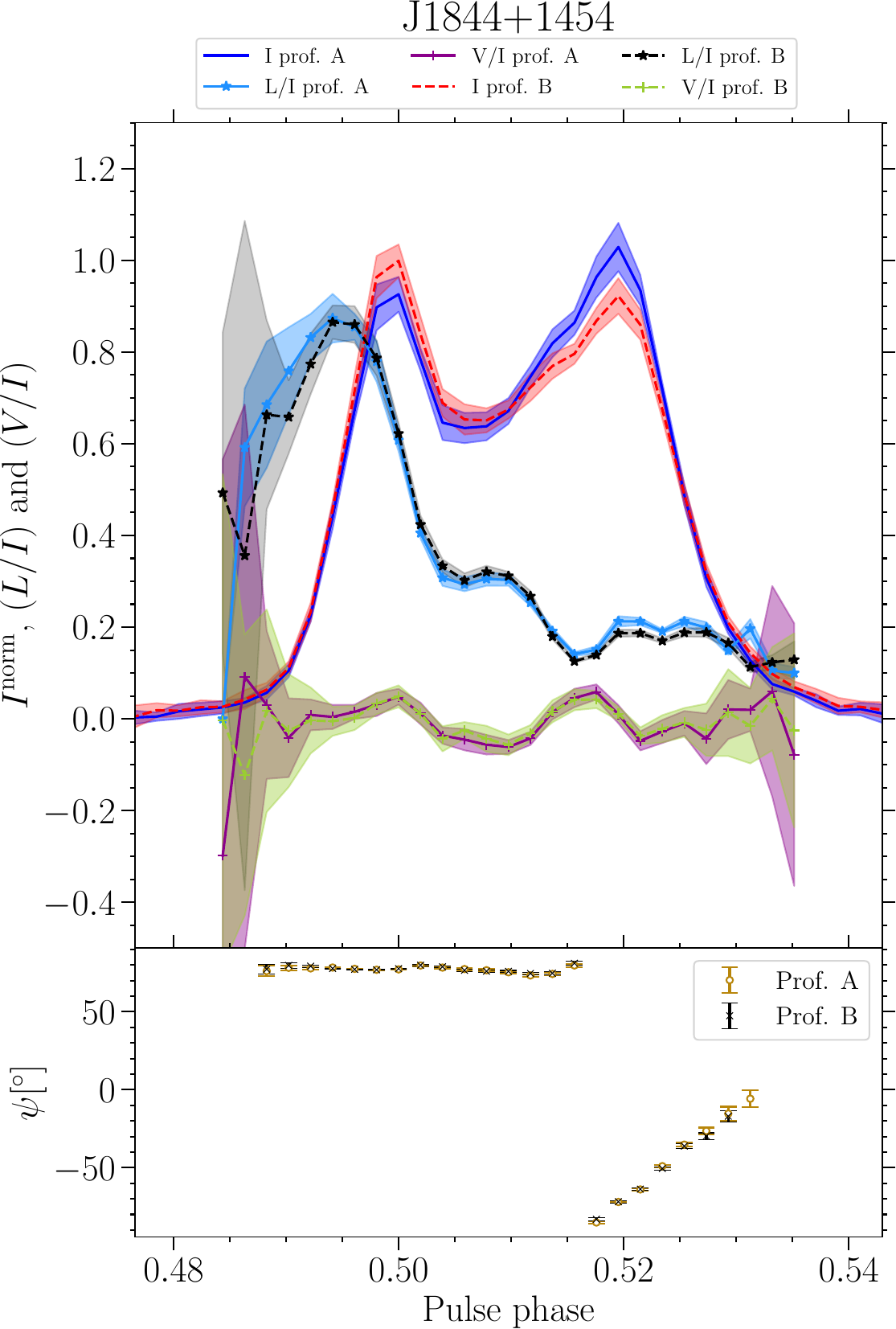}
    \caption{Provides the same information as in Fig. \ref{fig:0729diffmap}, but for the pulsar J1844+1454. For this pulsar, the amplitude ratio of the two components serves as the shape parameter. Observations between MJD 58850 and 59200 were combined to obtain profile A and observations between MJD 59400 and 59750 were combined together to obtain profile B.}
 \label{fig:1844diffmap}
\end{figure*}

As noted by \citep{Wu+1993}, a rapid PA swing occurs near the central component, suggestive of it being close to the beam axis. Therefore it could be attributed to core emission, with additional peripheral conal emission. This interpretation would be consistent with the weakening of the peripheral emission as observed at lower frequencies (see the 658 MHz profile in \citealt{MHQ+1998}). However, if both the leading and trailing components are originating from a cone, it is curious that the leading component is not participating in the variability. The absence of variability in the leading component implies that the anti-correlation in the variability in the two components is not due to localised emission changes in a single component.
%ALSO A GOOD POINT IS THAT IT IS NOT JUST CORE EMISSION WHICH IS CHANGING, AS OTHERWISE ALSO THE LEADING COMPONENT NEEDS TO VARY. NEED TO BE MADE FOR OTHER PULSARS? LEAVE FOR DISCUSSION?

%This pulsar has a complex profile shape (shown in Fig.~\ref{fig:1741diffmap}) the features in the leading and the trailing edge are much less prominent at a lower frequency (658 MHz) \citep{MHQ+1998}. This implies the outriding features as conal component which is consistent with their flatter spectral behaviour \citep{Rankin1993cone, Rankin+2022}. The rapid PA swing near the central component was interpreted to be originating much closer to the beam axis \citep{Wu+1993} and may be attributed to the core emission, making this source an example of a core-cone profile. A small feature near the extreme trailing edge around pulse phase $\sim0.525$ (shown in Fig.~\ref{fig:1741diffmap}) may be interpreted as the outer cone with a large overlap with the inner cone. Outer cones are described as the exterior most nested cones of the pulsar beam to explain the leading and trailing component of the pulse profile with more than three components \citep{Rankin1993cone}.
%If the component at phase $\sim 0.52$ is considered as the inner cone and $\sim 0.525$ as the outer cone, we do not find any correlated emission change between the inner and the outer cone. Whereas, the component to the left of the core component has very faint features, making it difficult 
%to attribute the variation in a deterministic way to any specific emission component.}

\subsection{PSR J1844+1454 (B1842+14)}\label{1844}
The difference map (shown in the left-hand side of the left panel of Fig. \ref{fig:1844diffmap}) clearly indicates profile evolution such that the relative intensity of the two profile components change by $\sim 10\%$.  This type of long-term evolution (long-timescale structures) is 
%absent 
much less pronounced in the pulse jitter simulated difference maps (Fig.~\ref{J1844alldiffs} of online supplementary material). Before MJD $\sim 59200$ the leading component is relatively faint compared to the trailing component,
%and the trailing component is relatively bright, 
after which for $\sim 150$ days the emission state is closer to the overall median pulse profile. However, after this,
%a halt for 150 days, 
the emission state evolves further such that the trailing component gets brighter (MJD $\sim59450$ and onwards).
%seems to reverse the pattern as seen at all the epochs before MJD 59252. 
Therefore, we combine all the epochs between MJD 58850 and 59200 to construct the average profile A as shown in the right-hand panels of Fig.~\ref{fig:1844diffmap}.  
The observations between MJD 59400 and 59750 were used for profile B.

There is no indication in the right-hand panels of Fig. \ref{fig:1844diffmap} for a change in the polarized component of the emission, apart from a lower $L/I$ at the trailing edge of 
%the profile for 
profile A. Indeed, in the second panel of Fig. \ref{fig:1844poldiff} of the online supplementary material there is a faint signature of $L/I$ changing (between pulse phase $\sim0.52$ and 0.53) simultaneously with the change in total intensity (top panel of the same figure).
   
The two components in the pulse profile are connected by a strong bridge of emission. To model the pulse profiles, three von Mises components are used.
The shape parameter, shown in Fig.~\ref{fig:1844diffmap}, is the amplitude ratio of the two components corresponding to the two profile peaks.
Its evolution has a correlation coefficient of $0.46\pm0.15$ with the spin-down rate. This reflects that after MJD $\sim 59400$ the shape parameter is on average larger, while $\dot\nu$ is less negative. If the profile evolution is cyclic, continued monitoring should reveal if this correlation is significant.

Although the single-pulse data reveals pulse shape variability, much of the variability is associated with the bridge region in between the two profile peaks (see Fig.~\ref{fig:singlepulse}). This region is relatively constant in the difference map shown in Fig. \ref{fig:1844diffmap}, making it unclear how and if the long-term variability is linked to changes at a single-pulse level.

The anti-correlated emission variability between the profile components, as shown in Fig.~\ref{fig:1844diffmap}, implies the variation is not from the two sides of the same emission cone. Therefore, one of the components could well be the core of the beam, most probably the component centred around $\sim 0.52$ in pulse phase. This coincides with the location where the PA swing is steepest. The leading component may then be conal emission, as it is absent at the low frequencies \citep{ORVW+2022}. 
%, making its behaviour consistent with the conal emission features \citep{Rankin1993cone}. 
This then suggest the trailing part of the cone might be weak or absent.
%The trailing end may have the other side of the conal component but is extremely weak to be detected.}

\subsection{PSR J1916+0951 (B1914+09)}\label{1916}
\begin{figure*}
    \centering
    \includegraphics[height=0.56\hsize,angle=90]{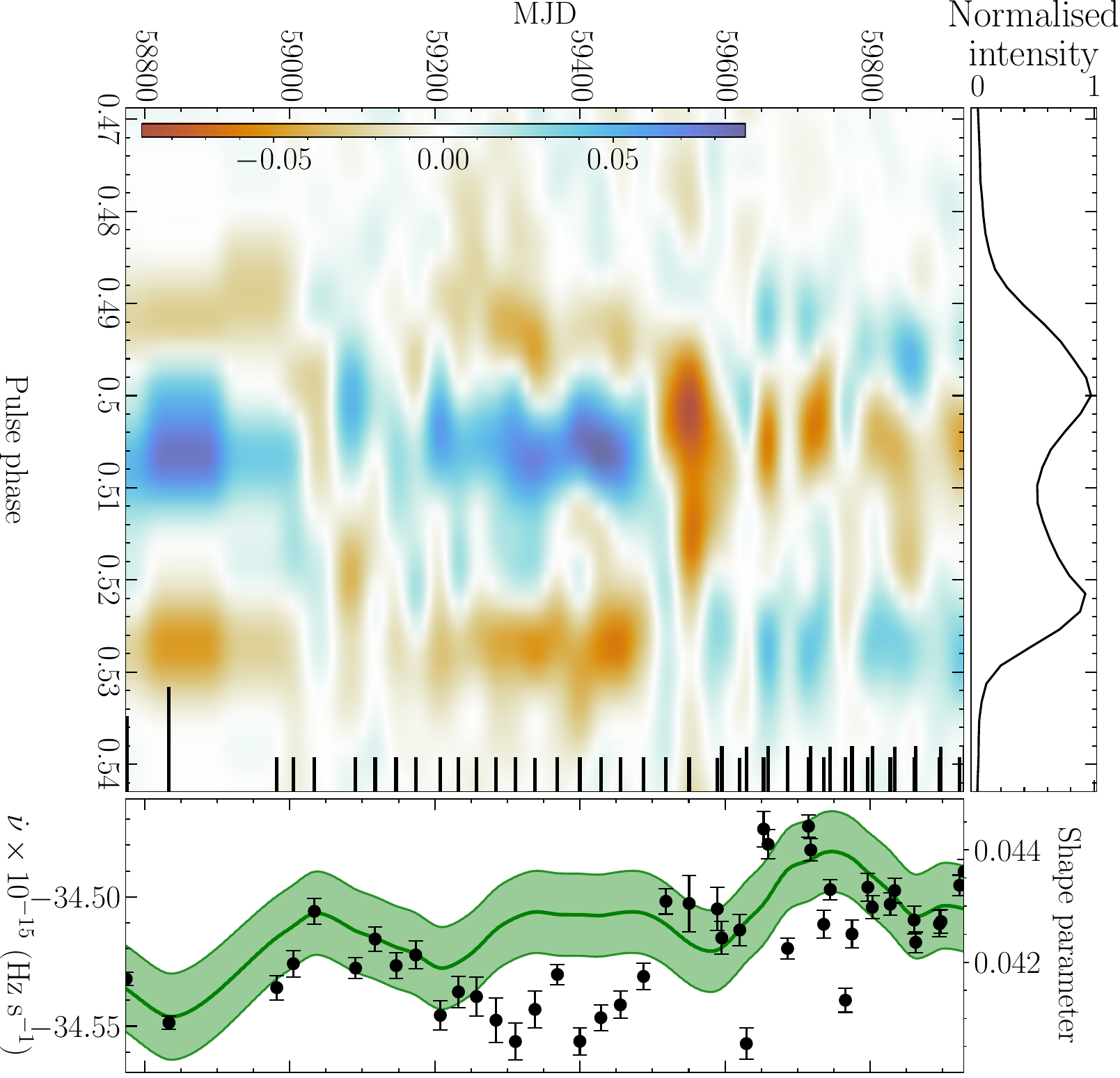}
    \hspace{0.02\hsize}
    \includegraphics[width=0.41\hsize,angle=0]{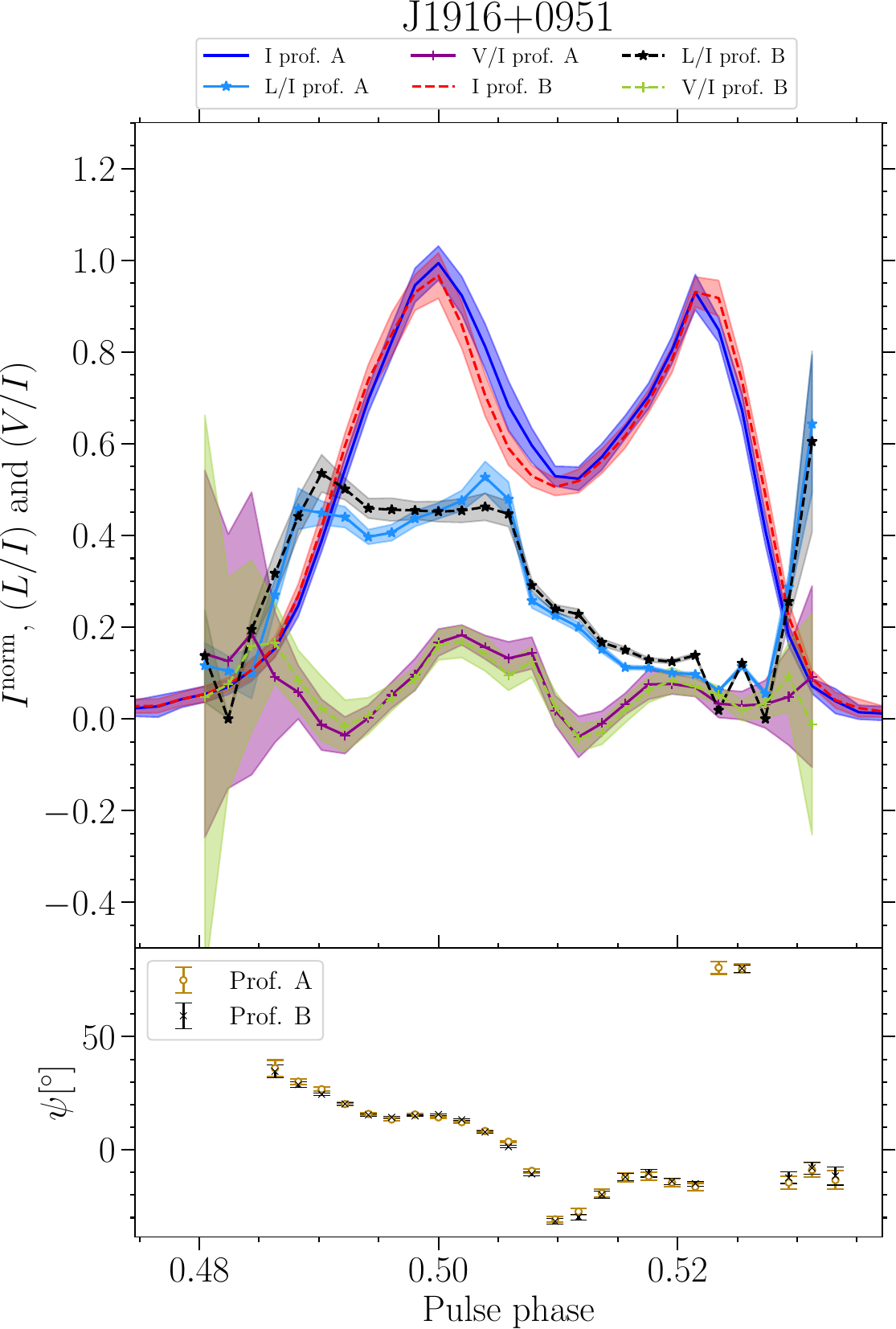}
    \caption{    \label{fig:1916diffmap}
Provides the same information as in Fig.\ref{fig:0729diffmap}, but for the pulsar J1916$+$0951. The shape parameter for this source corresponds to the width of the pulse profile at 25 per cent of the peak intensity. All profiles before MJD 59500 have been averaged to obtain profile A and the rest to obtain profile B.}
\end{figure*}

The variability in the emission of PSR J1916+0951 is evident from the difference map shown in Fig. \ref{fig:1916diffmap}. Such a systematic long-term pattern is absent in the 
%difference map obtained from the 
jitter simulations (Fig.~\ref{J1916alldiffs} of online supplementary material).
%, further strengthening our conclusion. 
The pulse profile has two distinct components
%, and both components show 
which changes in relative intensity by $\sim 10\%$. The single transition in
%from one 
emission state 
%to another 
occurs around MJD 59500, after which the trailing profile component is relatively bright.

In the right panels of Fig. \ref{fig:1916diffmap} profiles A and B correspond to profiles before and after the transition respectively. This figure furthermore reveals that $L/I$ is different in profile A such that it is weaker at the leading half of the first profile peak, but stronger at the trailing half. 
The difference map of $L/I$ (second panel of Fig.~\ref{fig:1916poldiff} in the online supplementary material) shows that indeed before MJD 59500
$L/I$ is relatively low at pulse phases just before 0.50, while it is relatively high just after this phase. There is no significant correlated change in $V/I$ and the polarisation position angle.  
The profile width, as measured at 25\% of the profile amplitude, is used as a shape parameter.
%We find the 25\% of the peak width to correlate with the $\dot \nu$, with 
The correlation coefficient of this with the $\dot\nu$ evolution is 0.4$\pm$0.1. With only one transition observed, continued monitoring is required to confirm the existence of such a correlation.

Apart from infrequent nulling, the single pulses reveal some marginal evidence for occasions where the emission appears to be shifted to slightly later phases (see Fig.~\ref{fig:singlepulse}). This might be related to the "flares" reported for PSR B0919+06 \citep{rrcw06}, although it is much less pronounced here. The spin-down of PSR B0919+06 is associated with long-term profile variation, but the association with the flares is unclear \citep{psw+15}. For PSR J1916+0951 we find no evidence the single-pulse properties change in tandem with the varying profile.

The correlated changes in the intensity in the leading and the trailing edge of the pulse profile, in combination with the anti-correlated changes in the bridge connecting the two profile component implies the profile shape may be interpreted as two sides of the same emission cone.
The central part of the profile would then correspond to core emission. Indeed \citet{ORVW+2022} proposed for this pulsar that the profile is conal double, where most of the core emission is missed by the line of slight.
%In this case, the emission variability may be pure conal or core but it is hard to disentangle which component is exactly variable.}

\subsection{PSR J1919+0021 (B1917+00)}\label{1919}
\begin{figure*}
    \centering
    \includegraphics[height=0.47\hsize,angle=90]{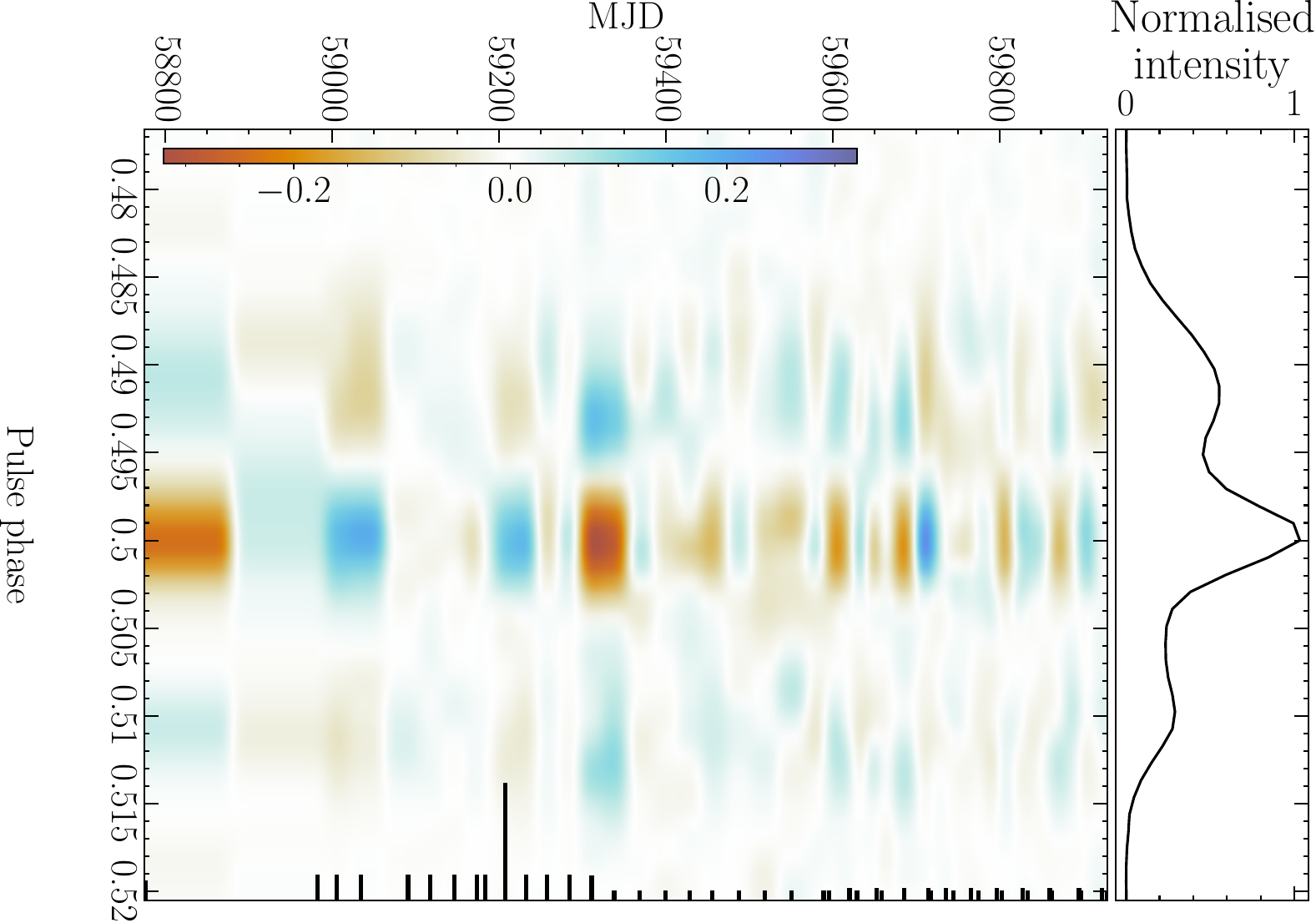}
    \hspace{0.02\hsize}
    \includegraphics[width=0.48\hsize,angle=0]{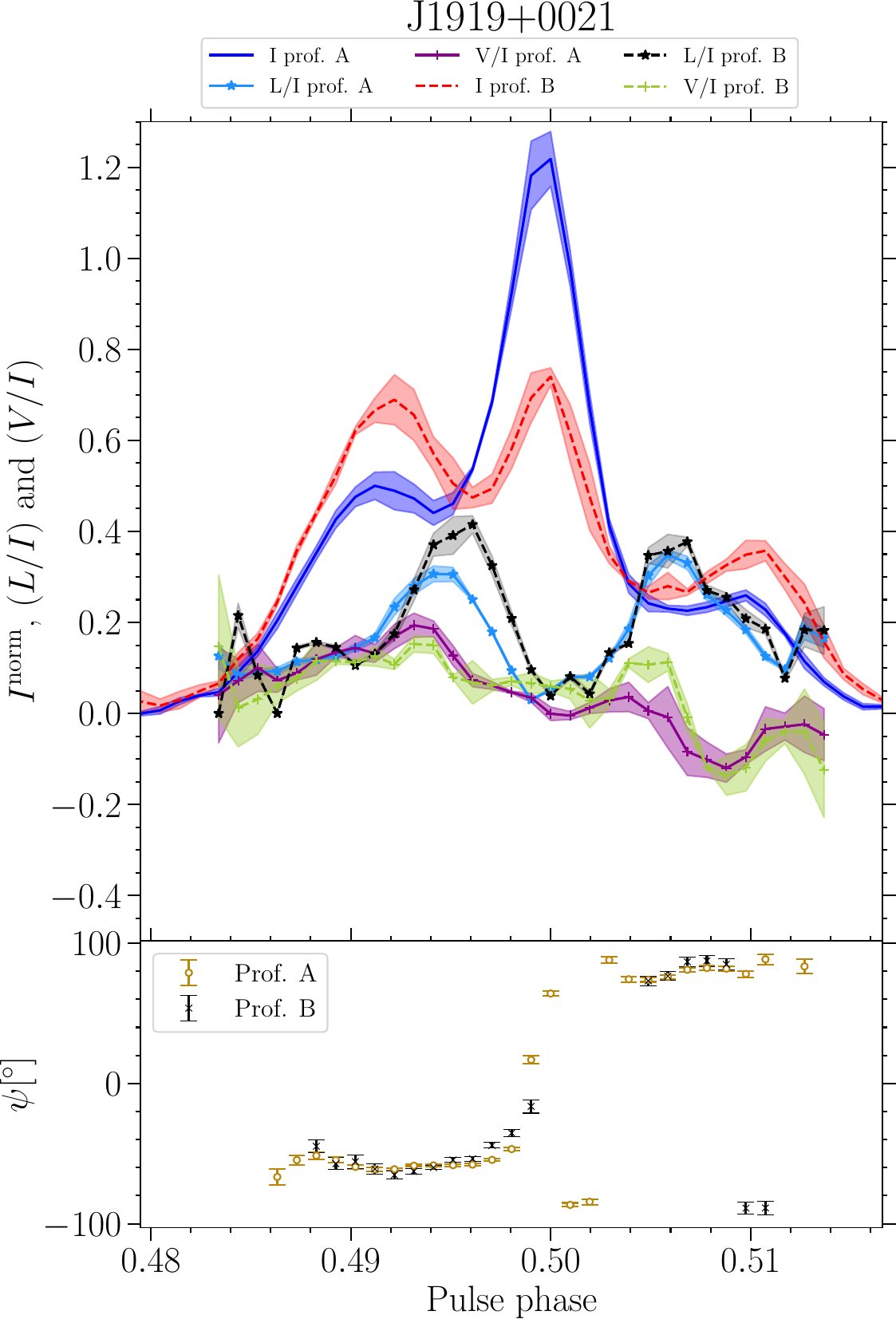}
    \caption{Provides the same information as in Fig.\ref{fig:0729diffmap} but for the pulsar J1919$+$0021.  Epochs when the central component was bright were combined to produce profile A and when dim were combined to produce profile B. In this case, no significant variability in $\dot\nu$ was detected, hence the panel showing its evolution was omitted.
    \label{fig:1919diffmap}
   }
 \end{figure*}

%{\bf [This text changed a lot, so make sure you agree with it.]}
The variability in the difference map shown in Fig.~\ref{fig:1919diffmap} shows slower variations until $\sim 59350$, after which the variations are 
%, exhibits a much systematic variation till MJD $\sim 59310$, after which the variations are much 
more stochastic. This is expected from the jitter simulation, as shown in Fig. \ref{J1919alldiffs} of the online supplementary material. 
%After MJD 59310 the integration time was reduced to half of its previous integration time. Hence, the stochastic variations are purely from the unstable profile. 
The reduced stability of the pulse profiles at later epochs is a result of a reduction in the duration of the observations.

The profile variations arise from a changing brightness of the central component relative to the outer components. The changes are very large, with the peak amplitude changing by $\sim50\%$. The origin of this variability can be traced to single-pulse variations. In Fig.~\ref{fig:singlepulse} it can be seen that the central component regularly shows burst of enhanced activity. The 
%periodic 
occasional
enhancement of the emission of the outer components in Fig.~\ref{J1919alldiffs} is the result of how the profiles are averaged, which makes the average flux density of the profiles constant. Analysis of the flux density distribution of the central component in single pulses does not reveal clear bi-modality. So there is no evidence the profile variability within observations represents mode changes where the emission state switches between two distinct states. This extreme single-pulse variability results in an exceptionally large modulation index, which peaks at 3 in the middle of the profile \citep{Song+2021}. 

The profile variability in Fig.~\ref{J1919alldiffs} is not well resolved with the cadence of our observations. This, in combination with the fact that the extreme single-pulse variability makes the profile of this pulsar exceptionally unstable, means that there is no evidence that there is long-term profile evolution of this source. This highlights the importance of considering the expected pulse jitter when establishing profile evolution. No significant variation of $\dot \nu$ is detected during the  $\sim3$ years of observations, hence it is omitted from the figure.

For completeness, the stochastic profile variation is highlighted by segregating the observations into two groups. Profiles A and B in the right panels of Fig.~\ref{fig:1919diffmap} correspond to epochs when the central component is relatively bright (profile A) and dim (profile B).
%respectively.
Given the stochastic nature of the profile variability, here the polarization evolution is not described in detail. It is noted, however, that there are relatively large ($\sim 30^\circ$) variations in the polarisation position angle difference map (see lower panel of Fig.~\ref{fig:1919poldiff} in the online supplementary material at a phase of 0.50). This is also evident in the $\psi$ swing of the mode-separated profile (shown in the lower-right panel of Fig.~\ref{fig:1919diffmap}).
Similar large deviations are seen in the jitter simulations.
Unlike PSR J1741$-$3927, there is no evidence of OPM activity. However, the PA swing is very steep at this phase. Therefore small offsets in the alignment of the profiles of different observations will have a large effect. 
%%%%This is not incorporated in the jitter simulations

We concluded that the profile variations arise from a changing brightness of the central component relative to the outer components. This suggests that the intensity of the core emission is changing relative to the conal emission. Indeed, the pulse profile is classified as a core-cone triple \citep{RWVO+2023}.
%The anti-correlated change in the emission between the central and the peripheral emission component makes it follow the classic definition of a core-cone triple \citep{RWVO+2023}. The short-term variability associated with this pulsar may be attributed to the core or correlated variation in the cone components.}
\section{Discussion}\label{discussion}
\subsection{$P- \dot P$ diagram}

\begin{figure}
    \centering
    \includegraphics[scale=0.4]{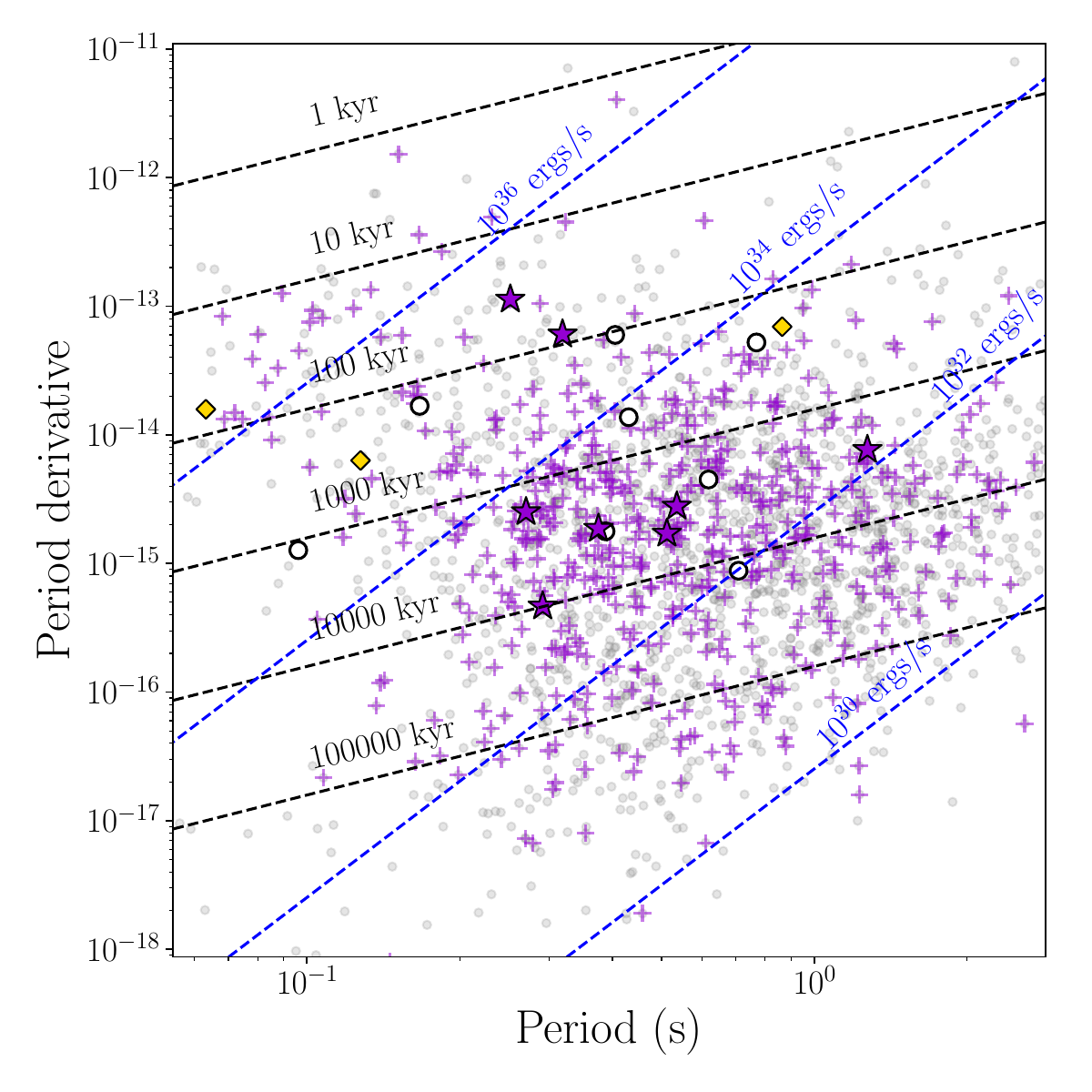}
    \caption{$P-\dot P$  diagram roughly centred around the normal pulsar population (shown in grey points) compared to the sample of TPA pulsars currently being monitored (``+'' markers), and the pulsars presented in this work (``$\star$'' markers). The hollow circle and the diamond points indicate the pulsars from \citet{Brook+2016} (PSRs J1105$-$6107, J1359$-$6038 and J1602$-$5100) and \citet{Shaw+2022} (PSRs J2043+2740, J2037+3631, J1830$-$1059, J1825$-$0935, J1645$-$0317, J1543$-$0620, J0922+0638 and J0742$-$2822). Lines of constant characteristic age ($P\dot{P}^{-1}/2$) and spin-down energy loss rate ($-4\pi^2 I\dot{P}P^{-3}$) are shown, where the moment of inertia $I$ of the pulsars is taken to be $10^{45}$~g~cm$^2$.
    \label{fig:ppdot}
}
\end{figure}

In Fig.~\ref{fig:ppdot}, we show the %zoomed-in version of the pulsar
location of pulsars in 
%their 
period ($P$) and period derivative ($\dot P$) space (commonly known as the $P-\dot P$ diagram).
%closely centred at 
The figure shows mainly pulsars which are part of the bulk 
%{\bf [As noted in the caption, this is not entirely true as it is.]} 
of the normal pulsar population. 
A comparison of the pulsars monitored by the TPA with  the known normal pulsar population 
%(shown by "+" symbols {\bf Can you use more contrasting colours? Maybe orange or something (which you also use for the stars, which are TPA as well)?}) 
shows that 
%this 
the TPA sample represents a largely unbiased sample of the known normal pulsar population. The same figure also highlights the position of pulsars with known correlated emission and rotation changes as reported in \citet{Brook+2016} and \citet{Shaw+2022}. 
As pointed out by e.g. \citet{Lyne+2013}, 
%It is clear most of the pulsars with such variabilities 
these are 
%belong to the group of 
associated mostly to the 
small characteristic age and large spin-down energy loss rate end of the distribution. The sample of pulsars we identify to have long-term profile variability occupy a similar region in the $P$-$\dot{P}$ diagram,
which hints towards common physical processes to operate in these pulsar magnetospheres.

\subsection{Total intensity profile variability}

We present the results from our analysis on eight sources after analysing the TPA data sets obtained from the monitoring observations of $\sim$ 500 pulsars. These observations span over $3$ years. 
Profile variability has been identified by comparing the difference maps generated by differencing the observed pulse profiles with a median profile. 
Pulsars with slow profile evolution were identified, that can be resolved by the $\sim$ monthly observing cadence. 
The eight most promising sources for which no slow profile evolution had been reported before in the literature were subjected to detailed analysis as reported here.

The significance of the slow profile variability was assessed by comparing the difference maps with maps generated from jitter-simulated data. This data, randomly generated from individual recorded pulses, shows the level of profile variability to expect from pulse-to-pulse variability (stochastic or otherwise).
The importance of jitter simulation has been demonstrated while identifying the profile variations in PSR~J1919+0021 (Sec.~\ref{1919}), which seem linked to burst-like activity at a single-pulse level rather than slow profile evolution.  So for seven out of eight sources we discover long-term profile evolution for the first time.

To highlight slow systematic variations, previous works (e.g. \citealt{Brook+2016, Brook+2018_ApJ, Shaw+2022}) have made use of Gaussian process regression (GPR). In contrast, we opted to use a 2D-Gaussian convolution to suppress stochastic variability
%improve the signal-to-noise ratio (S/N) 
in the pulse profile difference maps. An advantage is that this highlights variations of emission which are correlated both in pulse phase and from observation to observation. For instance, in PSR J1121$-$5444, the 2D Gaussian convolution smoothed map (Fig.~\ref{fig:1121diffmap}) captures the variations over a wider pulse phase range compared to 
%for instance 
the GPR smoothed difference map (Fig. \ref{J1121alldiffs} of the online supplementary material, see in particular phases around phase 0.50).
To allow comparison, the original, 2D-Gaussian convolution, and GPR-processed difference maps for all sources are presented in Sec.~\ref{apdxA} of the online supplementary material. 

As noted by \citet{Lyne+2010}, the long term profile variability of some pulsars is associated with the central component of the pulse profile. As suggested by \citet{Rankin+1986}, this is suggestive of the intensity of the profile component closest to the magnetic axis (core) to be changing relative to more outer components (conal). If such intensity ratio change is driving the profile variability, one could expect that after normalisation of the profiles, the difference maps as presented in the Sec. \ref{result} show correlated changes between outer components which are anti-correlated with variability in the central region. 
We systematically searched for such signatures, which were detected in two pulsars: PSRs J1916$+$0951 and J1919$+$0021, and possibly in a third (PSR J1121$-$5444, although not significant). However, the overall picture is inconclusive, with counter examples being PSRs J1141$-$3322 and J1741$-$3927, and for the remaining pulsars in our sample it seems that the conal emission is largely missing at one side of the profile. 

Two of the pulsars in our sample stand out as having a relatively large spin-down energy loss rate ($\dot E$). These are PSRs J0729$-$1448 and J1705$-$3950. In both cases the core emission might be blended with the trailing conal emission because of relativistic effects. \citet{Rankin_2020}, based on the theory of pair formation geometries \citep{Timokhin+harding2015}, have argued that $\dot E$ plays an important role in determining the profile structure, such that conal emission dominates the emission below $\dot E\approx 10^{32.5}$ ergs s$^{-1}$, and core emission dominates at higher values of $\dot E$. All pulsars with a signature of an intensity ratio change between core and conal emission have values of $\dot E$ relatively close to the transition $\dot E$ (between $10^{32}-10^{34}$ ergs s$^{-1}$).

%PLAN: Two highest Edot pulsars both inconclusive, with double looking profiles, because of em height/a/r effects? So that kind of makes sense. Some correlation with Edot might be expected, since Edot is related to core/cone intensity ratio stuff discussed in previous paragraph: Joanna stuff. Pulsars obeying correlated/anti-correlated expectations do not appear to correlate with Edot.

%{\bf The difference map also captures the variability associated with the various components of the pulse profiles, i.e. the core or cone components. The origin of the core/cone components can be tied to the physics of the plasma formation in the pulsar magnetosphere. A boundary line of spin-down energy of $\sim 10^{32.5}$ ergs s$^{-1}$ in the pulsars $P-\dot P$ diagram allows the classification of pulsar into core or cone component dominant emission process (for further details read \citet{Rankin_2020}).
%
%In the spirit of identifying if the long-term variability is associated with any particular component of the pulsars we systematically searched for such signatures as presented in the Sec. \ref{result}. However, we find the long-term variability may be associated with both components, in some cases it is hard to disentangle the variability between the components. Therefore the emission variability presented in this paper cannot be used to derive any constraints on the plasma physical origin of the long-term variability of the radio emission process.

Pulsars with correlated changes in their emission and spin-down rate operate on a wide range of timescales (e.g. \citealt{Shaw+2022}). This can be relatively short with a timescale of $\sim 50-100$ days 
such as for PSR B0740$-$28 \citep{Shaw+2022}. But it can also be much longer with a timescale of $\sim 1000-2000$ days for e.g. PSRs B1540$-$06, B1642$-$03 and J2043+2740.
The timescales we identify in this work
%find 
for the profile variability fall within this range.
For example, PSR J1705$-$3950 (see Fig.~\ref{fig:1705diffmap}) has a rapid transition between states with a time-scale of $\sim 50$ days.
On the other hand, PSR J0729$-$1448 (difference map shown in Fig.~\ref{fig:0729diffmap}) shows a single transition from one emission state to another. With a single transition, the time-scale is uncertain, but it will be at least 
%, from which we derive a time-scale of 
$\sim 600$ days.
%, but as we observe only a single transition this timescale serves a lower limit estimate.
Likewise, PSRs~J1121$-$5444
%, J1141$-$3322 
and J1916+0951 show a single transition,
%{\bf [Is that true?]}, 
hence have similarly long transition time-scales.
%share similar time scale around $\sim 500-700$ days (see Sec. \ref{result} for detailed discussion.). 

\citet{Lyne+2010} identified the transition of the magnetosphere from one emission state to the other to happen abruptly. However, the profile change seen in PSR J0738$-$4042 shows a gradual transition from one state to the other before it attains its new stable shape \citep{Karastergiou+2011, Brook+2014}.
We find a similar gradual transition taking place over $\sim 150$ days from one state to another in PSR J1844+1454 (shown in Fig. \ref{fig:1844diffmap}, for details see Sec. \ref{1844}).

\subsection{Polarized profile variability}\label{poldiss}

%Utilising 
The excellent polarization performance and sensitivity of MeerKAT allowed us to search for associated changes in the polarized component of the emission. Polarization variability is found to be common. We find a marginal detection of linear polarisation fraction changes associated with the total intensity changes in 5 out of the 7 sources with slow profile variations in total intensity.
Marginally significant changes in the circular polarisation fraction is detected in only 2 sources out of 7 sources. Typically the circular polarisation fraction is lower than the linear polarisation fraction and hence detection of significant variability in circular polarisation is expected to be less likely.

The change in the linear polarisation fraction is typically small ($\lesssim 10\%$).
%ranges between $\sim 3\%-7\%$, 
%which is much smaller compared to the 
Pulsars showing profile switches within the duration of individual observations can have polarization changes over a wide range of magnitudes.
Both the linear and circular polarization fraction can change between $\sim 1-30\%$ (e.g. \citealt{BMSH+1982, BM+2018, Brinkman+2019, SYW+2021, Miles+2022, WWD+2023}). Therefore, it seems likely that the same physics is governing polarization variability over a wide range of timescales.

Apart from the degree of polarization, the position angle (PA) of the linear polarization is an important property of the polarized emission of pulsars.
We find long-term PA variations of $\sim 10^\circ$  in PSR~J1741$-$3927 (shown in Fig.~\ref{fig:1741poldiff} of the online supplementary material).
Apart from PSR J1741$-$3927, we do not find any PA changes in tandem with total intensity variability for other pulsars. The fact that PA changes are observed shouldn't be a surprize, given PA variability is also seen in mode-changing pulsars which change emission mode within an observation (e.g. \citealt{Yan+2023}).
Although there are examples of pulsars with changing PA swings associated with mode changes, this is certainly not always the case (e.g. \citealt{Brinkman+2019, RBMM+2020}). 
Where PA variability is seen in short-term mode changes, they can sometimes be associated with the emergence or disappearance of OPMs (e.g.  \citealt{Karastergiou+2011, BPM+2019, Wen_2020, Sun+2022}). This is also the case for PSR~J1741$-$3927, where OPM activity appears to play a role.

The PA provides crucial information about the pulsar magnetosphere. An ``S'' shaped swing is observed in many pulsars (e.g. \citealt{JKK+2023}) and can be modelled using the Rotating Vector Model (RVM; \citealt{RC+1969}), which helps in inferring the orientation
%geometry 
of the pulsar's magnetic field structure
%magnetosphere 
%including its orientation 
with respect to the rotation axis.
%\textcolor{red}{and the }.
%The information can be further extended to infer the 
Emission heights can be inferred via the aberration and retardation effects \citep{BCW+91}. A difference in emission height between two emission states would imply a relative shift in phase between their PA swings. A change in polar-cap current density would lead to an offset in PA \citep{HA+2001}. Therefore variations in PA can be expected if the relative orientation of the magnetosphere changes as a function of time (for example due to precession),
%as discussed in Sec. \ref{intro}), 
or if the emission height changes between the various magnetospheric states, or if there are changes in the current density in the magnetosphere. 
%{\bf [Make sure you agree with next added sentence:]}
Apart from the OPM related activity in PSR~J1741$-$3927, no long-term changes in PA were detected.

A change in current density between states would result in spin-down changes. It could also lead to a change in the radio emission co-latitude \citep{Timokhin+2010}, hence emission height.
%{\bf [is this true?]}. 
In such a scenario one would expect PA variability in tandem with Stokes $I$ changes. Periodic variations in the PA can also arise if a pulsar undergoes precession \citep{WEC+2010}, which should also be accompanied by a periodic/quasi-periodic variation in $\dot \nu$. 
However, we do not detect quasi-periodic variations in $\dot \nu$ in our sample of pulsars, so it is not clear if precession or changes in current density apply. 
%If it is, then 
Therefore, the non-detection of PA changes implies that the changes in radio emission height, or orientation of the magnetic field relative to the line of sight needs to be relatively small to avoid the detection of associated PA variability. 

From the statistical uncertainty in the PA difference maps, we derive limits on the relative shifts of the  PA swing relative to the pulse profiles,  
%using the prescription of \citet{BCW+91},  
before detectable changes in the PA difference maps are expected. 
When deriving these limits, the smooth monotonic parts of the PA curve were considered which are potentially RVM-like. Discrete jumps arising from OPM transitions (or wraps in PA) were ignored.
These limits on the relative shift correspond to limits on the possible emission height changes associated with the profile changes according to the prescription of \citet{BCW+91}.
%If at all the state change is associated with the change in 
This shows that if the emission height changes, it should be within 0.2 to 1 per cent of the radius of the light cylinder of the pulsars. To put this into context, \citet{JK+2019} have shown that the radio emission from pulsars originates from a height of $\sim$ 200 to 400 km above the polar cap and this is independent of the pulse period. Taking typical emission heights to be 300 km, the relative change in the emission height $h_\mathrm{em}$ as presented in Table \ref{tabbetaH} are obtained. 
Therefore, for most pulsars, relatively large fractional changes in the emission height are required before a detectable signature in the PA variability can be detected. Here it should also be noted that any emission height changes associated with the profile variability are expected to be small, given the absence of large relative changes in the pulse profile width. 

We also explored the scenario where the evolution in the pulse profile is due to precession in the system, while the change in the $\dot \nu$ is too subtle to be detected. 
Precession does not change the angle between the pulsar's spin and magnetic axis, but the impact angle \footnote{The impact angle is defined as the angle between the line of sight when it is closest to the magnetic axis.} ($\beta$) evolves with time. According to the RVM, a change in $\beta$ should correspond to a change in the PA swing. Here we derive a limit on the change in $\beta$ before it would leave a detectable signature in the data.

According to the RVM, the steepest gradient $g$ of the PA swing obeys $g\propto1/\sin\beta\approx1/\beta$ (since radio beams, hence the impact angle, are relatively small). Therefore, if $\beta$ is perturbed by $\delta\beta$ it follows that the gradient gets perturbed as well: $\delta g/g=\left|\delta\beta/\beta\right|$.
To estimate a limit on the fractional change in $\beta$ we can therefore determine the maximum fractional change in the gradient of the PA swing before it becomes detectable. The effect of gradient changes is modelled by scaling the observed PA values of the PA swing (ignoring discontinuities such as OPMs).
The estimated limit on $\delta \beta / \beta$ is given in Table \ref{tabbetaH}.
%after which the variation in the PA difference map should be detectable.
This shows that in general relatively large fractional changes in the $\beta$ are required before significant  PA variability can be expected. Such large changes in the line of sight relative to the emission beam can be expected to result in much more significant changes in the pulse profile compared to what is observed.

Note that in Table~\ref{tabbetaH} PSR J1741$-$3927 is omitted. This is the pulsar for which PA variability correlated with total intensity variations has been detected (see Sec.~\ref{1741}). However, upon closer inspection, this PA variability is related to the pulse longitude regions where OPM transitions are observed. Therefore, what is observed is more complicated than a simple shift of a smooth PA curve in pulse longitude, or a RVM-like change because of a changing $\beta$. Therefore it is unclear how the observed PA variability could be related to a change in emission height or precession.

\begin{table}
\begin{center}
\begin{tabular}{llll}
\hline \hline
PSR J & PSR B&$\frac{\delta h_\mathrm{em}}{h_\mathrm{em}}$ ($\%$)& $\frac{\delta \beta}{\beta}$ (\%) \\ \hline \hline
J0729$-$1448 &     $-$           &10        & 10   \\
J1121$-$5444 &   B1119$-$54   &25        & 10   \\
J1141$-$3927 &   $-$       &3         & 6    \\
J1705$-$3950 &    $-$     &95        & 60   \\
%J1741$-$3927 &      &50        & 40   \\
J1844+1454 & B1842+14   &20        & 20   \\
J1916+0951   &  B1914+09  &30        & 10   \\
\hline
\end{tabular}
\end{center}
\caption{The table shows the limits on the relative change in the radio emission height ($h_\mathrm{em}$) in the third column and the relative change in the impact parameter angle ($\beta$) in the fourth column. Larger changes in these quantities are expected to produce detectable PA variability. The name of the pulsar is given in the first column (J Name) and the second column (B name).
These are the 6 pulsars with long-term profile evolution excluding PSR J1741$-$3927 which has more complicated  PA variation. See Sec.~\ref{poldiss} for a detailed discussion.}
\label{tabbetaH}
\end{table}

\subsection{Spin-down variability}\label{spindown_var_diss}

%\textcolor{red}{
Out of seven sources studied with long-term profile variability, only PSR J1141$-$3322 shows a hint of a possible correlated change in emission and spin-down rate.
%, apart from that in other sources we do not detect any significant correlation. 
%{\bf [Left out statement of 3 other pulsars. The conclusion is already in the previous sentence that there is only one "maybe" pulsar.]}
%Though the correlation coefficient indicates some marginal correlation between emission and spin-down in PSR J1121$-$5444, J1844+1454 and J1916+0951 yet they lack any kind of visual evidence for such correlation and large uncertainties on the measurements make these correlations much weak. 
Given the many examples of pulsars in the literature for which correlated changes in spin-down and profile changes have been established, 
%albeit on much longer time scales {\bf [Don't understand why this is true???]}, 
this raises the question if the current time span of our observations is sufficient to expect to see correlated changes.
%These observations bring a question, of how likely it is to find sources which show correlated changes in the profile and the spin-down rate. And how critically it depends on the observation time span?}

%\textcolor{red}{
%Previously 
\citealt{Lyne+2010} discovered correlated changes in only six sources, after selecting 17 sources showing strong 
%{\bf [you mean strong? or the strongest?]} 
timing noise (see also \citealt{Lyne+2013}) out of 366 pulsars reported in \citet{HLK+2010}. 
They also point out that because of observational limitations (such as sensitivity and the cadence of the observations), it well possible that all pulsars which display timing noise could have associated pulse shape variability. The profile variability was identified for two more sources
%Later the number increased to eight 
%after 
%reanalysing the sources with the 
in an extended 
%database
data set \citep{Shaw+2022}. 
%{\bf You had "On the other hand", but don't you mean similarly?} 
Similarly,  \citet{Brook+2016} analysed a sample of 168 pulsars and only 7 showed a noticeable long-term change in the pulse profile, out of which only 4 showed a marginal correlation between emission and spin-down rate. One additional source, PSR J1602$-$5100, exhibits strong correlated changes in emission and spin-down rate. This indicates the chance of detecting pulsars with substantial long-term variations in profile shape is low and establishing changes in spin-down associated with profile change is observationally challenging.
%far more rare. 

%\textcolor{red}{
Given the diverse nature of correlated changes between emission and spin-down, there is no unique minimum required length of an observing programme to establish such a correlation. Sources like PSR B1828$-$11, which show a strong quasi-periodic variability with the time scale of $\sim 500$ days \citep{Stairs+lyne+2000, Stairs+Lyne+2019, Shaw+2022} 
%ideally 
should be detectable within the data span (3 years) used in this paper. However, 3 years 
%time span 
may or may not be enough to detect correlated changes in sources like PSR B2035+36 which exhibit a large step change in emission and spin-down rate \citep{Shaw+2022}, but do not show any clear signature of correlated change in a window of $\sim 3$ years prior or posterior to a step change despite the presence of significant profile variability. 

It should also be noted that a good fraction of sources in \citet{Shaw+2022} and \citet{Brook+2016} show much bigger changes in their profile shape
%, which is correlated with the spin-down change, unlike our sample where 
compared to the subtle changes reported here. 
One could therefore expect that any associated correlated spin-down changes are also relatively small in our sample, making them more difficult to detect.

The relationship between the emission change and the spin-down rate is not direct and simple \citep{Brook+2016}. \citet{Lyne+2010} find sources exhibiting $\dot \nu$ change but no correlated change in the emission, which was interpreted as a possible emission change in a part of the beam not sampled by the line of sight between the Earth and the pulsar. 
Similarly, the non-correlation between the emission and spin-down may originate from a large non-linear response of the radio emission process due to a slight change in magnetospheric current which is not enough to 
%bring 
cause a significant change in the spin-down torque. 
So in conclusion, %Therefore based on these existing observations and our results, we conclude 
%the current sample does not have a new source which shows some correlated spin-down and emission change with a time scale of less than 3 years. However, 
an ongoing monitoring programme clearly has the potential 
%for future detection of sources not yet known to have 
to establish correlated emission and spin-down changes for the sources studied here, as well as for additional sources.
%}

\begin{figure}
    %\centering
    \includegraphics[scale=0.4]{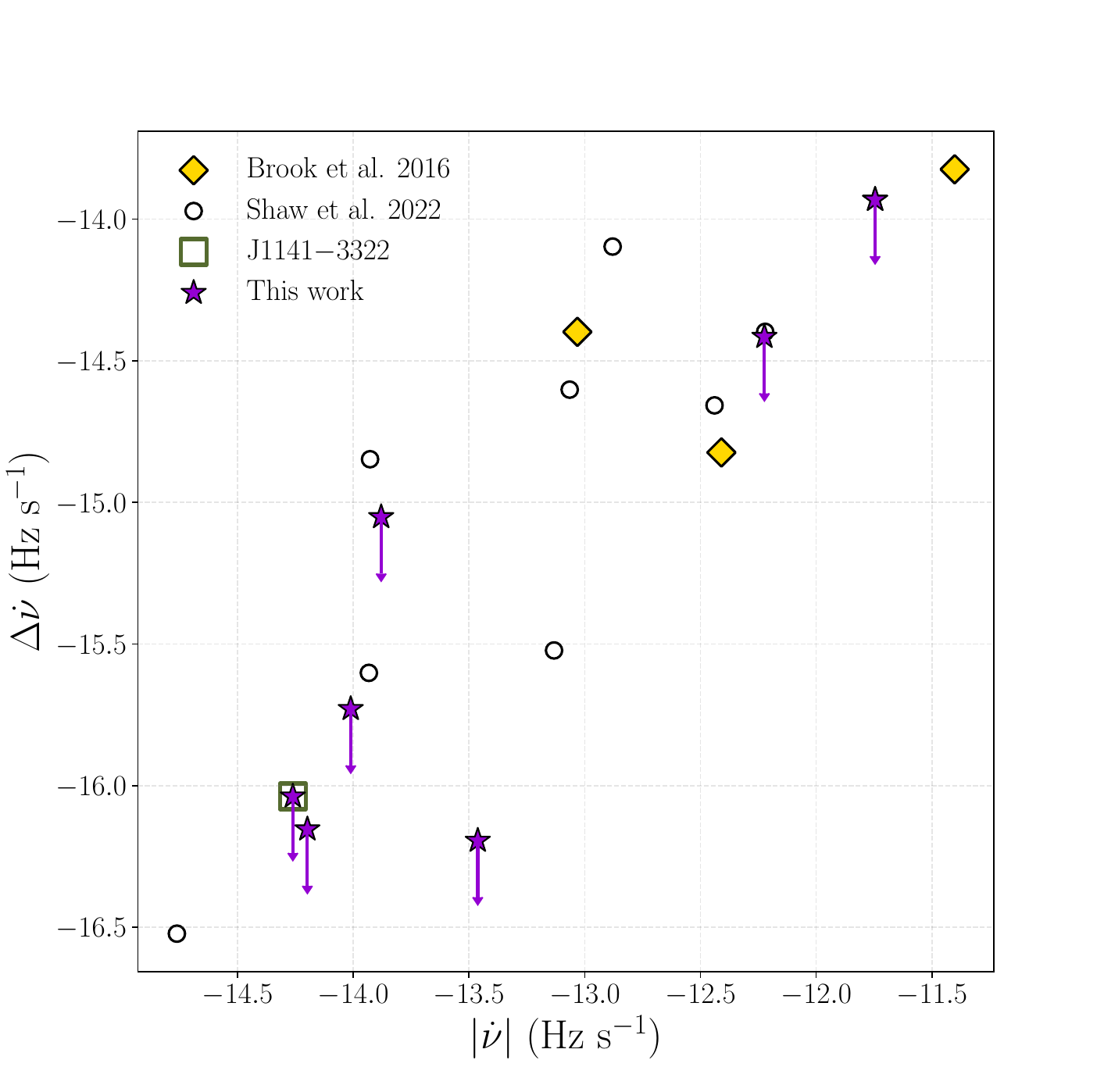}
    \caption{Diagram to show the trend between the changes in the spin-down rate ($\Delta \dot \nu$) and the spin-down rate ($\dot \nu$) itself. The hollow circles show the pulsars from \citet{Shaw+2022}  and the diamond symbols indicate the pulsars from \citet{Brook+2016}. In both cases, only pulsars for which evidence of correlated emission and $\dot\nu$ changes have been mentioned have been included. The stars show the seven pulsars presented in this paper for which long-term pulse profile variability has been established. All their measurements are treated as an upper limit, as no significant correlated emission and spin-down changes are observed. The exception is PSR J1141$-$3322, indicated by a square box with a star inside, for which there is an indication for a possible correlation.   \label{fig:nudot_deltanudot}
}
\end{figure}

%\textcolor{red}{
For all seven pulsars with long-term profile variability, we compute the peak-to-peak difference in spin-down rate $\Delta \dot \nu$. 
The relationship between the spin-down rate and its change is shown in Fig. \ref{fig:nudot_deltanudot}. 
In the absence of any clear 
%visual 
evidence of correlated spin down and emission changes, all the estimates of $\Delta \dot \nu$ are considered as an upper limit for the change in spin down associated with the profile variability.
%{\bf [1141 is not an exception, but a possible exception. With that in mind, I don't think to add that statement in the discussion. It is already in the caption.]}
%, with PSR J1141$-$3322 being a possible exception. 
These upper limits are compared to sources from \citet{Shaw+2022} and \citet{Brook+2016} for which there is evidence for correlated changes. 
%as followed by sources with correlated changes. 
As pointed out in \citet{Lyne+2010}, the largest spin-down rate changes are observed in sources with a large $|\dot\nu|$. Our upper limits follow a similar trend.
Since we only established upper limits, this suggests that if there are correlated emission changes, the magnitude of the correlated spin-down rate changes is likely to be on the small side.

%\subsection{$\Delta \dot \nu$ and charge density connection}
\begin{table}
\begin{center}
\begin{tabular}{llll} \hline \hline
PSR J & PSR B & $\Delta \dot \nu$ (Hz s$^{-1}$) & $\frac{\Delta \rho}{\rho_{\mathrm{GJ}}} (\%)$ \\
\hline  \hline
J0729$-$1448 & $-$  & 1.2$\times 10^{-14}$      & 1.3    \\
J1121$-$5444 &  B1119$-$54 & 1.9$\times 10^{-16}$     & 3.8    \\
J1141$-$3322 & $-$  & 9.1$\times 10^{-17}$      & 3.3    \\
J1705$-$3950 & $-$  & 3.8$\times 10^{-15}$      & 1.3    \\
%J1919+0021 &  & 2.0$\times 10^{-16}$      & 6.5    \\
J1741$-$3927 & B1737$-$39  & 7.0$\times 10^{-17}$      & 1.0    \\
J1844+1454 & B1842+14  & 8.9$\times 10^{-16}$      & 5.1    \\
J1916+0951 &  B1914+09 & 6.4$\times 10^{-17}$      & 2.7   \\
\hline
\end{tabular}
\end{center}
\caption{The first two columns tabulate the J- and B-name of the pulsars identified to have pulse-profile changes not caused by pulse-to-pulse jitter. 
%{\bf I added/changed, so make sure you agree:}
The third column shows the magnitude of the spin-down variation. 
Without evidence these are associated to pulse shape changes, $\Delta\dot\nu$ should be considered to be an upper limit on any spin-down rate variation correlated with the emission changes. The fourth column shows the corresponding upper limit on the change in the charge density obtained from Eq.~\ref {change_density}.\label{plasma_Den}
}
\end{table}

The correlated emission and spin-down behaviour of the intermittent pulsar  PSR B1931+24 (J1933+2421)
%, in first instance intermittent pulsar PSR B1931+24 (J1933+2421), 
%with the emission state 
was modelled by accounting for the difference between a charged depleted and charge-filled open-field line region of the magnetosphere 
%of the pulsar 
in its ``off'' and ``on'' state respectively \citep{Kramer+Lyne+2006}. This model also applies in cases where the radio emission changes are more subtle 
%rather 
than it switching off \citep{Lyne+2010}.
This model allows
%model was further used 
%generalised {\bf [I think I wrote down the equation in this form for that paper. I don't know why it is a generalisation. Isn't it just the same as in the Kramer paper? Should give the same answer as the equation in Kramer et al.]} 
the change in $\dot \nu$ to be associated with a change in the charge density of the magnetosphere. Following the parameterization of \citet{DJW+2018} this can be expressed as 
%associated with a change in the 
%by \citet{DJW+2018} to measure the change in the charge density associated with the measured change in the $\dot \nu$. 
\begin{equation}\label{change_density}
    \frac{\Delta \rho}{ \rho_{\mathrm{GJ}}} = -2 \left ( \frac{R}{10^4 \;\mathrm{m}} \right)^{-6} \left ( \frac{B_s}{10^8 \; \mathrm{T}} \right)^{-2} \left ( \frac{\nu}{1 \;\mathrm{Hz}} \right)^{-3} \left ( \frac{\Delta \dot \nu}{10^{-15} \;\mathrm{Hz\, s^{-1}}} \right).
\end{equation}
Here the change in the charge density is expressed relative to the Goldreich-Julian charge density $\rho_{\mathrm{GJ}}$ \citep{GJ+1969}. In the equation, $R$ is the radius of the neutron star. $B_s$ is the surface magnetic field strength which can be expressed in terms of the spin period ($P=1/\nu$) and spin period derivative $\dot P$ as $B_s = 10^{12} \sqrt{(\dot P /10^{-15}) (P/1\;\mathrm{s})} $ Gauss, and $\Delta \dot \nu$ is the change in the spin-down rate. 

From the estimates of the upper limit of $\Delta \dot \nu$ (which span between 1-5\%),
%(discussed in Sec.~\ref{spindown_var_diss}), 
we compute the upper limit on the percentage change of the charge density using the value of $R=10$ km. Since these are upper limits, the actual change can be much smaller than the quoted values. Using the same relations, \citet{Shaw+2022} found the percentage change in the plasma density ranges between 1-25\%. Our upper limits (Table~\ref{plasma_Den}),
%Therefore we detect the pulsars which 
belong to the lower end
%(in terms of change in the plasma density) 
of this range. 
%{\bf Happy with the following changes?: 
These stringent upper limits are possible because of the high-sensitivity observations from MeerKAT.
These allow the detection of minute changes in the pulse profile, which may correspond to small changes in the magnetospheric current and spin-down rate.

\section{Conclusion}\label{conclude}
In this paper, we have presented a detailed analysis of 8 radio pulsars out of $\sim 500$ sources monitored for $\sim$ 3 years using the MeerKAT telescope. Seven out of these eight sources are identified to have long-term pulse profile evolution. A simulation shows that the other is likely due to stochastic variability associated to single-pulse pulse-shape variability (jitter). 
%identified by comparing the difference map with their jitter-simulated counterparts. 
Profile variability is identified by making difference maps comparing pulse profiles of individual observations with a median profile.
The 
%signal-to-noise ratio of these 
difference maps are enhanced by 
%suppressing its stochastic variability 
using a 2D Gaussian convolution, which 
%further 
helps in highlighting the correlated variation both in pulse phase and from observing epoch to epoch.
The total intensity variations show quick changes from one emission state to another with the exception of PSR J1844+1454, which shows a slow transition over $\sim$ 150 days. 

%We have extended the study to 
We explore the emission variability in the polarization domain. Five out of seven sources exhibit a marginal detection of a change in the linear polarization fraction associated with the total intensity variability and only two out of seven exhibit an associated change in the circular polarisation fraction.  This shows that polarization variability is common, similar to mode-changing pulsars which switch at much shorter timescales.

None of the studied sources have PA variability correlated with the change in the total intensity, except for PSR J1741$-$3927 which shows changes associated with OPM activity.
Correlated changes can be expected if emission height changes or precession plays a role, however, the variability of PSR J1741$-$3927 is too complex to relate them to these effects.
For the remaining sources without detectable PA variability, it is found that the required fractional emission height changes or changes in the line of sight should be large before a detectable PA variability signature is expected. However, such large changes would imply larger changes in the pulse profile compared to what is observed.

We do not detect any correlated changes in the emission and the spin-down rate. Only a hint of such variation can be seen in the case of PSR J1141$-$3322. The magnitude of the observed variability in $\dot \nu$ is therefore considered to be an upper limit for any underlying correlated spin-down rate switching. These limits follow the same trend with $\dot\nu$ as seen for pulsars with detected 
correlated changes in spin-down and emission.
This suggests that the variation in spin-down seen in these sources may have a common physical origin. These variations can be attributed to 
small (upper limit between $1-5\%$) changes in the plasma density in the magnetosphere compared to the Goldreich-Julian charge density.

These observations showcase the potential of MeerKAT to detect relatively small changes happening in the pulsar magnetosphere, giving insight in 
%and help in an in-depth study of
the associated dynamical magnetospheric processes.
%Clearly, 
These observations also demonstrate the importance of upcoming sensitive telescopes such as the SKA in identifying even more subtle emission variability, thereby expanding the sample of pulsars known to exhibit magnetospheric changes over long timescales.
%capability of the upcoming telescopes like SKA in identifying much more subtle changes in the pulsar magnetosphere and help in advancing the field of pulsar astrophysics.

\section*{Acknowledgements}
The MeerKAT telescope is operated by the South African Radio Astronomy Observatory, which is a facility of the National Research Foundation, an agency of the Department of Science and Innovation. Pulsar research at Jodrell Bank Centre for Astrophysics and Jodrell Bank Observatory is supported by a consolidated grant from
the UK Science and Technology Facilities Council (STFC). AB acknowledge the discussion with Benjamin Shaw, Paul R. Brook and Andrew G. Lyne. LSO acknowledges the support of Magdalen College, Oxford. MeerTime data is housed and processed on the OzSTAR supercomputer at Swinburne University of Technology. Parts of this research were conducted by the Australian Research Council Centre of Excellence for Gravitational Wave Discovery (OzGrav), through project number CE170100004.

%%%%%%%%%%%%%%%%%%%%%%%%%%%%%%%%%%%%%%%%%%%%%%%%%%
\section*{Data Availability}
The data underlying this article will be shared on reasonable request
to the corresponding author.
 
%The inclusion of a Data Availability Statement is a requirement for articles published in MNRAS. Data Availability Statements provide a standardised format for readers to understand the availability of data underlying the research results described in the article. The statement may refer to original data generated in the course of the study or to third-party data analysed in the article. The statement should describe and provide means of access, where possible, by linking to the data or providing the required accession numbers for the relevant databases or DOIs.

%%%%%%%%%%%%%%%%%%%% REFERENCES %%%%%%%%%%%%%%%%%%

% The best way to enter references is to use BibTeX:

\bibliographystyle{mnras}
\bibliography{refppv} % if your bibtex file is called example.bib

% Alternatively you could enter them by hand, like this:
% This method is tedious and prone to error if you have lots of references
%\begin{thebibliography}{99}
%\bibitem[\protect\citeauthoryear{Author}{2012}]{Author2012}
%Author A.~N., 2013, Journal of Improbable Astronomy, 1, 1
%\bibitem[\protect\citeauthoryear{Others}{2013}]{Others2013}
%Others S., 2012, Journal of Interesting Stuff, 17, 198
%\end{thebibliography}

%%%%%%%%%%%%%%%%%%%%%%%%%%%%%%%%%%%%%%%%%%%%%%%%%%

%%%%%%%%%%%%%%%%% APPENDICES %%%%%%%%%%%%%%%%%%%%%
\clearpage

\onecolumn
\iffalse
\section*{ONLINE MATERIAL}

{\Large \bf For the benefit for the referee, the online material has been made as part of a single document. For publication, this online material will be separated in a separate document.}

\centerline{}

%{\bf Note: the following doesn't seem A specific, so why was it in A? }
This document provides figures supporting the main journal manuscript ``The Thousand-Pulsar-Array programme on MeerKAT - XII: Discovery of long-term pulse profile evolution in 7 young pulsars'' by {\it Basu et al.} published in MNRAS in XXXX. In this document, we present the 
%Appendix A of the main journal displays the 
difference maps of the total intensity for all 8 sources presented in the paper. Out of these eight, for seven we discovered long-term evolution in the pulse profile. For the other (PSR J1919+0021) the variability has been identified to be associated with short-term (at the single-pulse level) changes in the emission. 

%{\bf Seems to make sense to explain first why there are two parts.}
In Sec. \ref{apdxA} figures are presented to show the effect of smoothing on the total intensity data. In addition, the results of the jitter simulations are presented to assess the effect of single-pulse variability. The figures illustrating the profile variability in polarization can be found in Sec.~\ref{apdxI}.

\fi
\appendix

\section{Figures of total intensity difference maps}\label{apdxA}

For each of the 8 sources there is one figure, each having 5 panels. All panels are resampled uniformly in time. For details please refer to the main journal.
The top panel shows the difference map obtained from the observations without any smoothing. The second panel from the top displays the smoothened version of the same map after a 2D Gaussian convolution is applied. The middle panel shows the difference map uniformly spaced in time with each epoch predicted 
%but at all the intervening epochs where no observations were available are predicted 
using a Gaussian process regression algorithm. The second to last panel shows the not smoothed difference map 
%uniformly resampled in the time 
showing the expected variability to originate from jitter at the single pulse level. This is for the simulation assuming that there is no memory effect such that the single-pulse variability is independent from pulse to pulse: 
%. This method of simulation is called as 
the individual pulse method (IPM). The last panel shows 
corresponds to the Block method (BM) simulation of pulse-jitter, which takes memory as could arise from
%another difference map obtained from the simulation (called as Block method (BM)) capturing the structures expected to originate from the short-term variations like 
mode changing, nulling and drifting if present.

\begin{figure*}
    %\centering
         \includegraphics[scale=0.56]{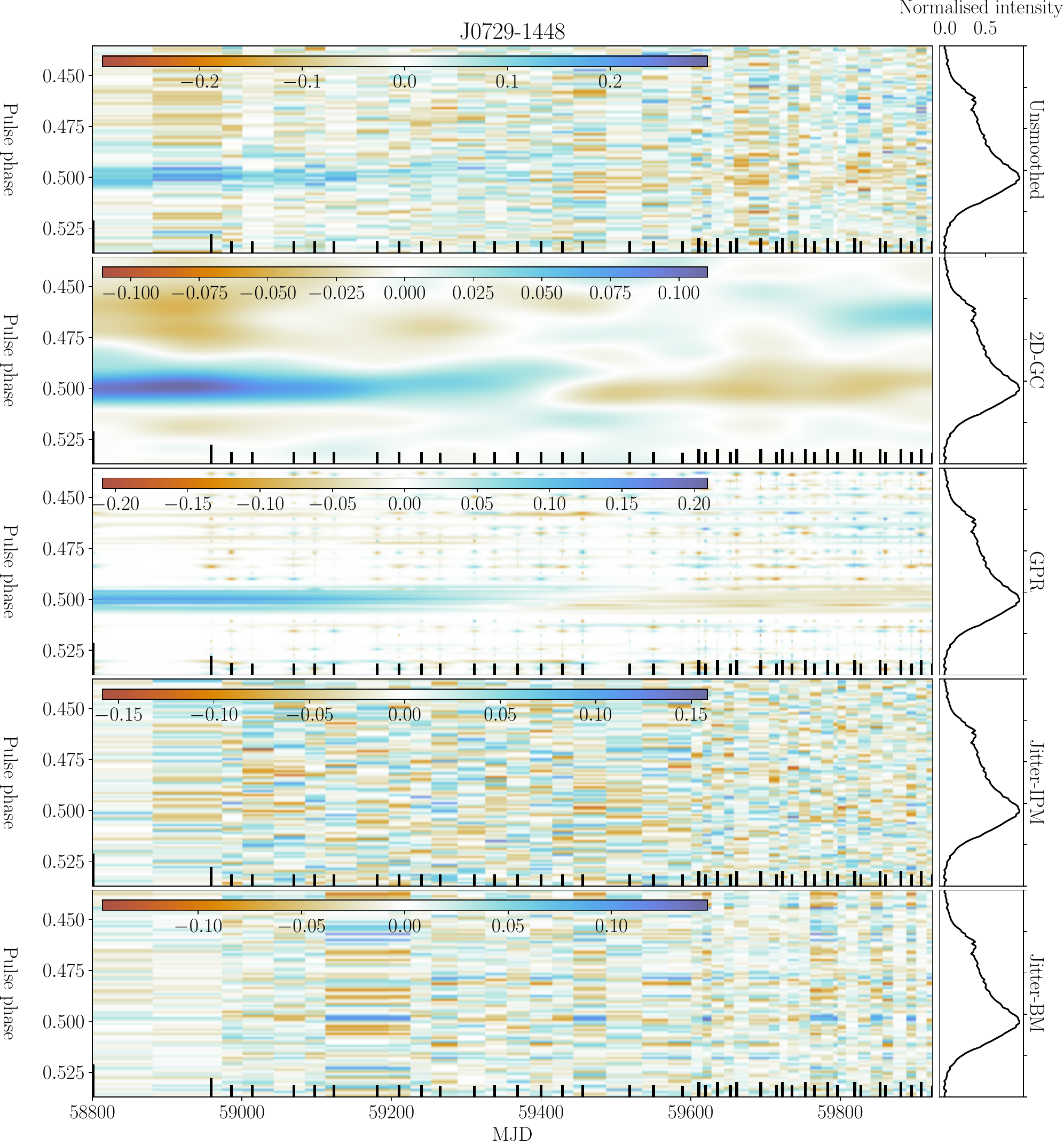}
         \caption{The top panel of the figure shows the difference map obtained after re-sampling the data uniformly in time (see Sec.~\ref{diffsmooth}) for pulsar J0729$-$1448. The second panel from the top shows the smoothed version of a difference map obtained after convolving the map from the upper panel with a 2D Gaussian kernel. The third panel from the top shows another smoothed version of the difference map obtained by the technique of  Gaussian process regression as discussed in Sec.~\ref{diffsmooth}. The fourth panel from the top shows the re-sampled version of a difference map obtained from the jitter simulation IPM method (refer to the Sec.~\ref{IPmethod}), similarly, the last panel also shows the difference map obtained from the block method of the jitter simulation as discussed in the Sec.~\ref{Bmethod}. The position of the vertical tick marks indicates the observation epoch and their length corresponds to the length of observations.}
         \label{J0729alldiffs}
\end{figure*}

\begin{figure*}
    %\centering
         \includegraphics[scale=0.56]{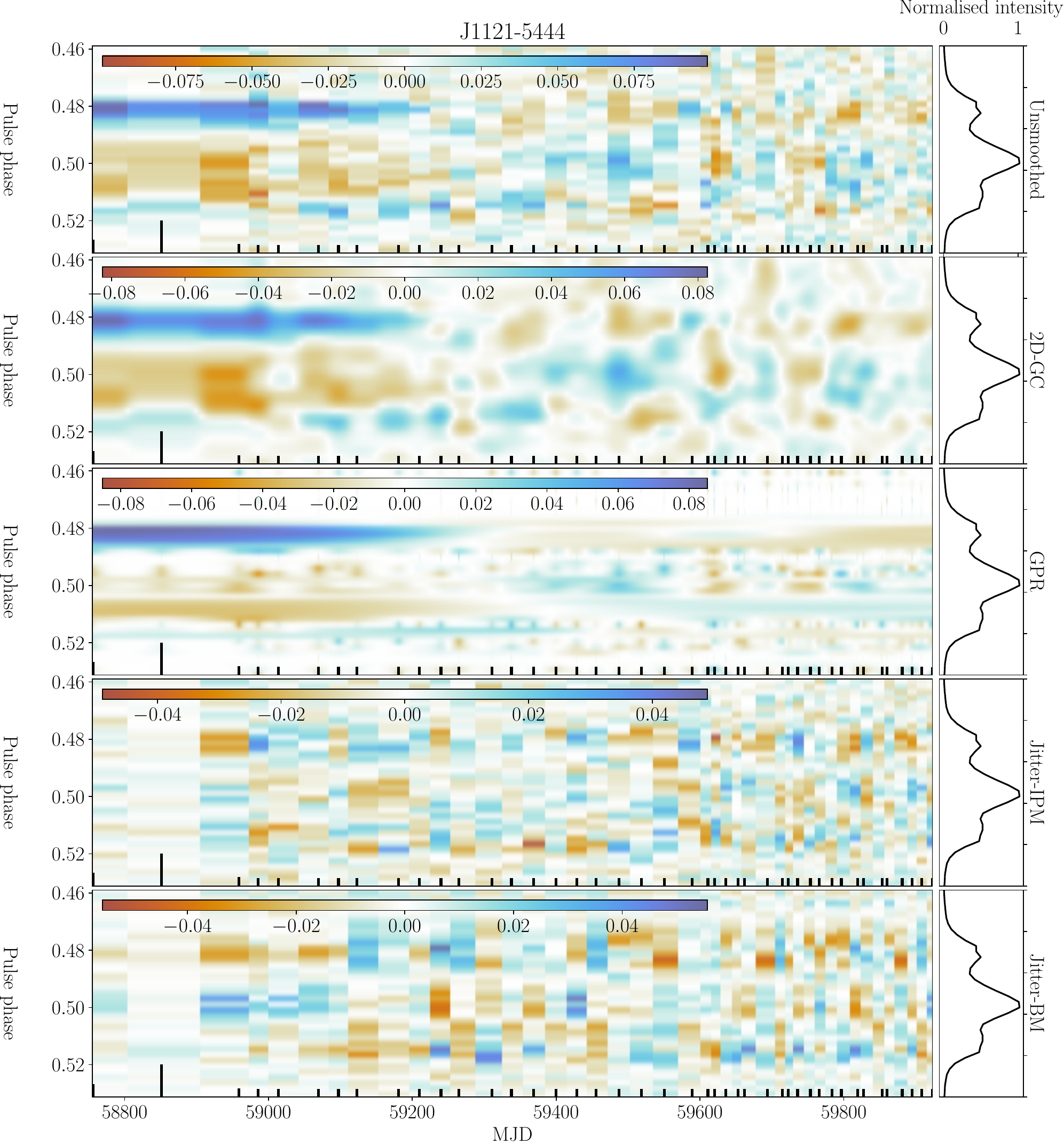}
         \caption{The figure provides the similar information as conveyed in Fig.~\ref{J0729alldiffs}, but for PSR J1121$-$5444.}
        \label{J1121alldiffs}
\end{figure*}

\begin{figure*}
    %\centering
         \includegraphics[scale=0.56]{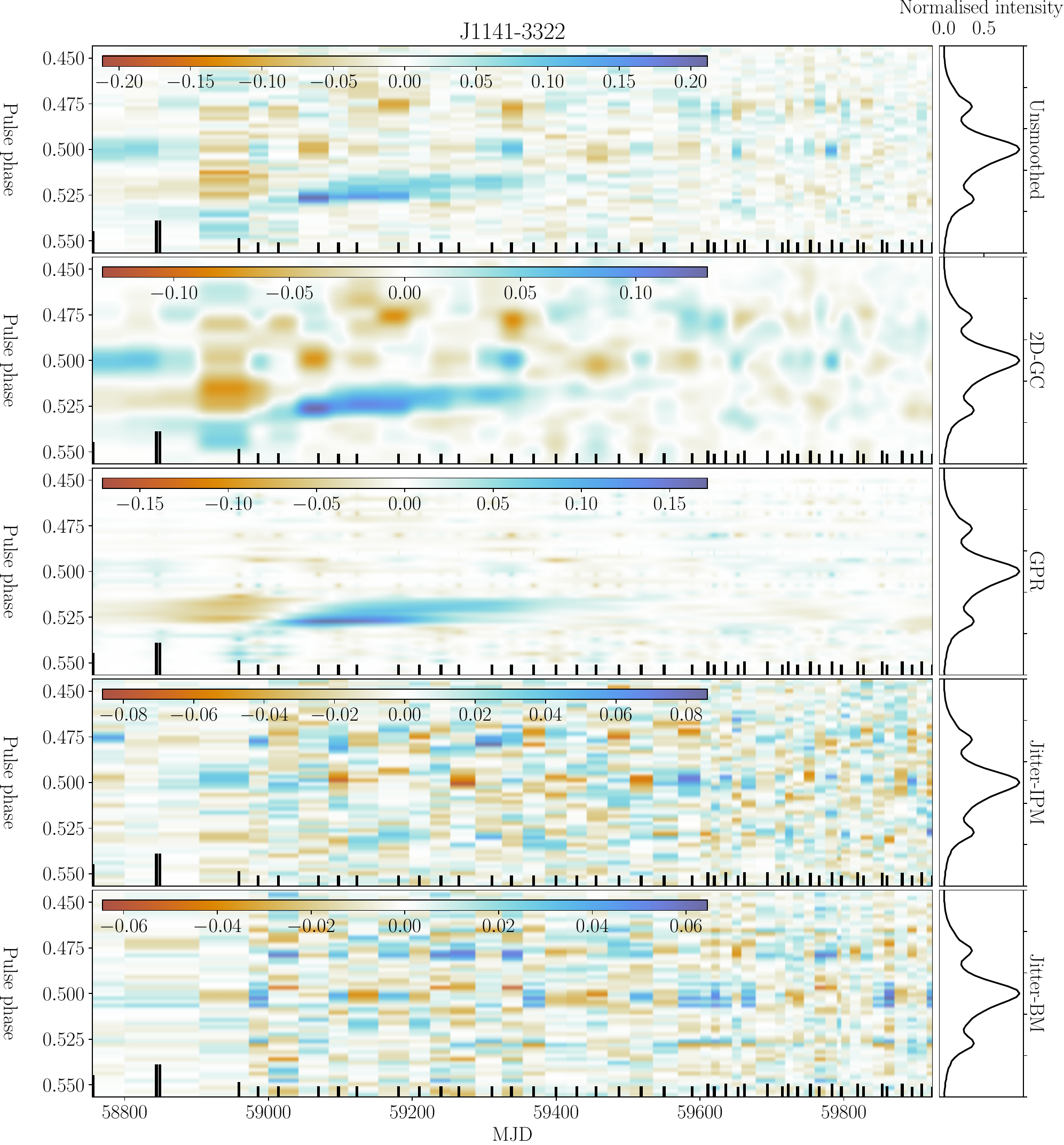}
         \caption{The figure provides the similar information as conveyed in Fig.~\ref{J0729alldiffs}, but for PSR J1141$-$3322.}
        \label{J1141alldiffs}
\end{figure*}

\begin{figure*}
    %\centering
         \includegraphics[scale=0.56]{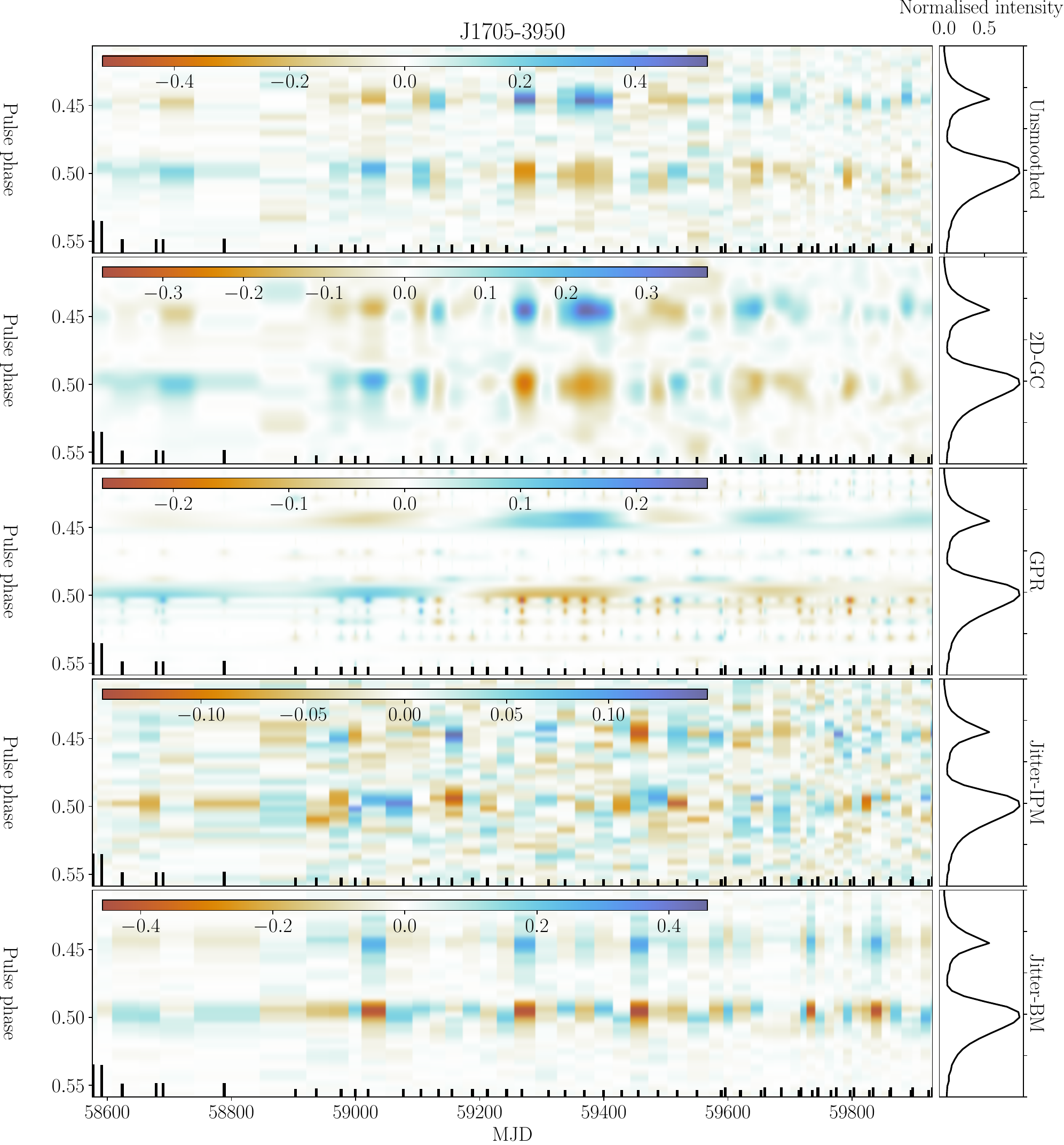}
         \caption{The figure provides the similar information as conveyed in Fig.~\ref{J0729alldiffs}, but for PSR J1705$-$3950}
         \label{J1705alldiffs}
\end{figure*}

\begin{figure*}
    %\centering
         \includegraphics[scale=0.56]{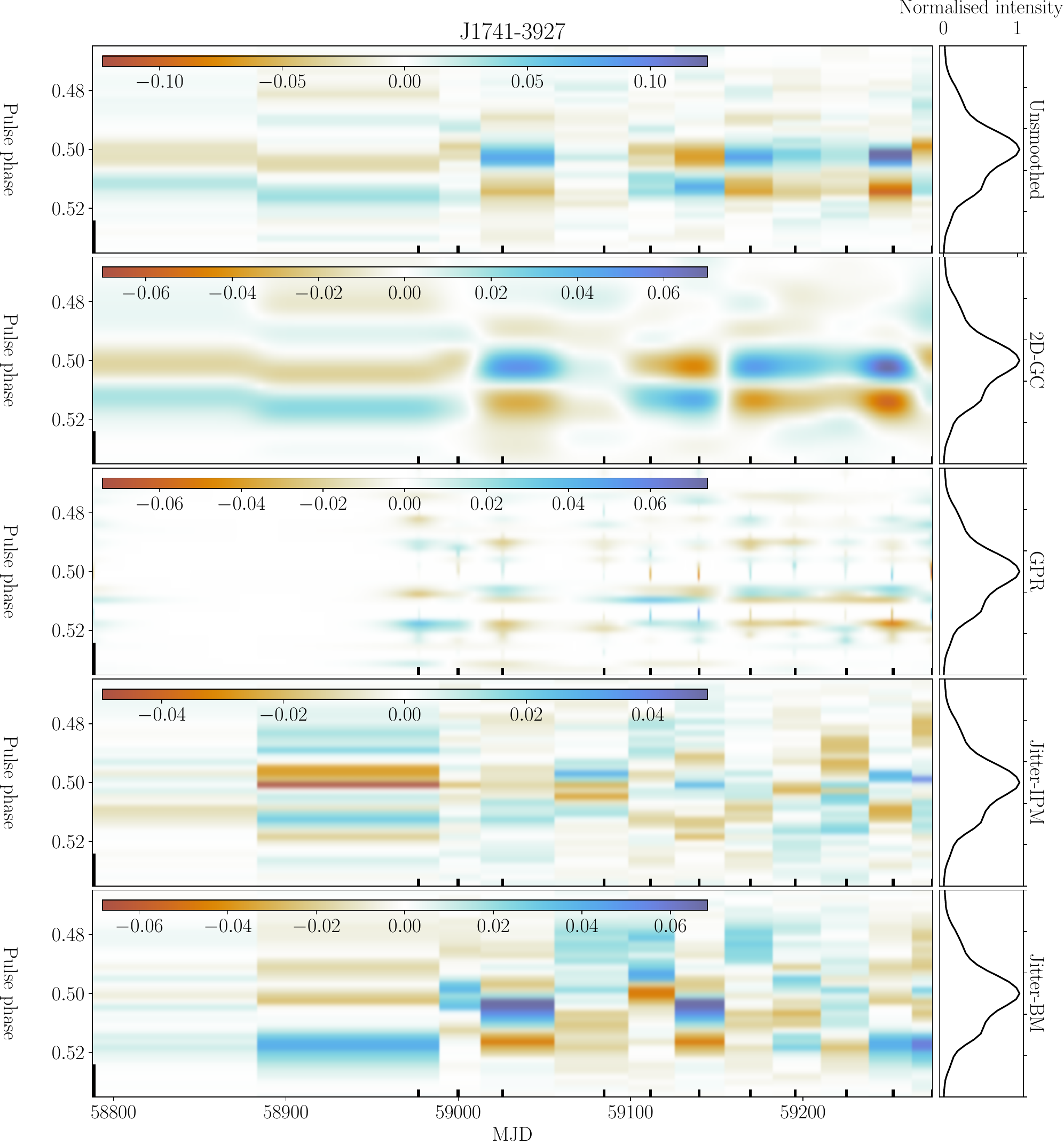}
         \caption{The figure provides the similar information as conveyed in Fig.~\ref{J0729alldiffs}, but for PSR J1741$-$3927}
         \label{J1741alldiffs}
\end{figure*}

\begin{figure*}
    %\centering
         \includegraphics[scale=0.56]{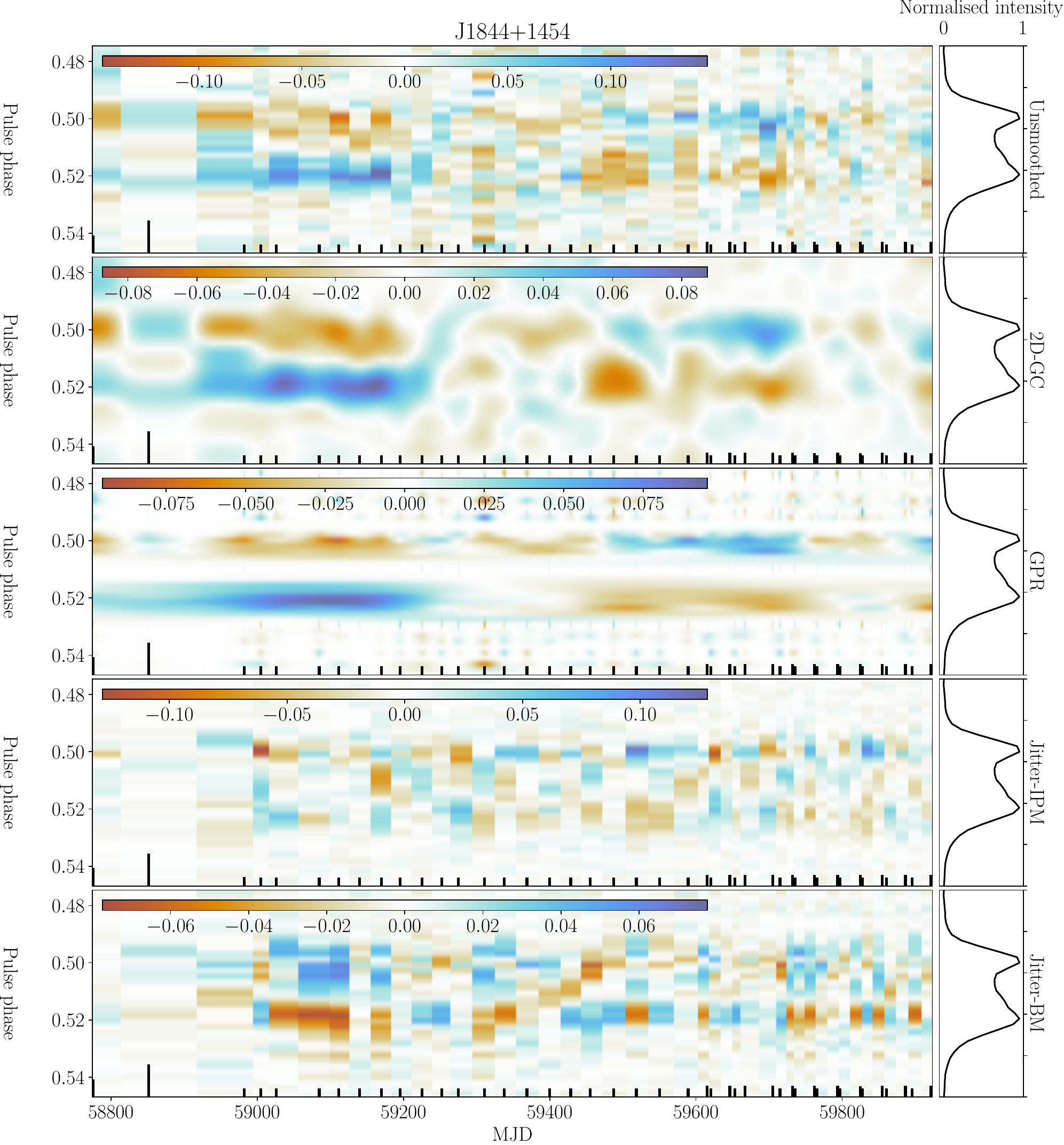}
         \caption{The figure provides the similar information as conveyed in Fig.~\ref{J0729alldiffs}, but for PSR J1844+1454}
         \label{J1844alldiffs}
\end{figure*}

\begin{figure*}
    %\centering
         \includegraphics[scale=0.56]{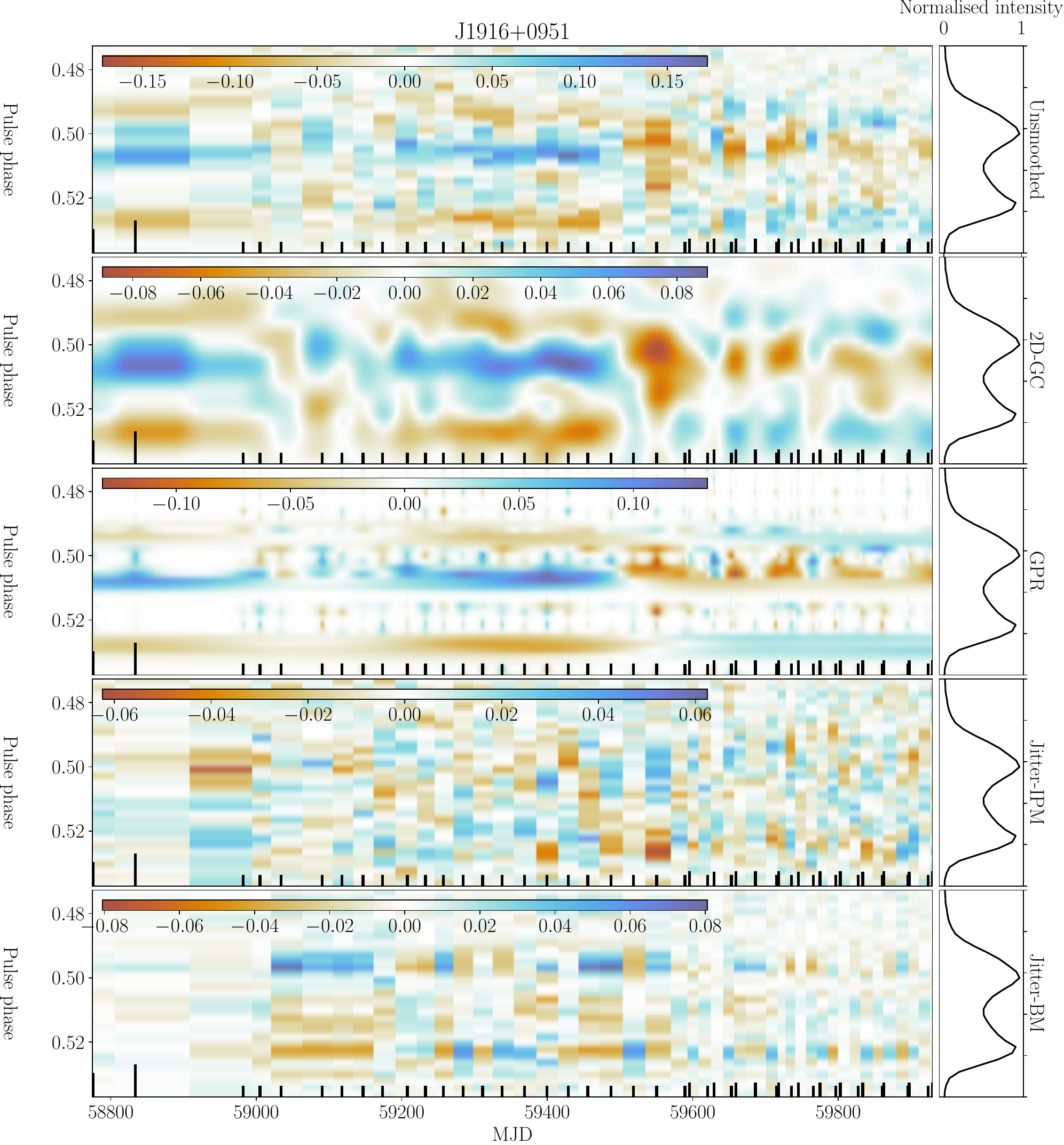}
         \caption{The figure provides the similar information as conveyed in Fig.~\ref{J0729alldiffs}, but for PSR J1916+0951}
         \label{J1916alldiffs}
\end{figure*}

\begin{figure*}
    %\centering
         \includegraphics[scale=0.56]{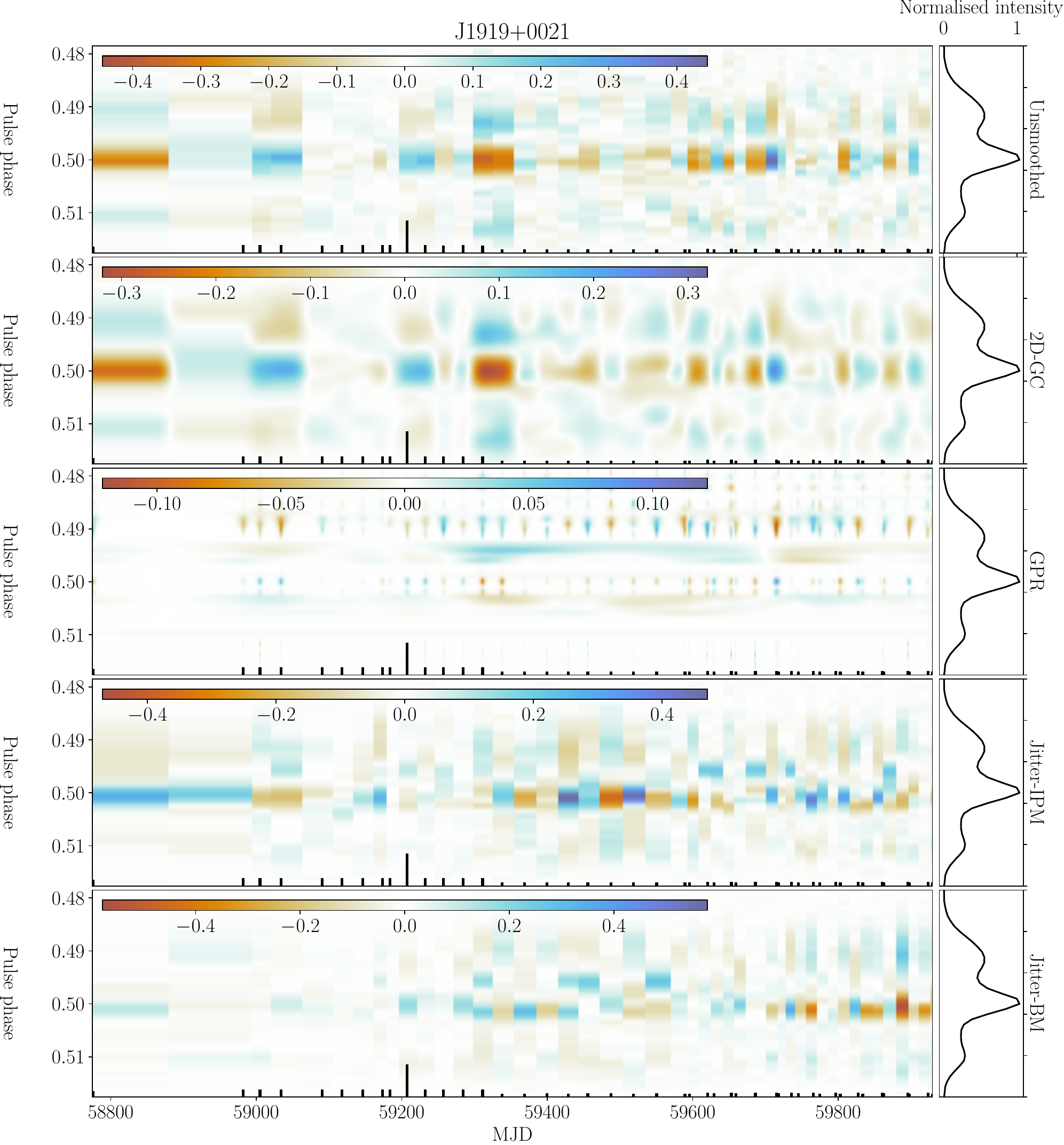}
         \caption{The figure provides the similar information as conveyed in Fig.~\ref{J0729alldiffs}, but for PSR J1919+0021}
         \label{J1919alldiffs}

\end{figure*}

\clearpage

\section{Figures of polarization fraction and polarization position angle difference maps} \label{apdxI}
This section provides the difference maps of various polarization products for all pulsars. There are three figures for every pulsar, and every figure has four panels. 
The first figure for each pulsar is derived from the actual data. 
The second and last figure displays the outcome of the IPM and BM jitter simulation respectively. All these difference maps are smoothed by convolution with a 2D Gaussian kernel. No smoothing is applied to the jitter-simulated total intensity difference maps. The details on the results of the polarization analysis are discussed in Sec.~\ref{polanalysis} in the main paper.

In each of these figures, the top panel represents the total intensity difference map. The second and third panel show the difference map of the linear and circular polarization fraction respectively.
%, third panel shows the difference map of the circular polarisation fraction and 
The bottom panel shows the polarization position angle (PA or $\psi$) difference map.

\begin{figure*}
   \centering
\includegraphics[scale=0.56]{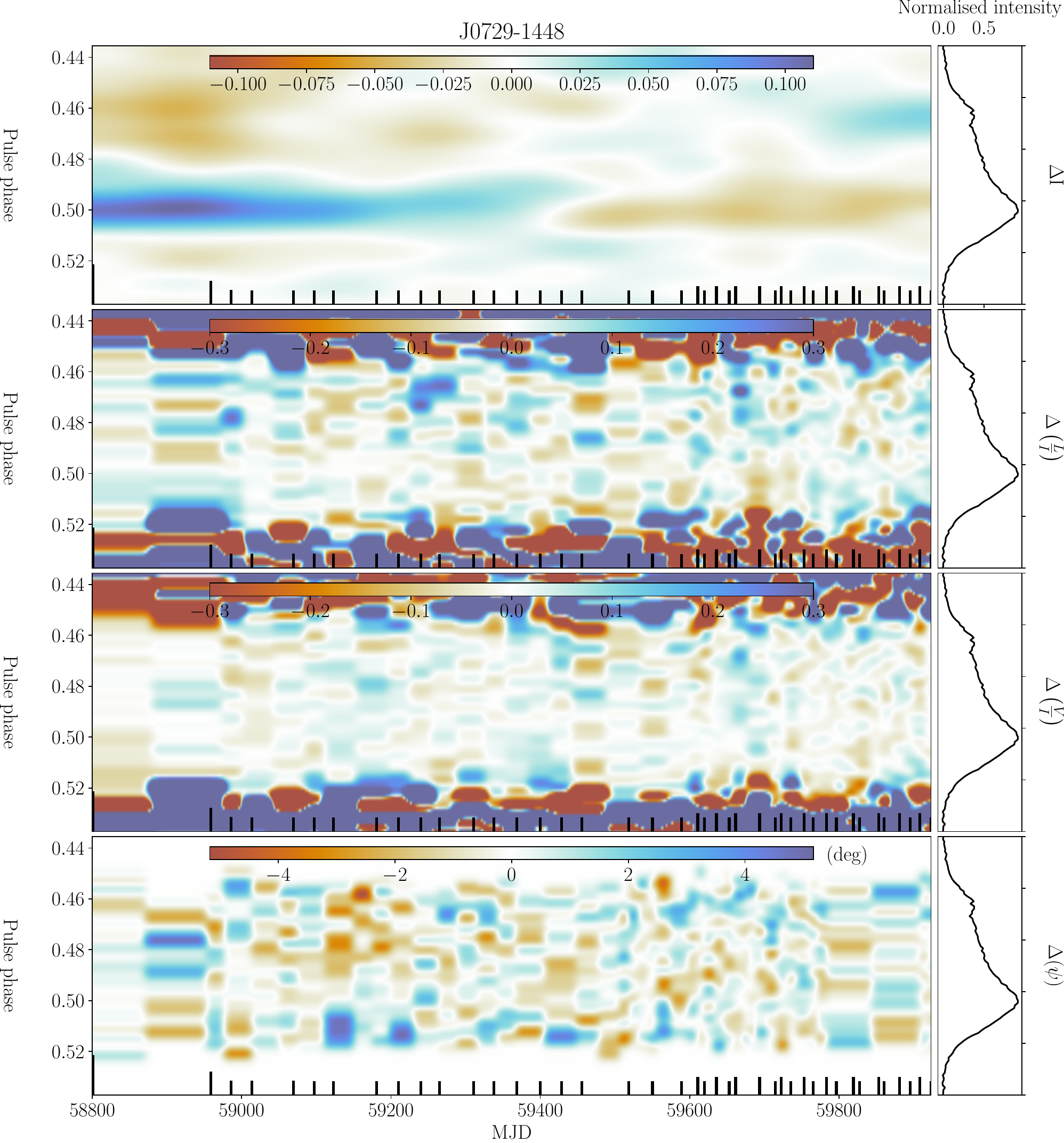}
\caption{This figure captures the emission variability for PSR J0729$-$1448. The top panel of the figure shows the 2D convolution smoothed version of the difference map for total intensity, zoomed in on the on-pulse phase range. The second and the third panel from the top shows the smoothed version of the difference map for the $L/I$ and $V/I$. The bottom panel shows the difference map for the polarization position angle. The position of the vertical tick marks indicates the observation epoch and their length corresponds to the length of observations. Details of the analysis are presented in Sec.~\ref{polanalysis}. }
\label{fig:0729poldiff}
\end{figure*}

\begin{figure*}
   \centering
\includegraphics[scale=0.56]{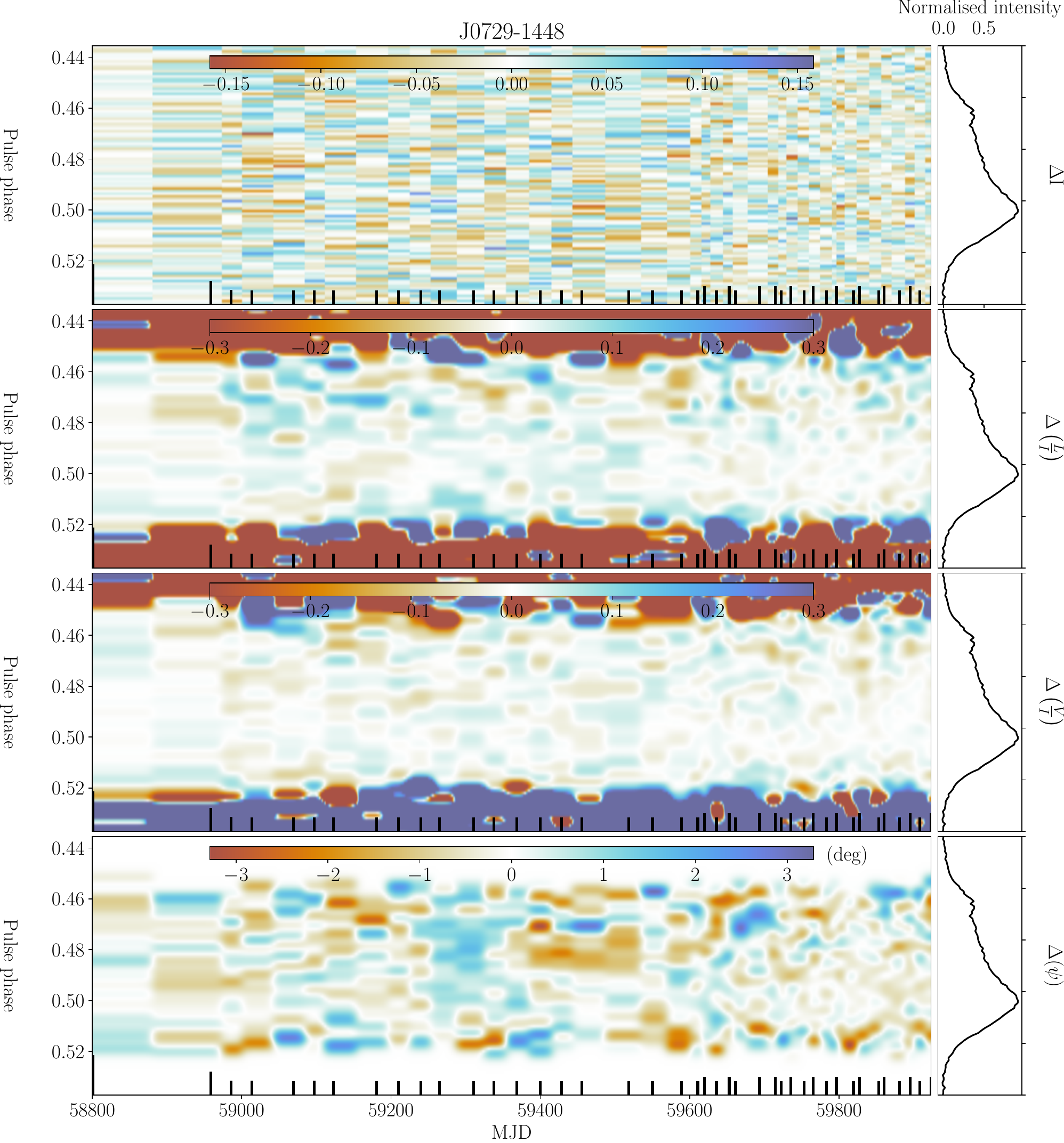}
\caption{This figure captures the same information as shown in Fig.~\ref{fig:0729poldiff}, but in this case, the data was generated using the jitter simulation using IPM, discussed in Sec.~\ref{IPmethod}.}
\label{fig:0729poldiff_ipm}
\end{figure*}

\begin{figure*}
   \centering
\includegraphics[scale=0.56]{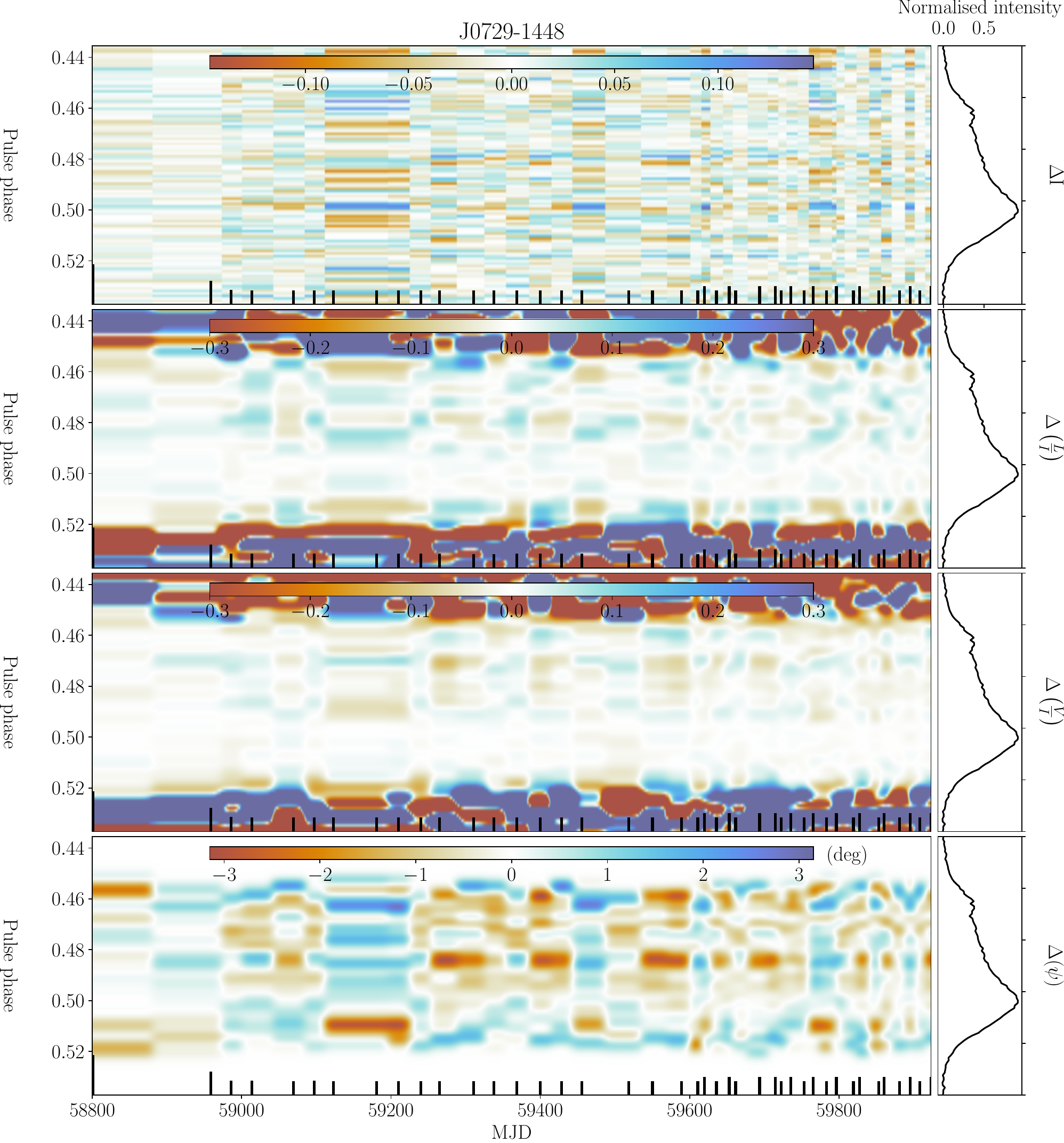}
\caption{This figure captures the same information as shown in Fig.~\ref{fig:0729poldiff}, but in this case, the data was generated using the jitter simulation using BM as described in Sec.~\ref{Bmethod}.}
\label{fig:0729poldiff_bm}
\end{figure*}

\begin{figure*}
   \centering
\includegraphics[scale=0.56]{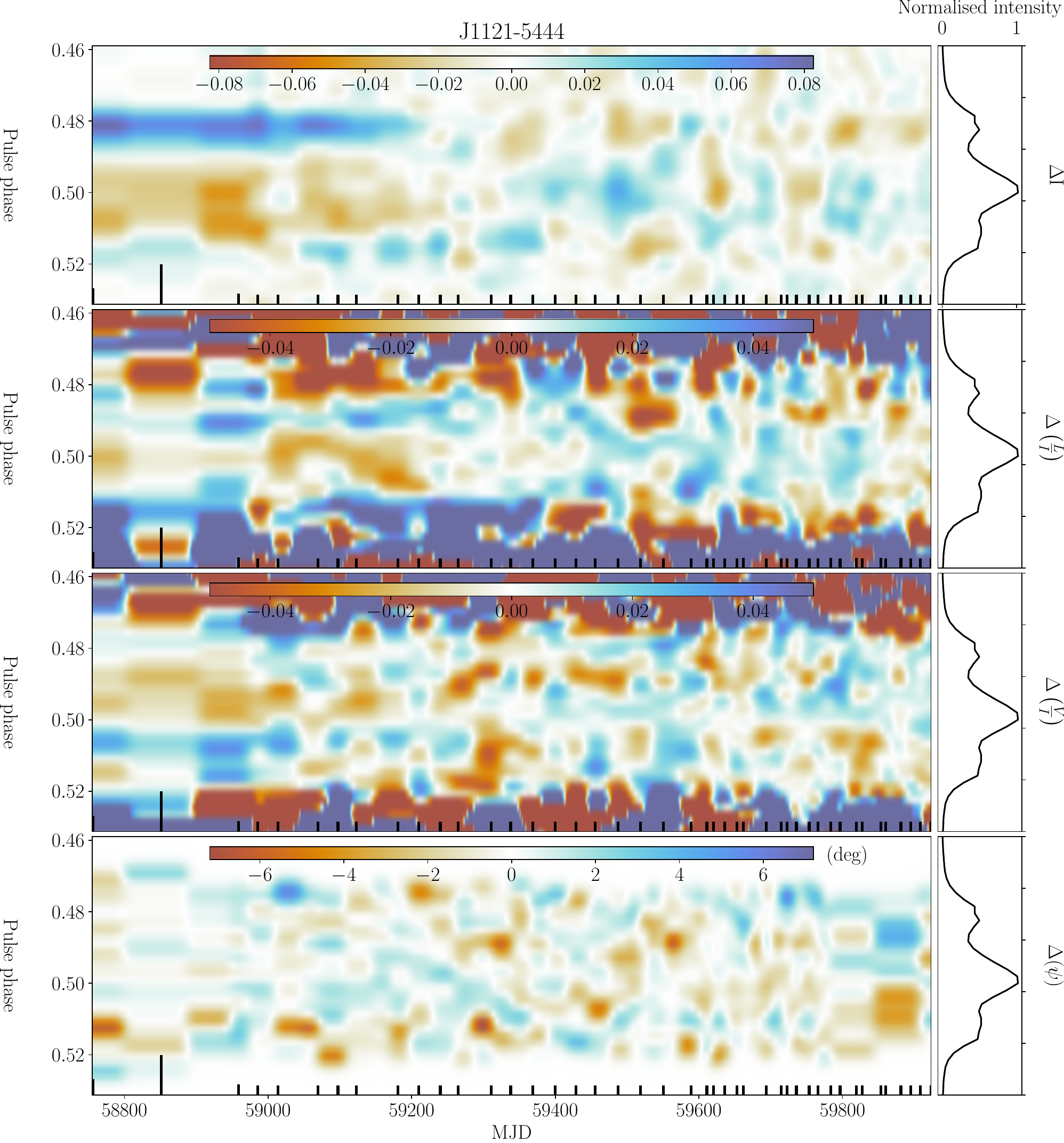}
\caption{ This figure conveys the same information as that of Fig.\ref{fig:0729poldiff} but for PSR J1121$-$5444.}
\label{fig:1121poldiff} % I can do without the label too
\end{figure*}

\begin{figure*}
   \centering
\includegraphics[scale=0.56]{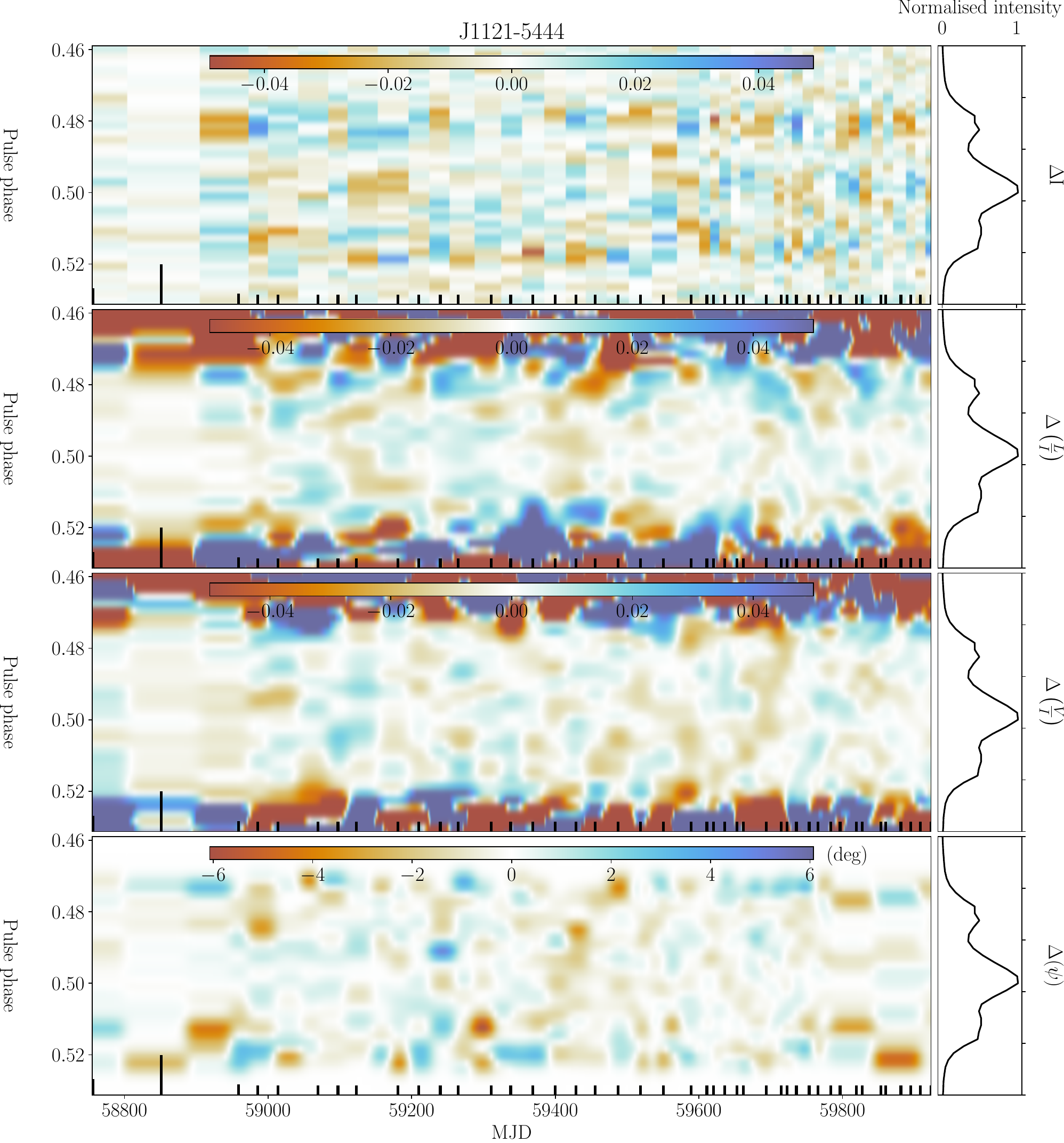}
\caption{ The figure captures the same information as shown in Fig.~\ref{fig:0729poldiff_ipm} but for PSR J1121$-$5444.}
\label{fig:1121poldiff_ipm} % I can do without the label too
\end{figure*}

\begin{figure*}
   \centering
\includegraphics[scale=0.56]{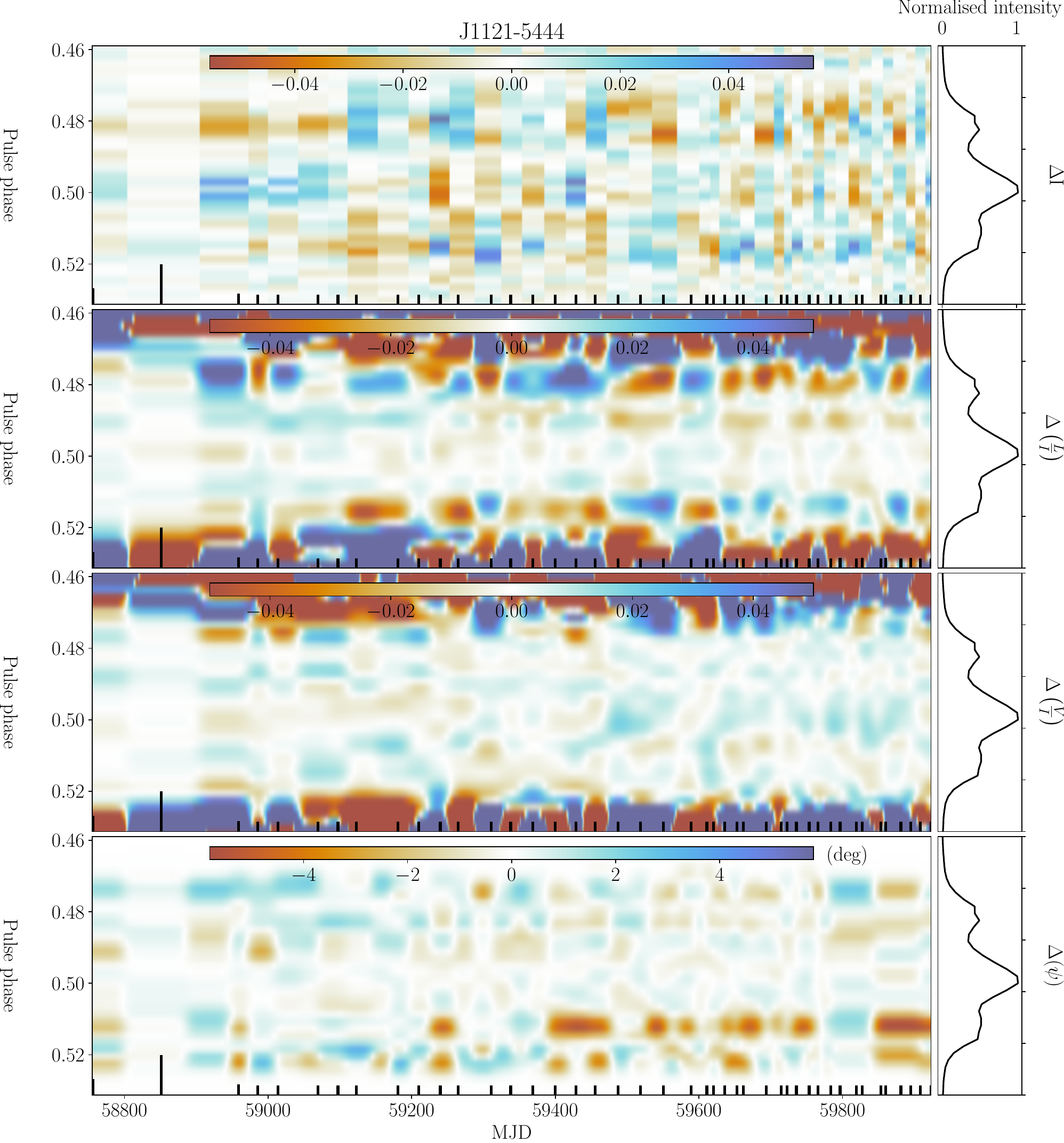}
\caption{ The figure captures the same information as shown in Fig.~\ref{fig:0729poldiff_bm} but for PSR J1121$-$5444.}
\label{fig:1121poldiff_bm} % I can do without the label too
\end{figure*}

\begin{figure*}
   \centering
\includegraphics[scale=0.56]{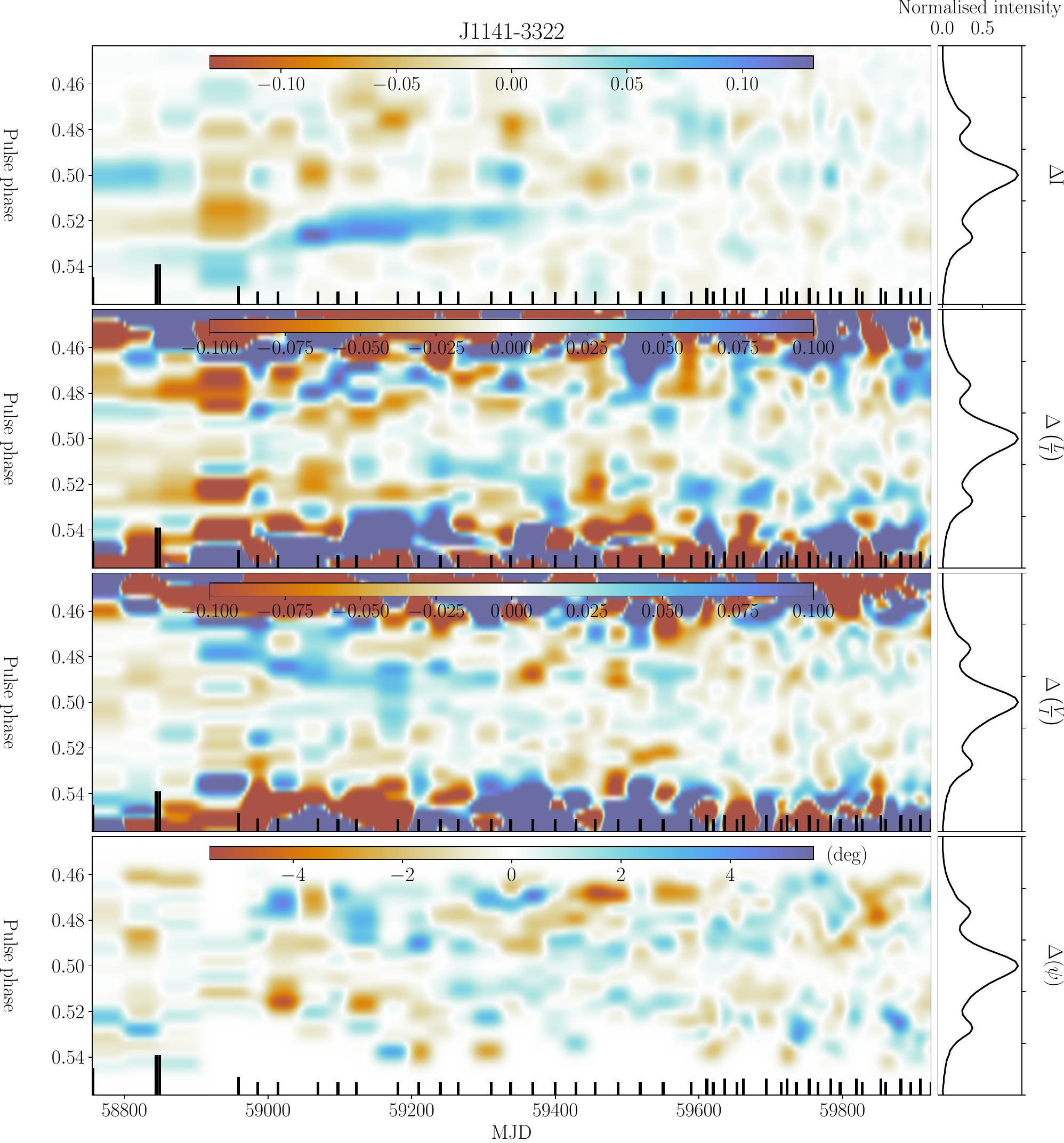}
\caption{This figure conveys the same information as that of Fig.\ref{fig:0729poldiff} but for PSR J1141$-$3322.}
\label{fig:1141poldiff} % I can do without the label too
\end{figure*}

\begin{figure*}
   \centering
\includegraphics[scale=0.56]{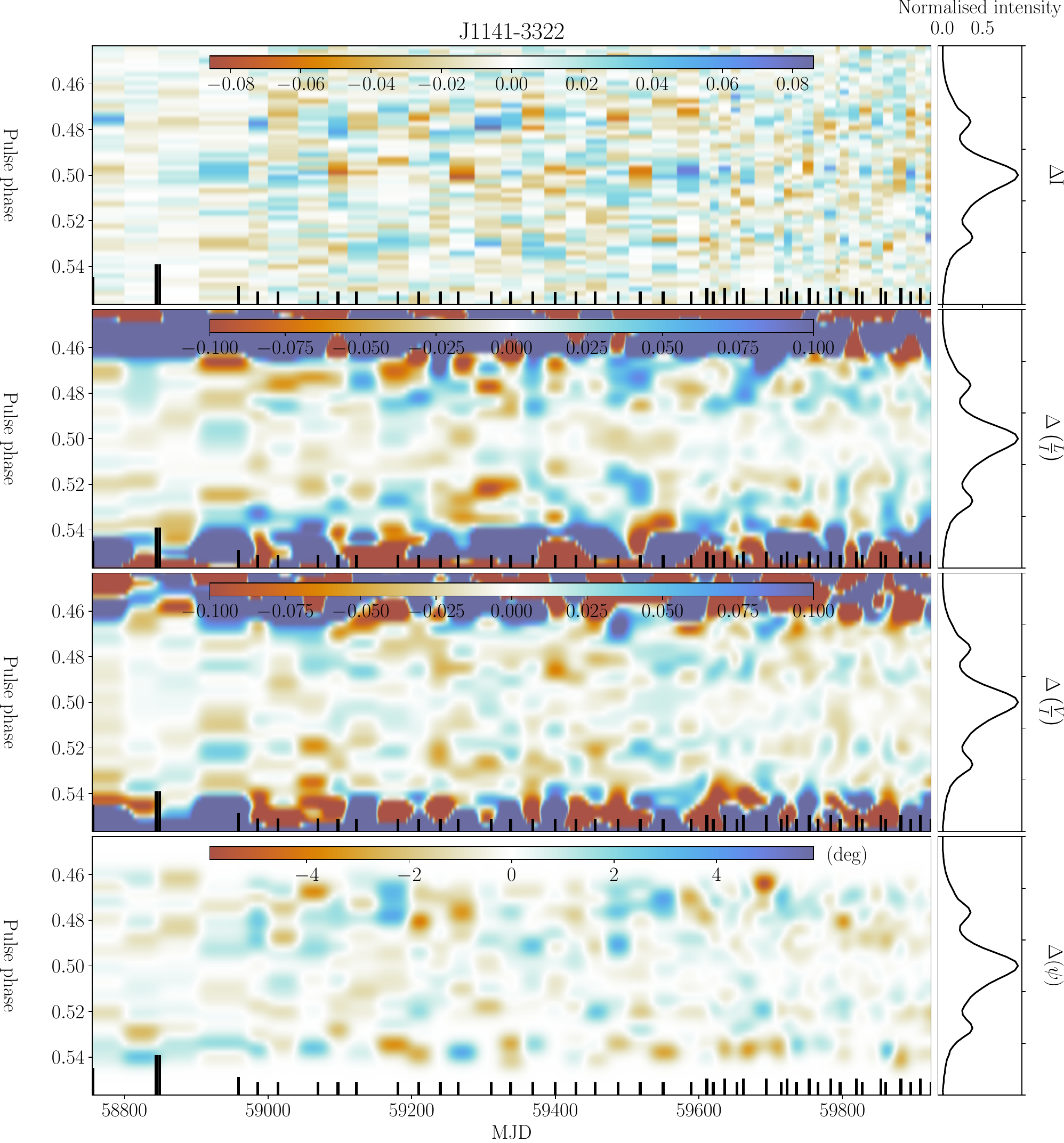}
\caption{The figure captures the same information as shown in Fig.~\ref{fig:0729poldiff_ipm} but for PSR J1141$-$3322.}
\label{fig:1141poldiff_jit1} % I can do without the label too
\end{figure*}

\begin{figure*}
   \centering
\includegraphics[scale=0.56]{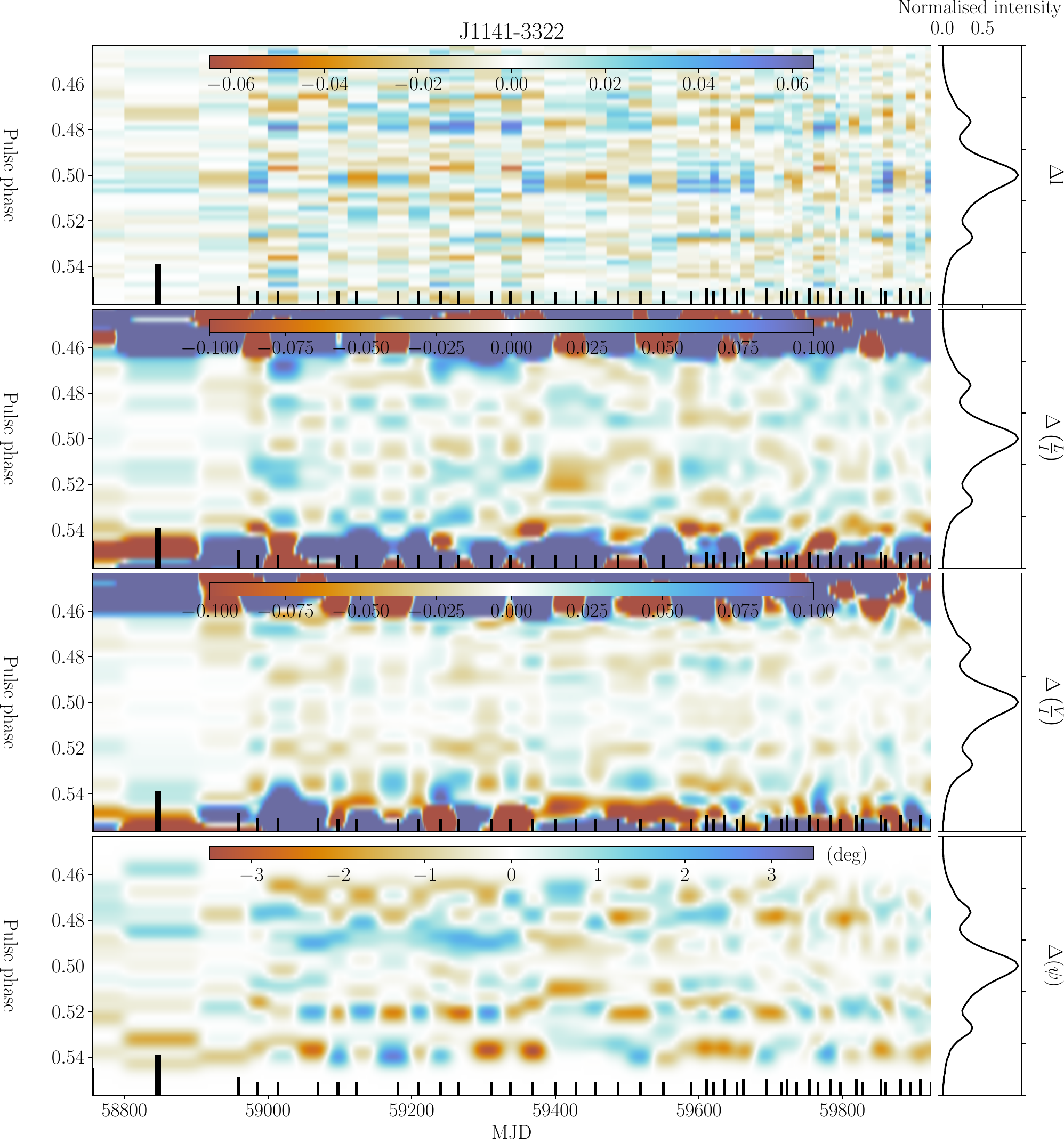}
\caption{The figure captures the same information as shown in Fig.~\ref{fig:0729poldiff_bm} but for PSR J1141$-$3322.}
\label{fig:1141poldiff_jit2} % I can do without the label too
\end{figure*}

\begin{figure*}
   \centering
\includegraphics[scale=0.56]{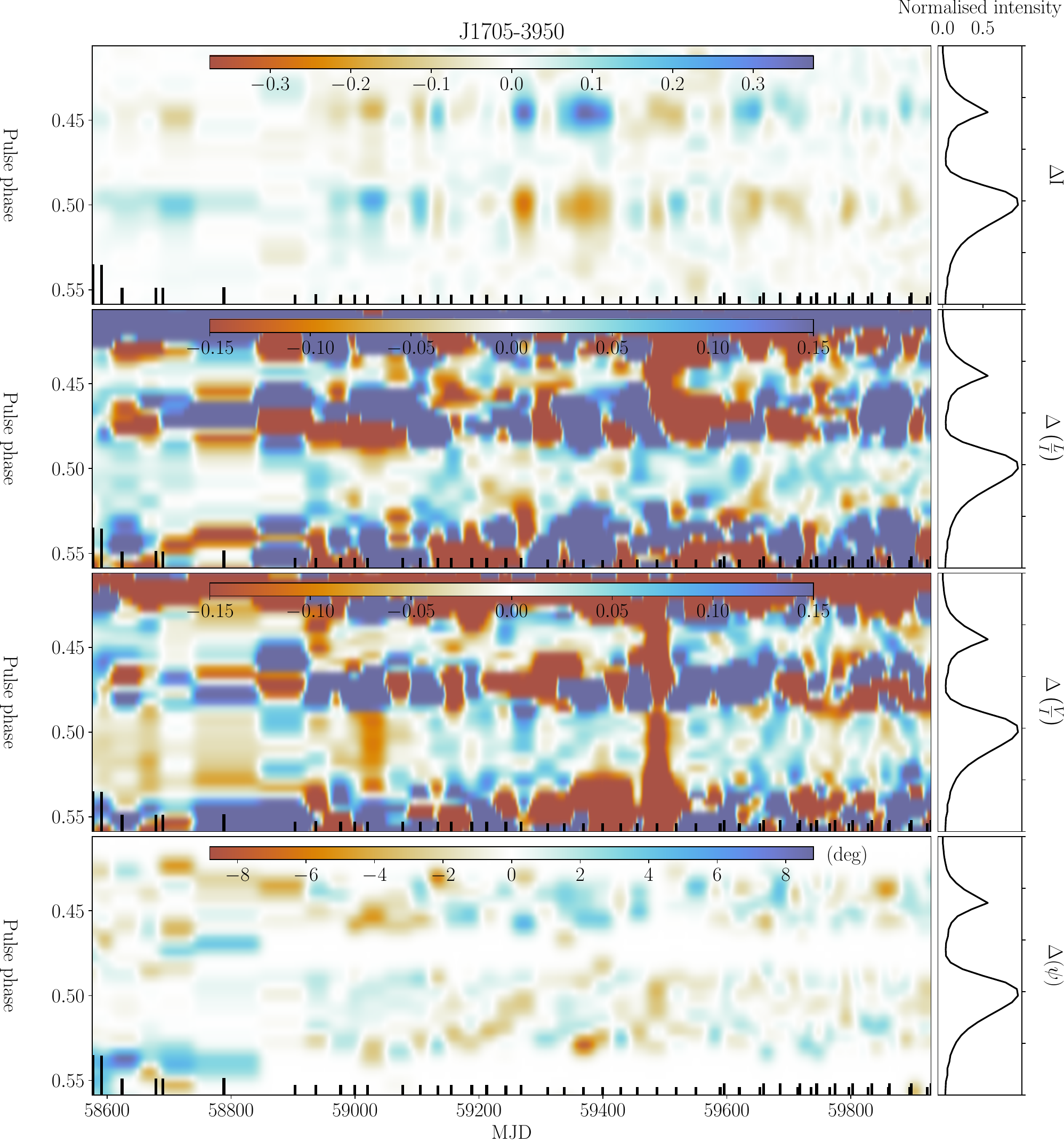}
\caption{This figure conveys the same information as that of Fig.\ref{fig:0729poldiff} but for PSR J1705$-$3950.}
\label{fig:1705poldiff} % I can do without the label too
\end{figure*}

\begin{figure*}
   \centering
\includegraphics[scale=0.56]{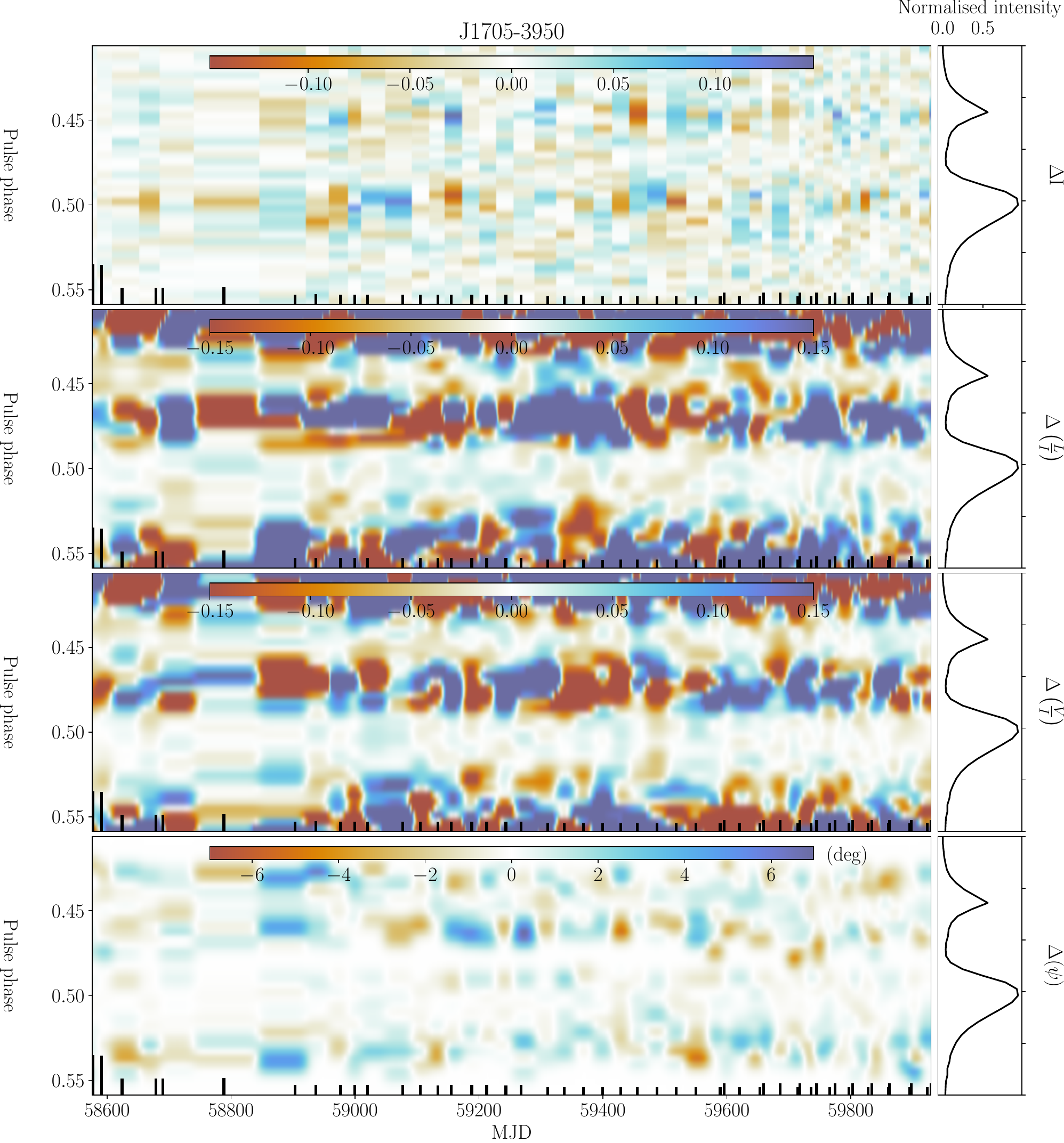}
\caption{The figure captures the same information as shown in Fig.~\ref{fig:0729poldiff_ipm} but for PSR J1705$-$3950.}
\label{fig:1705poldiff_ipm} % I can do without the label too
\end{figure*}

\begin{figure*}
   \centering
\includegraphics[scale=0.56]{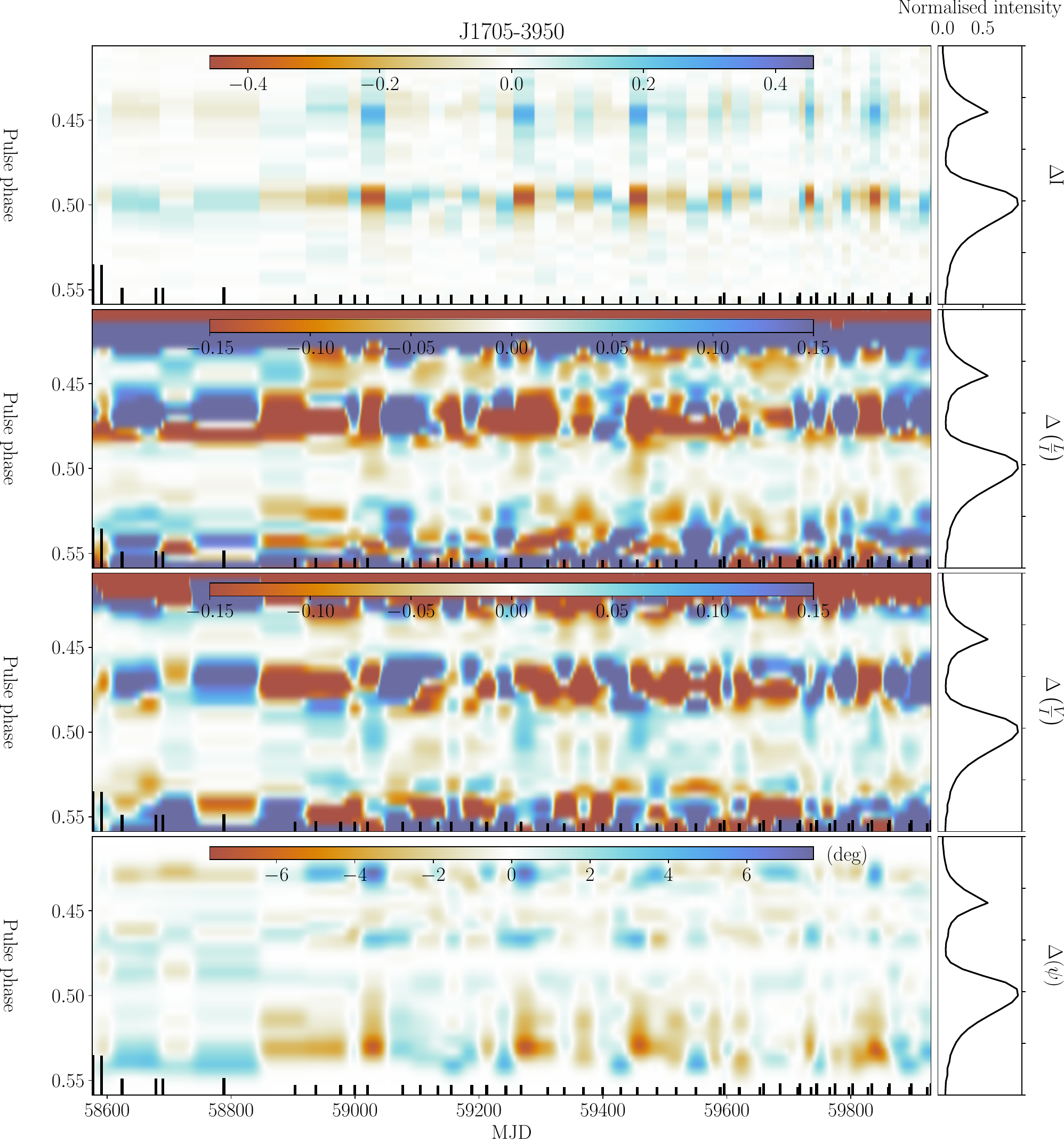}
\caption{The figure captures the same information as shown in Fig.~\ref{fig:0729poldiff_bm} but for PSR J1705$-$3950.}
\label{fig:1705poldiff_bm} % I can do without the label too
\end{figure*}

\begin{figure*}
   \centering
\includegraphics[scale=0.56]{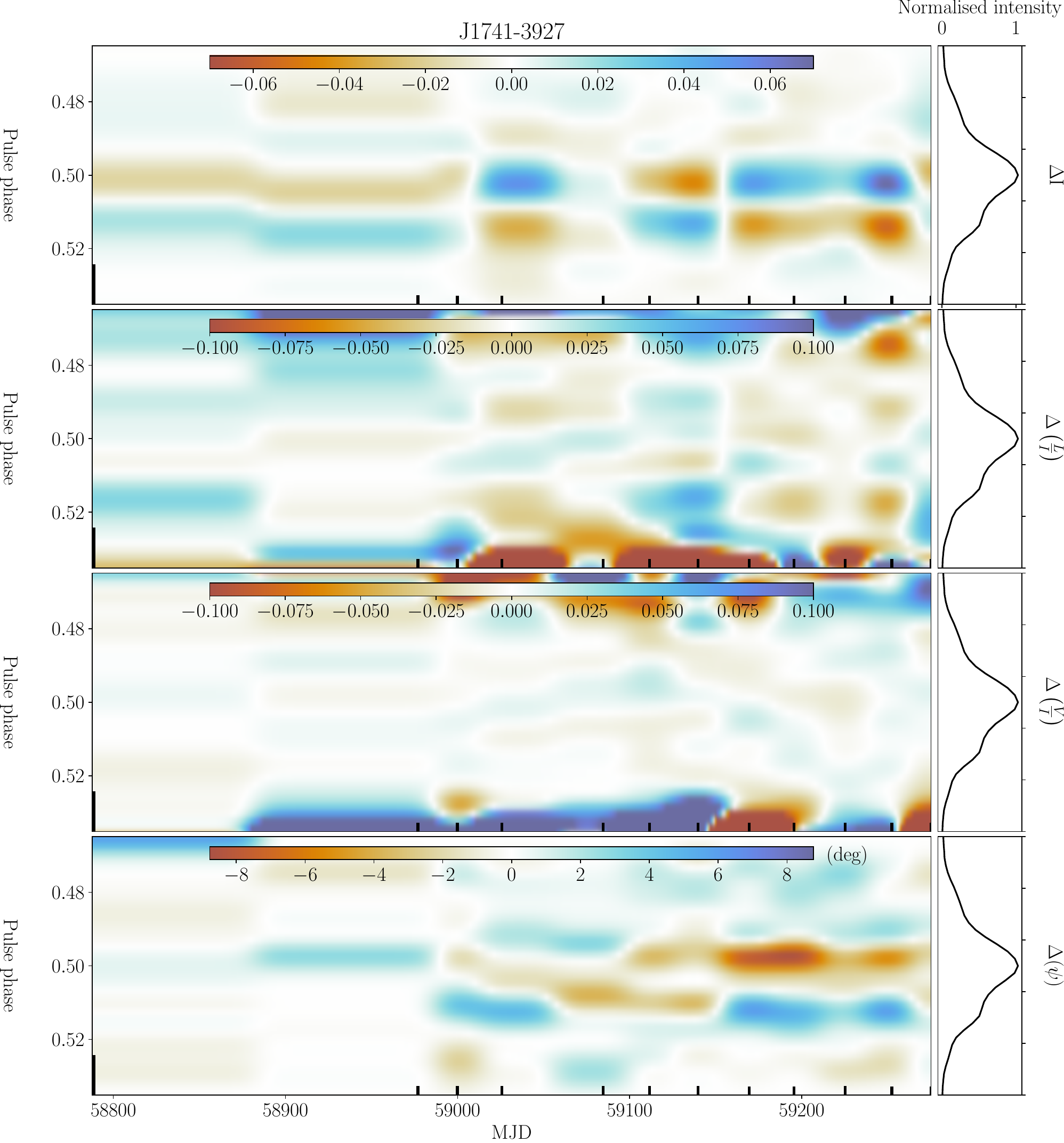}
\caption{This figure conveys the same information as that of Fig.\ref{fig:0729poldiff} but for PSR J1741$-$3927.}
\label{fig:1741poldiff} % I can do without the label too
\end{figure*}

\begin{figure*}
   \centering
\includegraphics[scale=0.56]{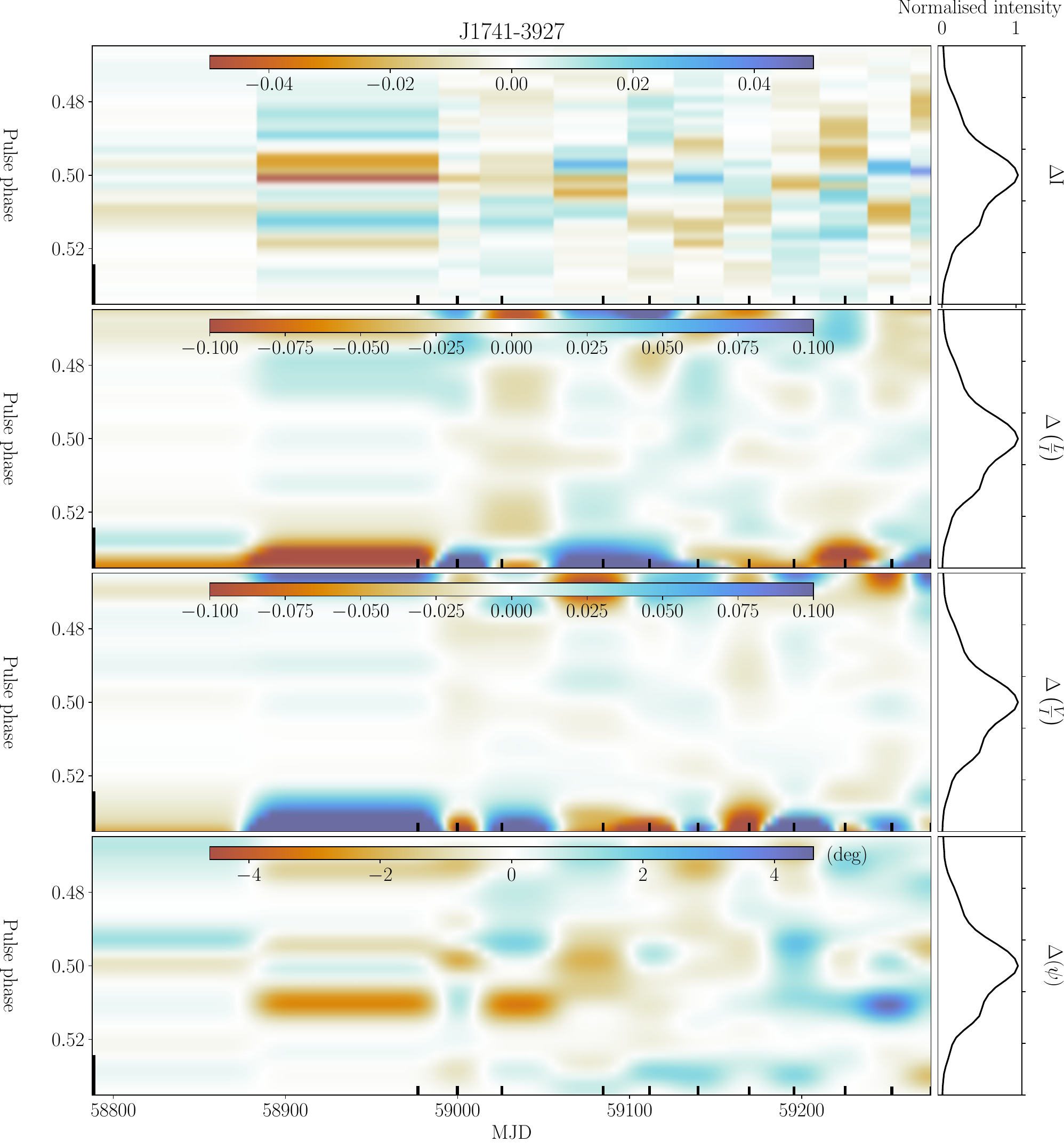}
\caption{The figure captures the same information as shown in Fig.~\ref{fig:0729poldiff_ipm} but for PSR J1741$-$3927.}
\label{fig:1741poldiff_ipm} % I can do without the label too
\end{figure*}

\begin{figure*}
   \centering
\includegraphics[scale=0.56]{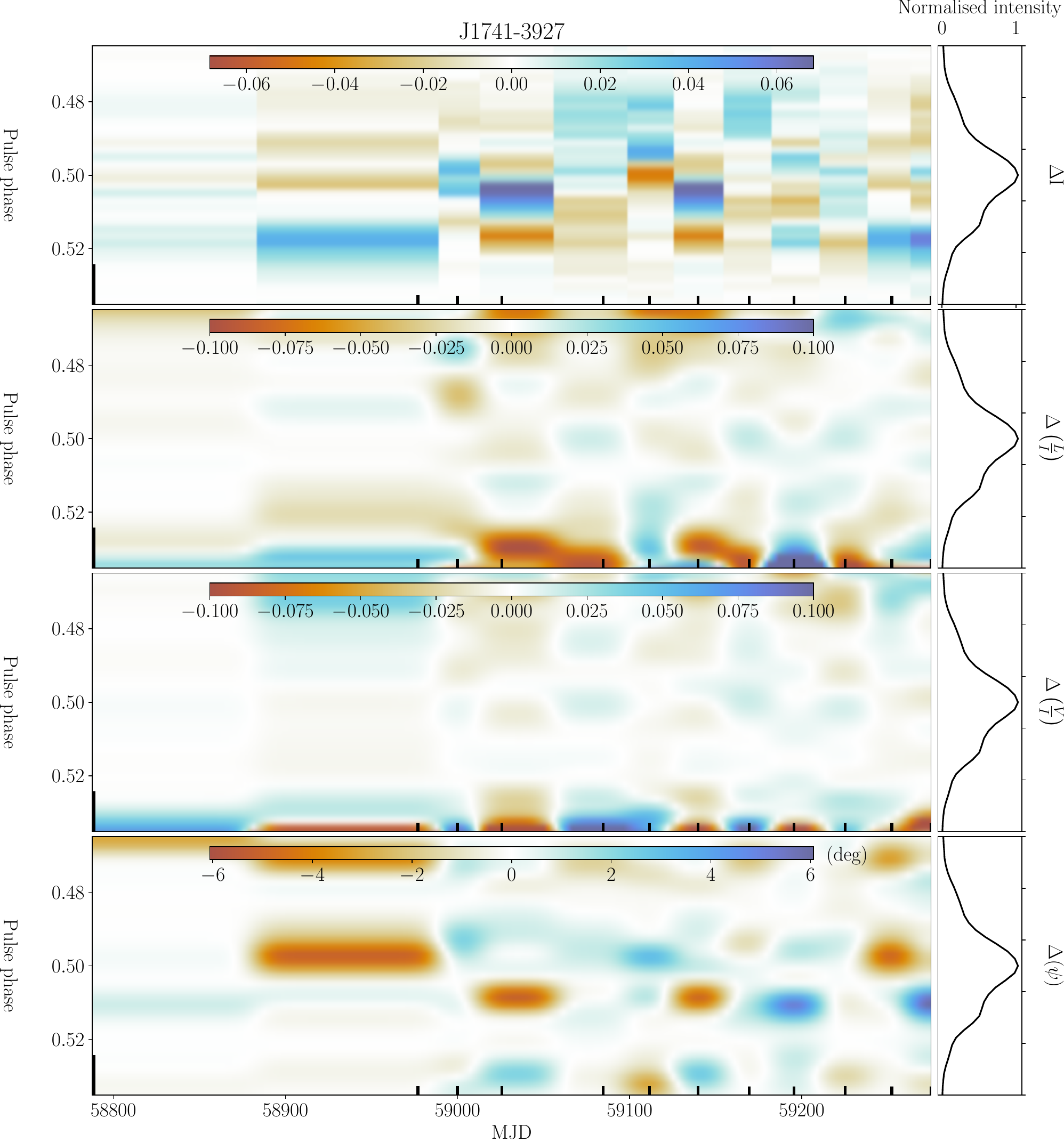}
\caption{The figure captures the same information as shown in Fig.~\ref{fig:0729poldiff_bm} but for PSR J1741$-$3927.}
\label{fig:1741poldiff_bm} % I can do without the label too
\end{figure*}

\begin{figure*}
   \centering
\includegraphics[scale=0.56]{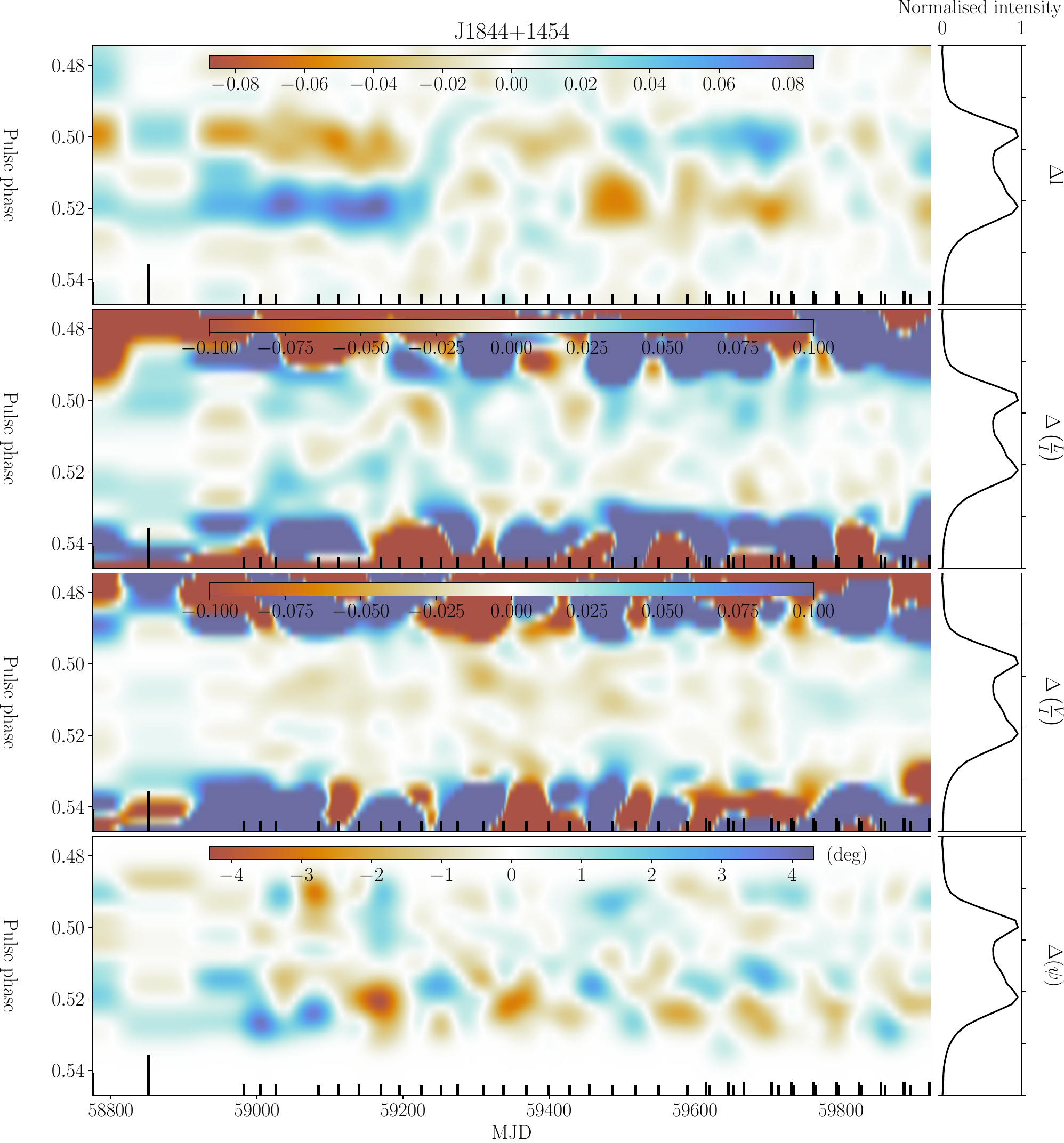}
\caption{This figure conveys the same information as that of Fig.\ref{fig:0729poldiff} but for PSR J1844$+$1454.}
\label{fig:1844poldiff} % I can do without the label too
\end{figure*}

\begin{figure*}
   \centering
\includegraphics[scale=0.56]{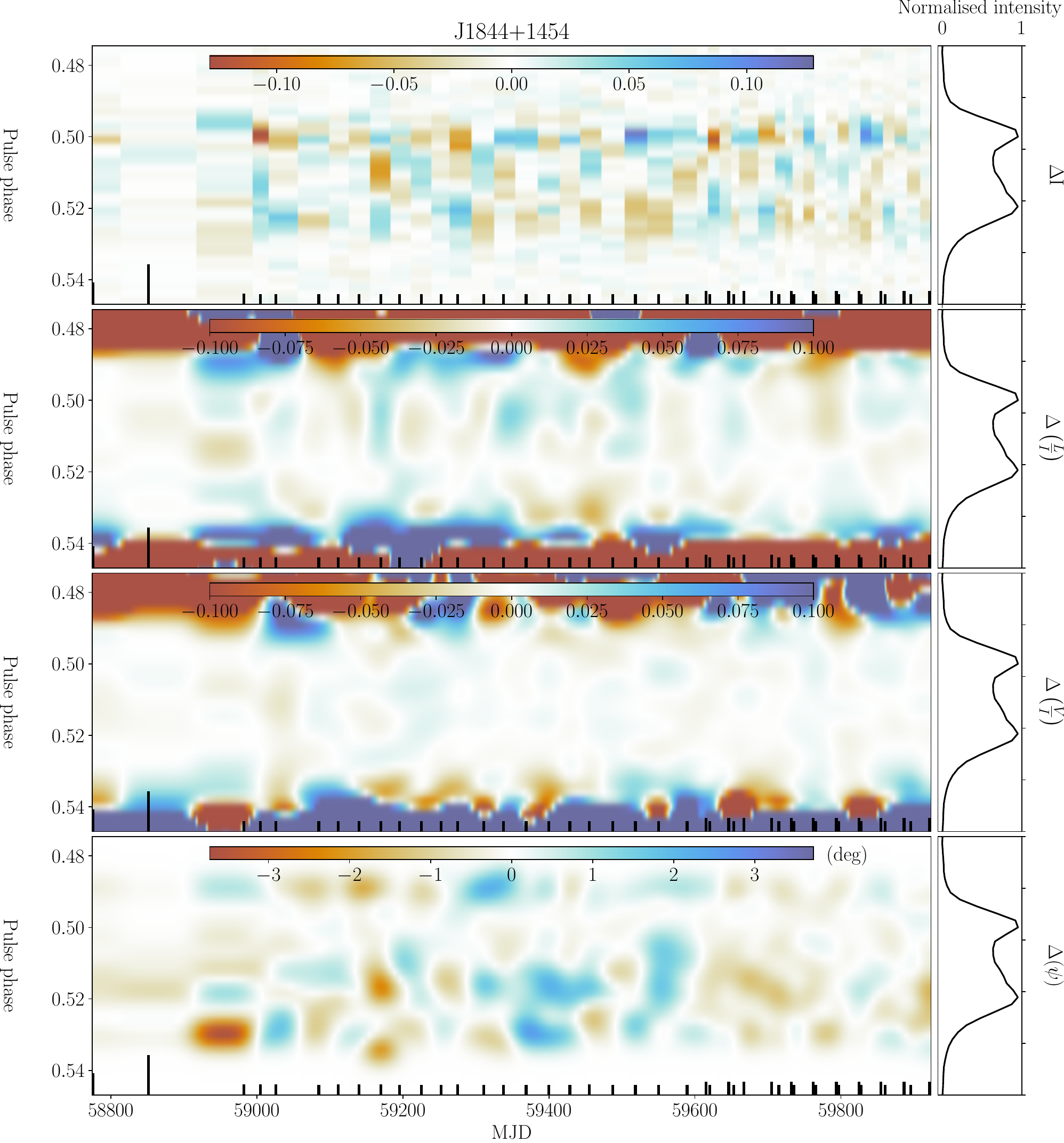}
\caption{The figure captures the same information as shown in Fig.~\ref{fig:0729poldiff_ipm} but for PSR J1844$+$1454.}
\label{fig:1844poldiff_ipm} % I can do without the label too
\end{figure*}

\begin{figure*}
   \centering
\includegraphics[scale=0.56]{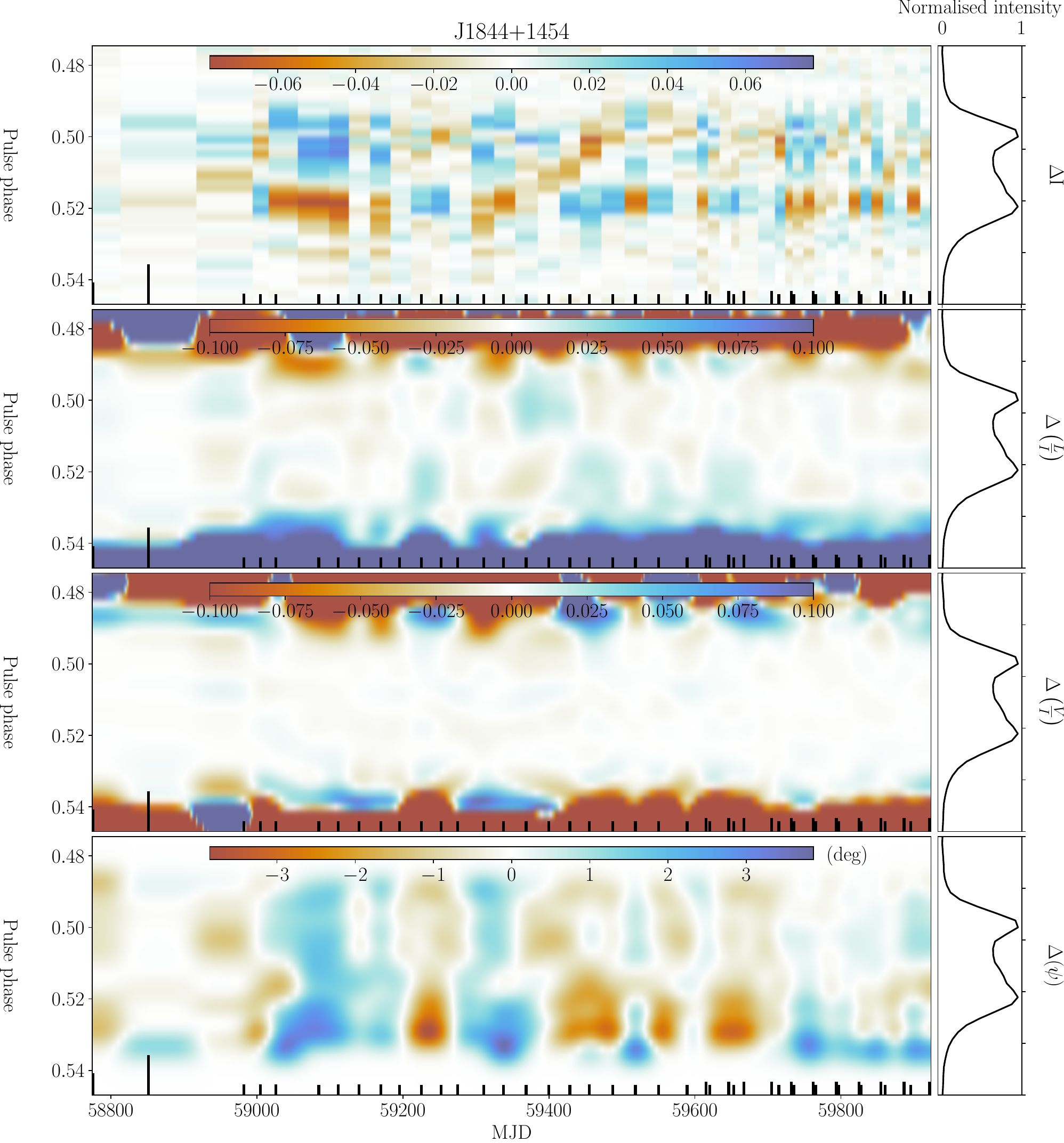}
\caption{The figure captures the same information as shown in Fig.~\ref{fig:0729poldiff_bm} but for PSR J1844$+$1454.}
\label{fig:1844poldiff_bm} % I can do without the label too
\end{figure*}

\begin{figure*}
   \centering
\includegraphics[scale=0.56]{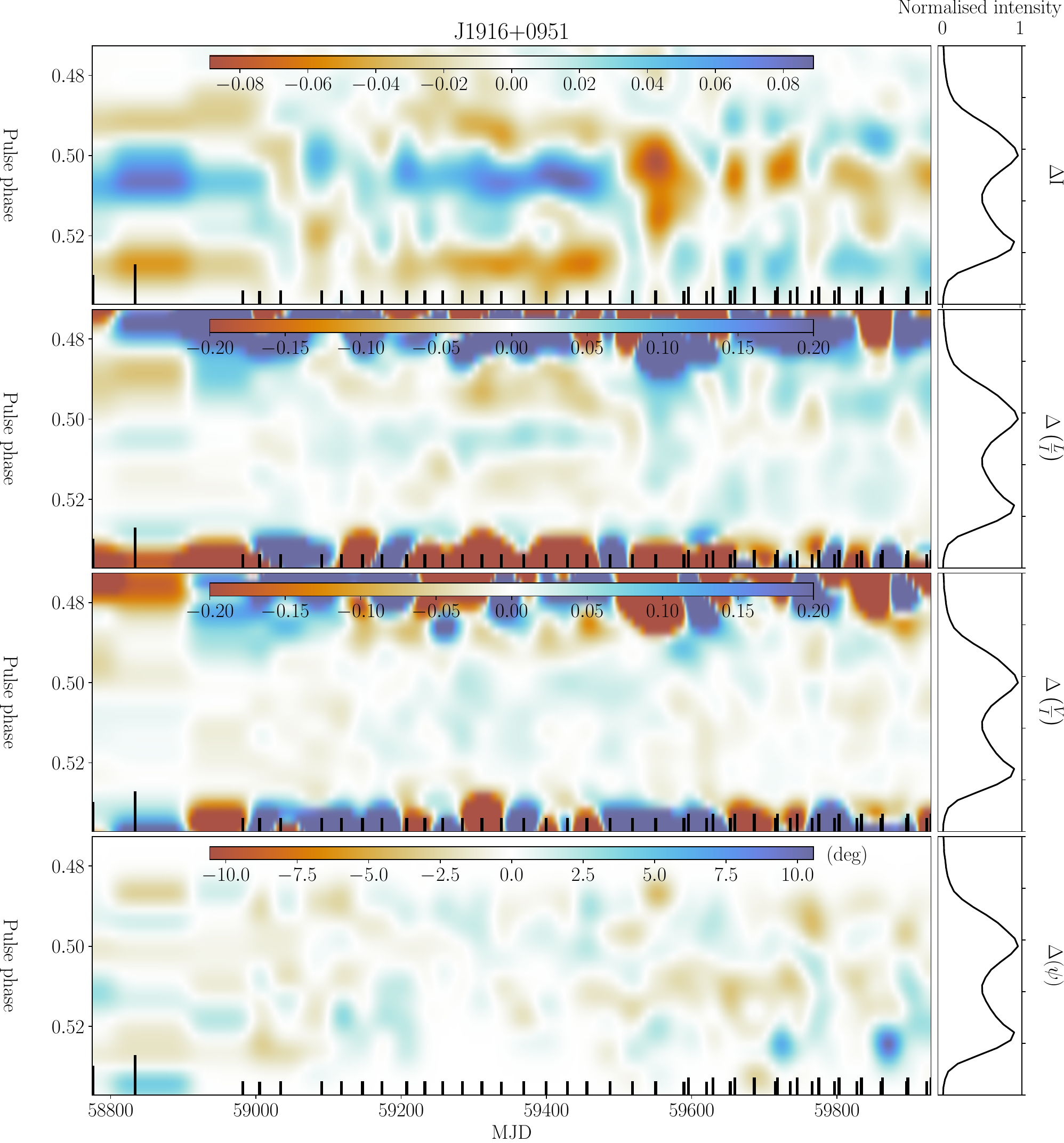}
\caption{This figure conveys the same information as that of Fig.\ref{fig:0729poldiff} but for PSR J1916+0951. }
\label{fig:1916poldiff} % I can do without the label too
\end{figure*}

\begin{figure*}
   \centering
\includegraphics[scale=0.56]{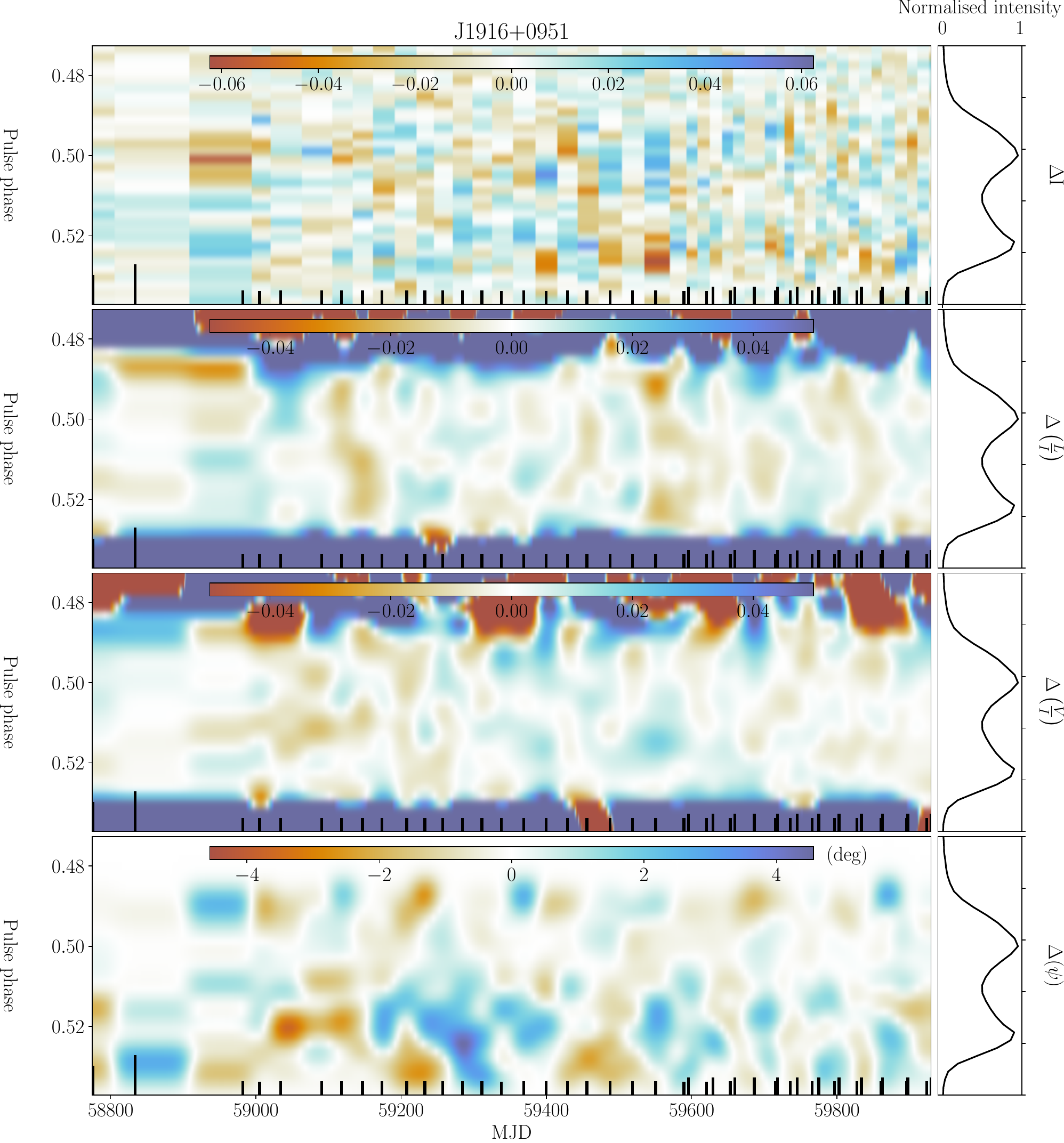}
\caption{The figure captures the same information as shown in Fig.~\ref{fig:0729poldiff_ipm} but for PSR J1916$+$0951.}
\label{fig:1916poldiff_ipm} % I can do without the label too
\end{figure*}

\begin{figure*}
   \centering
\includegraphics[scale=0.56]{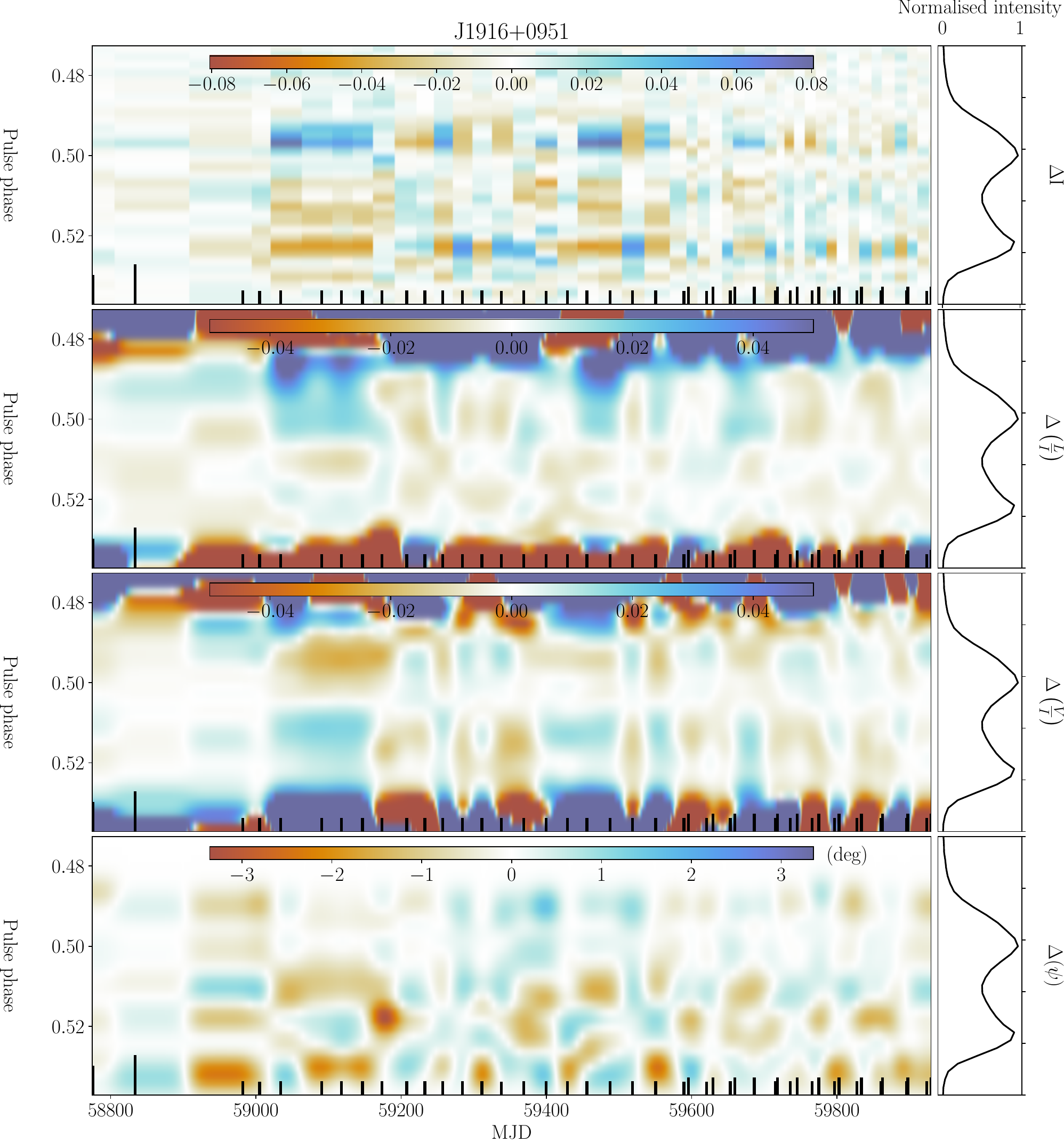}
\caption{The figure captures the same information as shown in Fig.~\ref{fig:0729poldiff_bm} but for PSR J1916$+$0951.}
\label{fig:1916poldiff_bm} % I can do without the label too
\end{figure*}

\begin{figure*}
   \centering
\includegraphics[scale=0.56]{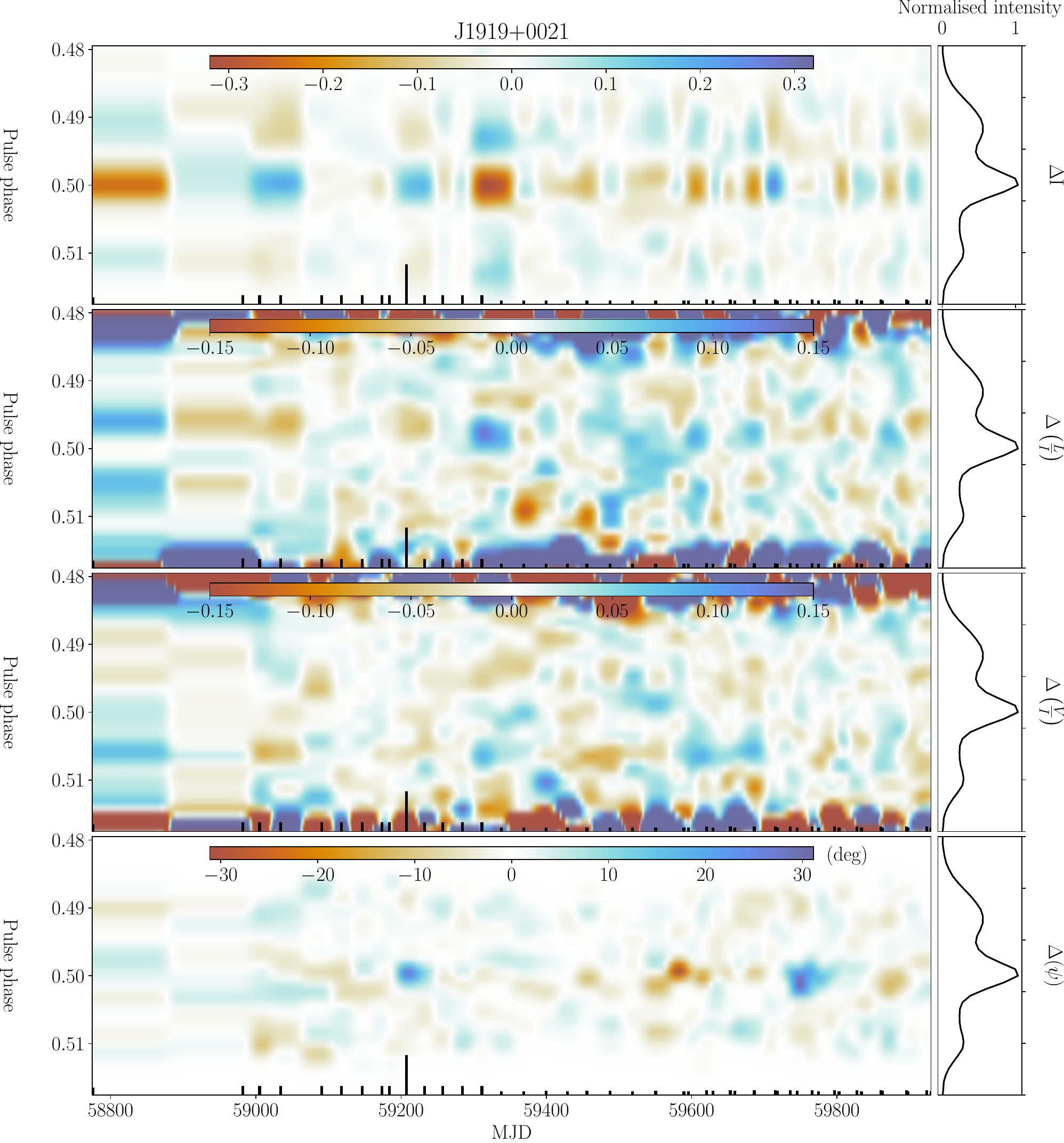}
\caption{This figure conveys the same information as that of Fig.\ref{fig:0729poldiff} but for PSR J1919$+$0021.}
\label{fig:1919poldiff} % I can do without the label too
\end{figure*}

\begin{figure*}
   \centering
\includegraphics[scale=0.56]{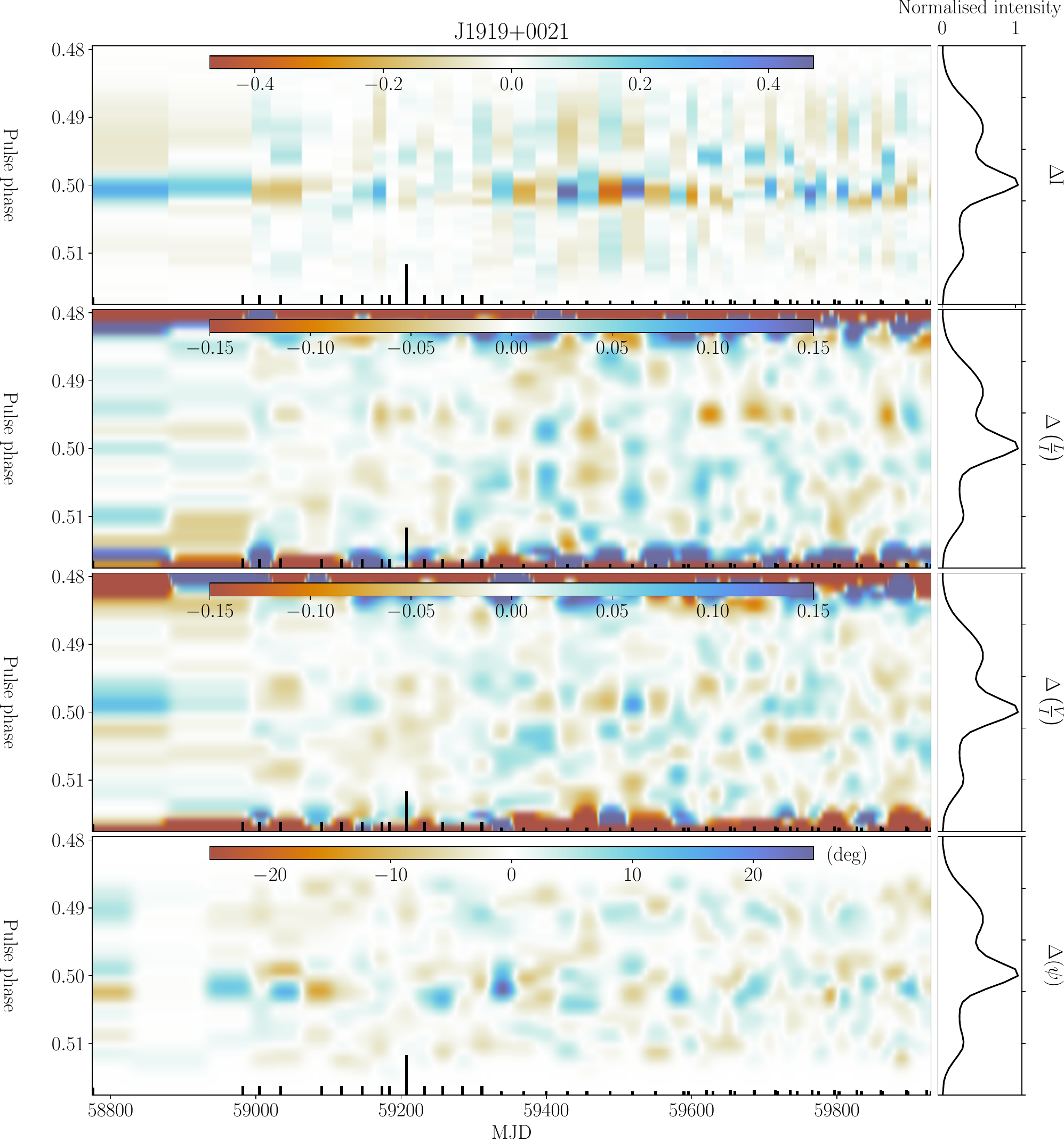}
\caption{The figure captures the same information as shown in Fig.~\ref{fig:0729poldiff_ipm} but for PSR J1919$+$0021.}
\label{fig:1919poldiff_ipm} % I can do without the label too
\end{figure*}

\begin{figure*}
   \centering
\includegraphics[scale=0.56]{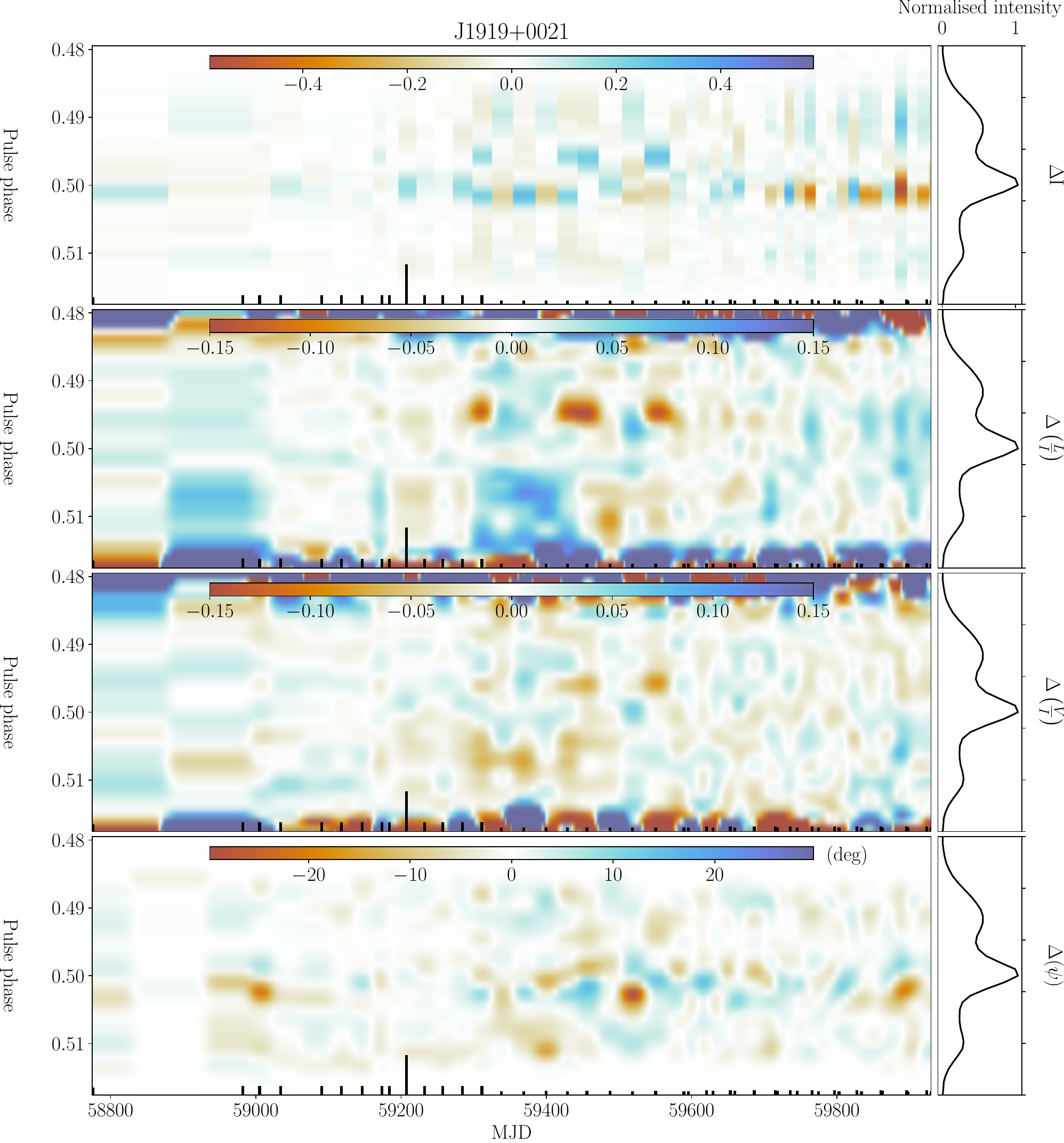}
\caption{The figure captures the same information as shown in Fig.~\ref{fig:0729poldiff_bm} but for PSR J1919$+$0021.}
\label{fig:1919poldiff_bm} % I can do without the label too
\end{figure*}

%%%%%%%%%%%%%%%%%%%%%%%%%%%%%%%%%%%%%%%%%%%%%%%%%%

% Don't change these lines
\bsp % typesetting comment
\label{lastpage}
\end{document}